\documentclass[twocolumn]{aastex631}

\expandafter\def\csname editcolor1\endcsname{magenta}
\expandafter\def\csname editcolor2\endcsname{blue}  
\expandafter\def\csname editcolor3\endcsname{violet} 

\usepackage{newtxtext,newtxmath}
\usepackage[T1]{fontenc}
\usepackage{graphicx}	
\usepackage{amsmath}	
\usepackage{float}
\usepackage{bookmark} 
\bookmarksetup{numbered, open}
\usepackage{enumitem}
\setlist[enumerate]{itemsep=0mm}
\usepackage{hyperref}
\usepackage{subfigure}
\usepackage{xspace}
\usepackage{xcolor} 
\usepackage{multirow}
\usepackage{tabularx}
\usepackage[normalem]{ulem}

\newcommand{\eg}{e.g.\xspace}

\newcommand{\Mch}{M$_{\rm Ch}$\xspace}
\newcommand{\Msun}{M$_{\odot}$\xspace}

\newcommand{\kms}{km~s\ensuremath{^{-1}}\xspace}

\newcommand{\gcm}{g cm\ensuremath{^{-3}}\xspace}

\newcommand{\Fefs}{\ensuremath{^{56}}Fe\xspace}
\newcommand{\Cofs}{\ensuremath{^{56}}Co\xspace}
\newcommand{\Nifs}{\ensuremath{^{56}}Ni\xspace}
\newcommand{\Nife}{\ensuremath{^{58}}Ni\xspace}

\newcommand{\sBV}{s\ensuremath{_{BV}}\xspace}

\newcommand{\mic}{\ensuremath{\mu}m\xspace}
\newcommand{\DmB}{\ensuremath{\Delta{\rm m_{15}(B)}}\xspace}
\newcommand{\DmV}{\ensuremath{\Delta{\rm m_{15}(V)}}\xspace}
\newcommand{\snia}{SN~Ia\xspace}
\newcommand{\sneia}{SNe~Ia\xspace}
\newcommand{\aefx}{SN~2021aefx\xspace}
\newcommand{\xkq}{SN~2022xkq\xspace}
\newcommand{\subl}{91bg-like\xspace}

\makeatletter

\newcommand\Autoref[1]{\@first@ref#1,@}
\def\@throw@dot#1.#2@{#1}
\def\@set@refname#1{
    \edef\@tmp{\getrefbykeydefault{#1}{anchor}{}}%
    \xdef\@tmp{\expandafter\@throw@dot\@tmp.@}%
    \ltx@IfUndefined{\@tmp autorefnameplural}%
         {\def\@refname{\@nameuse{\@tmp autorefname}s}}%
         {\def\@refname{\@nameuse{\@tmp autorefnameplural}}}%
}
\def\@first@ref#1,#2{%
  \ifx#2@\autoref{#1}\let\@nextref\@gobble
  \else%
    \@set@refname{#1}
    \@refname~\ref{#1}
    \let\@nextref\@next@ref
  \fi%
  \@nextref#2%
}
\def\@next@ref#1,#2{%
   \ifx#2@ and~\ref{#1}\let\@nextref\@gobble
   \else, \ref{#1}
   \fi%
   \@nextref#2%
}

\makeatother

\received{October 13, 2023}
\revised{November 3, 2023}
\accepted{November 6, 2023}

\submitjournal{ApJ}

\graphicspath{{./}{figs/}}

\begin{document}

\title{JWST MIRI/MRS Observations and Spectral Models of the Under-luminous Type Ia Supernova 2022xkq}

\correspondingauthor{James M DerKacy}
\email{jmderkacy@vt.edu}

\author[0000-0002-7566-6080]{J.~M.~DerKacy}
\affiliation{Department of Physics, Virginia Tech, Blacksburg, VA 24061, USA}

\author[0000-0002-5221-7557]{C.~Ashall}
\affiliation{Department of Physics, Virginia Tech, Blacksburg, VA 24061, USA}

\author[0000-0002-4338-6586]{P.~Hoeflich}
\affiliation{Department of Physics, Florida State University, 77 Chieftan Way, Tallahassee, FL 32306, USA}

\author[0000-0001-5393-1608]{E.~Baron}
\affiliation{Planetary Science Institute, 1700 East Fort Lowell Road, Suite 106, Tucson, AZ 85719-2395, USA}
\affiliation{Hamburger Sternwarte, Gojenbergsweg 112, D-21029 Hamburg, Germany}

\author[0000-0002-9301-5302]{M.~Shahbandeh}
\affiliation{Space Telescope Science Institute, 3700 San Martin Drive, Baltimore, MD 21218-2410, USA}

\author[0000-0003-4631-1149]{B.~J.~Shappee}
\affiliation{Institute for Astronomy, University of Hawai'i at Manoa, 2680 Woodlawn Dr., Hawai'i, HI 96822, USA }

\author[0000-0003-0123-0062]{J.~Andrews}
\affiliation{Gemini Observatory/NSF’s NOIRLab, 670 North A`ohoku Place, Hilo, HI 96720-2700, USA}

\author[0000-0003-1637-9679]{D.~Baade}
\affiliation{European Organization for Astronomical Research in the Southern Hemisphere (ESO), Karl-Schwarzschild-Str. 2, 85748 Garching b. M\"unchen, Germany}

\author{E.~F~Balangan}
\affiliation{Department of Physics, Florida State University, 77 Chieftan Way, Tallahassee, FL 32306, USA}

\author[0000-0002-4924-444X]{K.~A.~Bostroem}
\altaffiliation{LSSTC Catalyst Fellow}
\affiliation{Steward Observatory, University of Arizona, 933 North Cherry Avenue, Tucson, AZ 85721-0065, USA}

\author[0000-0001-6272-5507]{P.~J.~Brown}
\affiliation{George P. and Cynthia Woods Mitchell Institute for Fundamental Physics and Astronomy, Texas A\&M University, Department of Physics and Astronomy, College Station, TX 77843, USA}

\author[0000-0003-4625-6629]{C.~R.~Burns}
\affiliation{Observatories of the Carnegie Institution for Science, 813 Santa Barbara Street, Pasadena, CA 91101, USA}

\author[0000-0002-5380-0816]{A.~Burrow}
\affiliation{Homer L. Dodge Department of Physics and Astronomy, University of Oklahoma, 440 W. Brooks, Rm 100, Norman, OK 73019-2061, USA}

\author[0000-0001-7101-9831]{A.~Cikota}
\affiliation{Gemini Observatory/NSF's NOIRLab, Casilla 603, La Serena, Chile}

\author[0000-0001-6069-1139]{T.~de~Jaeger}
\affiliation{Institute for Astronomy, University of Hawai'i at Manoa, 2680 Woodlawn Dr., Hawai'i, HI 96822, USA }

\author[0000-0003-3429-7845]{A.~Do}
\affiliation{Institute for Astronomy, University of Hawai'i at Manoa, 2680 Woodlawn Dr., Hawai'i, HI 96822, USA }

\author[0000-0002-7937-6371]{Y.~Dong}
\affiliation{Department of Physics, University of California, 1 Shields Avenue, Davis, CA 95616-5270, USA}

\author[0000-0002-3827-4731]{I. Dominguez}
\affiliation{Universidad de Granada, 18071, Granada, Spain}

\author[0000-0003-2238-1572]{O.~Fox}
\affiliation{Space Telescope Science Institute, 3700 San Martin Drive, Baltimore, MD 21218-2410, USA}

\author[0000-0002-1296-6887]{L.~Galbany}
\affiliation{Institute of Space Sciences (ICE, CSIC), Campus UAB, Carrer de Can Magrans, s/n, E-08193 Barcelona, Spain}
\affiliation{Institut d’Estudis Espacials de Catalunya (IEEC), E-08034 Barcelona, Spain}

\author[0000-0003-2744-4755]{E. T. Hoang}
\affiliation{Department of Physics and Astronomy, University of California, Davis, 1 Shields Avenue, Davis, CA 95616-5270, USA}

\author[0000-0003-1039-2928]{E.~Y.~Hsiao}
\affiliation{Department of Physics, Florida State University, 77 Chieftan Way, Tallahassee, FL 32306, USA}

\author[0000-0003-0549-3281]{D. Janzen}
\affiliation{Department of Physics and Engineering Physics, University of Saskatchewan, 116 Science Place, Saskatoon, SK S7N 5E2, Canada}

\author[0000-0001-5754-4007]{J. E.\ Jencson}
\affiliation{Department of Physics and Astronomy, The Johns Hopkins University, 3400 North Charles Street, Baltimore, MD 21218, USA}

\author[0000-0002-6650-694X]{K.~Krisciunas}
\affiliation{George P. and Cynthia Woods Mitchell Institute for Fundamental Physics and Astronomy, Texas A\&M University, Department of Physics and Astronomy, College Station, TX 77843, USA}

\author[0000-0001-8367-7591]{S.~Kumar}
\affiliation{Department of Physics, Florida State University, 77 Chieftan Way, Tallahassee, FL 32306, USA}

\author[0000-0002-3900-1452]{J.~Lu}
\affiliation{Department of Physics, Florida State University, 77 Chieftan Way, Tallahassee, FL 32306, USA}

\author[0000-0001-9589-3793]{M. Lundquist}
\affiliation{W.~M.~Keck Observatory, 65-1120 M\=amalahoa Highway, Kamuela, HI 96743-8431, USA}

\author[0000-0001-5888-2542]{T.~B.~Mera~Evans}
\affiliation{Department of Physics, Florida State University, 77 Chieftan Way, Tallahassee, FL 32306, USA}

\author[0000-0003-0733-7215]{J.~R.~Maund}
\affiliation{Department of Physics and Astronomy, University of Sheffield, Hicks Building, Hounsfield Road, Sheffield S3 7RH, U.K.}

\author[0000-0001-6876-8284]{P.~Mazzali}
\affiliation{Astrophysics Research Institute, Liverpool John Moores University, UK}
\affiliation{Max-Planck Institute for Astrophysics, Garching, Germany}

\author[0000-0001-7186-105X]{K.~Medler}
\affiliation{Astrophysics Research Institute, Liverpool John Moores University, UK}

\author[0000-0002-7015-3446]{N.~E.\ Meza Retamal}
\affiliation{Department of Physics and Astronomy, University of California, Davis, 1 Shields Avenue, Davis, CA 95616-5270, USA}

\author[0000-0003-2535-3091]{N.~Morrell}
\affiliation{Las Campanas Observatory, Carnegie Observatories, Casilla 601, La Serena, Chile}
	
\author[0000-0002-0537-3573]{F.~Patat}
\affiliation{European Organization for Astronomical Research in the Southern Hemisphere (ESO), Karl-Schwarzschild-Str. 2, 85748 Garching b. M\"unchen, Germany}

\author[0000-0002-0744-0047]{J. Pearson}
\affiliation{Steward Observatory, University of Arizona, 933 North Cherry Avenue, Tucson, AZ 85721-0065, USA}

\author[0000-0003-2734-0796]{M.~M.~Phillips}
\affiliation{Las Campanas Observatory, Carnegie Observatories, Casilla 601, La Serena, Chile}

\author[0000-0002-4022-1874]{M. Shrestha}
\affiliation{Steward Observatory, University of Arizona, 933 North Cherry Avenue, Tucson, AZ 85721-0065, USA}

\author[0000-0001-5570-6666]{S.~Stangl}
\affiliation{Homer L. Dodge Department of Physics and Astronomy, University of Oklahoma, 440 W. Brooks, Rm 100, Norman, OK 73019-2061, USA} 

\author[0000-0003-0763-6004]{C.~P.~Stevens}
\affiliation{Department of Physics, Virginia Tech, Blacksburg, VA 24061, USA}

\author[0000-0002-5571-1833]{M.~D.~Stritzinger}
\affiliation{Department of Physics and Astronomy, Aarhus University, Ny Munkegade 120, DK-8000 Aarhus C, Denmark}

\author[0000-0002-8102-181X]{N.~B.~Suntzeff}
\affiliation{George P. and Cynthia Woods Mitchell Institute for Fundamental Physics and Astronomy, Texas A\&M University, Department of Physics and Astronomy, College Station, TX 77843, USA}

\author[0000-0002-0036-9292]{C.~M.~Telesco}
\affiliation{Department of Astronomy, University of Florida, Gainesville, FL 32611 USA}

\author[0000-0002-2471-8442]{M.~A.~Tucker}
\altaffiliation{CCAPP Fellow}
\affiliation{Center for Cosmology and AstroParticle Physics, The Ohio State University, 191 W. Woodruff Ave., Columbus, OH 43210, USA}

\author[0000-0001-8818-0795]{S.~Valenti}
\affiliation{Department of Physics, University of California, 1 Shields Avenue, Davis, CA 95616-5270, USA}

\author[0000-0001-7092-9374]{L.~Wang}
\affiliation{Department of Physics and Astronomy, Texas A\&M University, College Station, TX 77843, USA}

\author[0000-0002-6535-8500]{Y.~Yang}
\altaffiliation{Bengier-Winslow-Robertson Postdoctoral Fellow}
\affiliation{Department of Astronomy, University of California, Berkeley, CA 94720-3411, USA}

\begin{abstract}
We present a \textit{JWST} mid-infrared spectrum of the
under-luminous Type Ia Supernova (\snia) 2022xkq, obtained
with the medium-resolution spectrometer on the Mid-Infrared
Instrument (MIRI) $\sim130$ days post-explosion. We identify
the first MIR lines beyond 14 \mic in \snia observations.
We find features unique to under-luminous \sneia, including:
isolated emission of stable Ni, strong blends of [\ion{Ti}{2}],
and large ratios of singly ionized to doubly ionized species
in both [Ar] and [Co]. Comparisons to normal-luminosity \sneia
spectra at similar phases show a tentative trend between the
width of the [\ion{Co}{3}] 11.888 \mic feature and the SN light
curve shape. Using non-LTE-multi-dimensional radiation hydro
simulations and the observed electron capture elements we
constrain the mass of the exploding white dwarf. The
best-fitting model shows that \xkq is consistent with an
off-center delayed-detonation explosion of a near-Chandrasekhar
mass WD (M$_{\rm ej}$ $\approx 1.37$~\Msun) of high-central density
($\rho_c \geq 2.0\times10^{9}$ \gcm) seen equator on, which
produced M(\Nifs) $= 0.324$~\Msun and
M(\Nife) $\geq 0.06$~\Msun. The observed line widths are
consistent with the overall abundance distribution; and the
narrow stable Ni lines indicate little to no mixing in the
central regions, favoring central ignition of sub-sonic carbon
burning followed by an off-center DDT beginning at a single point.
Additional observations may further constrain the physics revealing
the presence of additional species including Cr and Mn. Our work
demonstrates the power of using the full coverage of MIRI in
combination with detailed modeling to elucidate the physics of
\sneia at a level not previously possible.
\end{abstract}


\section{Introduction} \label{sec:intro}

The use of Type Ia supernovae (\sneia) as cosmological distance indicators 
belies their diverse nature. Through the luminosity-width relation 
\citep{Phillips1993}, both bright, slow-declining and dimmer, fast-declining 
\snia can be standardized for cosmological analyses, which have revealed the 
accelerating expansion of the universe \citep{Riess1998,Perlmutter1999}. 
Spectroscopic diversity among \sneia has also been observed, with multiple schemes 
to understand this diversity, based solely on spectroscopic information 
\citep{Branch2006, Wang2009}, or combined photometric and spectroscopic 
measurements \citep{Benetti2005}, having been developed.
Evidence suggests that much of the observed diversity originates from 
differences in the radioactive Ni mass (e.g. \Nifs) 
\citep{Nugent1995,Pinto2000a,Pinto2000b}, resulting in several sub-types of 
\sneia with unique photometric and spectroscopic properties including 
(but not limited to):
91T-like objects \citep{Filippenko1992b,Phillips1992,Phillips2022,Yang2022},
91bg-like objects \citep{Filippenko1992a,Leibundgut1993,Galbany2019,Hoogendam2022}
02cx-like (aka SN Iax) \citep{Li2003,Foley2013,Jha2017},
and 03fg-like (formerly known as super-Chandrasekhar SNe Ia) 
\citep{Howell2006,Hicken2007,Scalzo2010,Hsiao2020,Ashall2021}.
 
A major outstanding question is whether individual \sneia subgroups (e.g 
under-luminous or 91bg-like objects) arise from different progenitor systems 
and/or explosion mechanisms (or particular combinations thereof) than other 
subgroups. \sneia are known to originate from the 
thermonuclear explosion of a carbon-oxygen (C/O) white dwarf (WD) in a multi-star 
system \citep{Hoyle1960,Bloom2012}. The range of possible progenitor systems 
includes: the single degenerate (SD) scenario in which the companion is a 
main-sequence star or an evolved, non-degenerate companion such as a red giant 
or He-star \citep{Whelan1973}; the double degenerate scenario (DD) in which the 
companion is also a WD \citep{Iben1984,Webbink1984}, or a triple system
in which at least two of the bodies are C/O WDs 
\citep{Thompson2011,Kushnir2013,Shappee2013,Pejcha2013}.

Within the context of the above progenitor scenarios, multiple explosion
mechanisms exist \citep[\eg][]{Benz1990,Hoeflich1996,Rosswog2009,Pakmor2012,Pakmor2013,Kushnir2013,Soker2013,Garcia-Berro2017,Lu21}.  
Currently, two of the leading explosion models are the detonation of a sub-\Mch 
WD or the explosion of a near-\Mch WD. In sub-\Mch explosions, He on the WD surface 
(which may be accreted from a degenerate or non-degenerate companion) detonates, 
driving a shockwave into the WD which triggers a second, interior
detonations which disrupts the whole WD 
\citep{Nomoto1984,Woosley1994,Livne1995,Hoeflich1996,Shen2018,Polin2019,Boos2021}.
In contrast, for near-\Mch explosions, H, He, and/or C material is accreted from a
companion star (which may again be degenerate or non-degenerate) onto the surface of 
the WD, until compressional heating triggers an explosion near the WD center
\citep{1976Ap&SS..39L..37N,Iben1984,Hoeflich1996,Diamond_etal_2018}. The flame front may propagate
as either a deflagration, detonation, or both via a deflagration-to-detonation 
transition (\citealp[DDT;][]{1991A&A...245..114K,Hoeflich1996,Gamezo2003,Poludnenko2019}). 

Nebular-phase spectra of \sneia in the mid-infrared (MIR) are key to distinguishing
between different explosion models. As the location of the photosphere is 
wavelength dependent, different spectral lines are revealed in the MIR 
\citep{Meikle1993,Hoeflich_etal_2002,Wilk2018} which are better at constraining the physics 
of SN explosions than their optical counterparts \citep{Diamond2015,Diamond_etal_2018}. 
For example, the amount of stable Ni (e.g. \Nife) and other iron group elements (IGEs) 
serve as a direct indicator of the central density ($\rho_{c}$) at the time of explosion
\citep{Hoeflich1996,Seitenzahl2017,Blondin2022}.
Optical lines from stable Ni are only identified as weak components of heavily blended 
features \citep{Maguire2018,Mazzali2020} and the 1.94~\mic [\ion{Ni}{2}] NIR line 
lies directly adjacent to (and often overlapping) a telluric region; making its 
identification difficult and sensitive to reduction methodology in high-quality
data \citep{Friesen2014,Dhawan2018,Diamond_etal_2018,Hoeflich2021}.
However, in the MIR, stable Ni can be seen even in low S/N observations 
\citep{Gerardy2007,Telesco2015}, with multiple strong lines in different ionization
stages seen in high S/N JWST observations \citep{Kwok_etal_2023_21aefx,DerKacy2023}.

To date, the complete published sample of MIR nebular phase spectra of \sneia 
contains nine spectra of five normal luminosity objects\footnote{We note that 
\citet{Leloudas2009} found SN~2003hv was a normal-luminosity SN Ia that obeys 
the Phillips relation, with M(B) $= -19.13$~mag and $\DmB = 1.61 \pm 0.02$~mag.}, 
and one spectrum of a 2003fg-like SNe Ia. This data includes single epoch observations 
of SNe 2003hv, 2005df \citep{Gerardy2007}, and 2006ce \citep{Kwok_etal_2023_21aefx} with 
the Spitzer Space Telescope; plus a {\it JWST} observation of the 03fg-like SN 2022pul
\citep{Siebert23,Kwok_etal_2023_23pul}. Two time series data sets exist; four observations 
of SN 2014J with CarnariCam on the Gran Telescopio de Canarias (GTC) \citep{Telesco2015}, 
and two observations of SN 2021aefx \citep{Kwok_etal_2023_21aefx,DerKacy2023} with the Low Resolution 
Spectrograph (\citealp[LRS;][]{Kendrew2015,Rigby2023}) mode of the Mid-Infrared 
Instrument (\citealp[MIRI;][and references therein]{Rieke2015}). All of these published 
MIR spectra have resolutions of R~$\lesssim$~200. With an average resolution 
R~$\sim 2700$ from 5-25~\mic, observations of \sneia with JWST/MIRI using the 
Medium Resolution Spectrograph (\citealp[MRS;][]{Rieke2015,Wells2015,Argyriou2023}) 
can make more precise measurements of the line velocities and line profiles,
and provide coverage in the 14-25~\mic region unobtainable with MIRI/LRS. 
To date, no spectral observations of \sneia from MIRI/MRS have been published 
in literature. 

In this work, we present a medium-resolution nebular-phase spectrum of the under-luminous 
\xkq taken $\sim$~130 days after explosion with MIRI/MRS. \autoref{sec:obs} describes 
our observations and reduction procedures. Line identifications are made in 
\autoref{sec:line_ids}, and important aspects of the spectrum are characterized 
in \autoref{sec:characterization}, including discussions of the differences in 
the MIR spectra of \xkq versus normal-luminosity \sneia observed at similar epochs. 
Interpretation of our observations with the use of radiative hydrodynamic models 
are presented in \autoref{sec:models}. We summarize our results in \autoref{sec:summary}
and discuss the implications of this work in \autoref{sec:disc}. We conclude
in \autoref{sec:conc}.

\section{Observations} \label{sec:obs}

\begin{deluxetable}{lcc}
  \tablecaption{Properties of SN~2022xkq and NGC 1784 \label{tab:props}.} 
  \tablehead{\colhead{Parameter} & \colhead{Value} & \colhead{Source}}
  \startdata
    \multicolumn{3}{c}{SN 2022xkq} \\
    \hline
    R.A. & 05$^{h}$05$^{m}$23$^{s}$.70 & (1) \\
    Dec. & $-$11\arcdeg52\arcmin56\arcsec.13 & (1) \\
    Last Non-Detection (MJD) & 59864.25 & (1) \\
    Discovery (MJD) & 59865.37 & (1) \\
    $t_{\rm exp}$ (MJD) & $59865.0 \pm 0.3$ & (2) \\
    $t_{\rm max}$ (MJD) & $59879.03 \pm 0.34$ & (2) \\
    $M_{B,{\rm max}}$ (mag) & $-18.01 \pm 0.15$ & (2) \\
    \DmB (mag) & $1.65 \pm 0.03$ & (2) \\
    \sBV & $0.63 \pm 0.03$ & (2) \\
    $E(B-V)_{MW}$ (mag) & $0.116 \pm 0.002$ & (3) \\ 
    \hline
    \multicolumn{3}{c}{NGC 1784} \\
    \hline
    R.A. & 05$^{h}$05${^m}$27$^{s}$.10 & (4) \\
    Dec. & $-$11\arcdeg52\arcmin17\arcsec.50 & (4) \\
    Morphology & SB(r)c & (4) \\
    $v_{\rm helio}$ (\kms) & $2291 \pm 11$ & (5) \\
    $v_{\rm rot}$ (\kms) & $41 \pm 20$ & (4) \\
    $z$ & 0.0077 & (5) \\
    $\mu$ & $32.46 \pm 0.15$ & (2) \\
  \enddata
  \tablecomments{\citet{Pearson_etal_2023} find negligible host 
  extinction at the site of \xkq.}
  \tablerefs{(1) \citet{Janzen2022}, (2) \citet{Pearson_etal_2023}, 
  (3) \citet{Schlafly2011}, (4) \citet{Ratay2004}, 
  (5) \citet{Koribalski2004}, (6) NED\footnote{http://ned.ipac.caltech.edu/}}
\end{deluxetable}

\subsection{Early Observations of SN 2022xkq}

\xkq was discovered on 2022 October 13.4 UT (MJD=59865.4) by the Distance Less 
Than 40 Mpc Survey (\citealp[DLT40;][]{DLT40}), with a last non-detection
on 2022 October 12.3 UT (MJD$=$59864.3). Initially classified as a Type I 
\citep{Chen2022a} or Type Ic \citep{Hosseinzadeh2022} supernova, further 
observations revealed \xkq to be a \subl \snia \citep{Chen2022b}. 

A detailed analysis of \xkq can be found in \citet{Pearson_etal_2023}. 
We briefly summarize some of their key findings, with important measurements of 
\xkq and its host NGC~1784 shown in \autoref{tab:props}. \xkq is a transitional 
SNe~Ia similar to SNe~1986G \citep{Phillips87,Ashall16}, 2005ke \citep{Patat_etal_2012}, 
2007on, 2011iv \citep{gall2018,Ashall18}, and 2012ij \citep{Li2022}. 
\texttt{SNooPy} \citep{Burns2011,Burns2014} fits to the early light 
curve show that \xkq has \DmB~$=1.65 \pm 0.03$~mag and \sBV~$= 0.63 \pm 0.03$.
Using Arnett's Rule \citep{Arnett1982,Arnett1985}, the
pseudo-bolometric light curves, and the peak bolometric luminosity,
\citet{Pearson_etal_2023} derived a \Nifs mass of $0.22 \pm 0.03$~\Msun in \xkq.
Early photometry revealed \xkq to be red in color, with a flux excess in the 
redder bands relative to power-law fits in the first few days after explosion. 
In addition, densely sampled spectra showed strong, persistent C lines 
(particularly \ion{C}{1} $\lambda 1.069$~\mic), similar to SN~1999by.
Radio observations taken on 2022 October 15.9 with the Australia Telescope 
Compact Array (ATCA) found no radio emission at the site of \xkq, placing 
3-$\sigma$ upper limits of 0.07~mJy at 5.5~GHz and 0.04~mJy at 9.0~GHz
\citep{Ryder2022}.

The host galaxy of \xkq is NGC~1784, a barred spiral
SB(r)c galaxy at a distance of roughly 31~Mpc \citep{deVaucouleurs1991,Pearson_etal_2023}. 
\xkq is located near the edge of a weak spiral arm, 49\arcsec.91 W and
38\arcsec.63 S of the center of its host galaxy (see Figure 1 in
\citealp{Pearson_etal_2023}). Underluminous \sneia are typically found
in older stellar populations, and thus more likely to occur in elliptical
galaxies \citep{Hamuy1996,Branch1996,Howell2001,Sullivan2010,Ashall2016,Nugent2023};
although old stellar populations can also be found in late-type spirals
such as NGC~1784. Interestingly, the under-luminous SN~1999by was also
discovered in a late-type spiral \citep{Howell2001,Hoeflich2002,Garnavich2004},
and shows some properties similar to those of \xkq \citep{Pearson_etal_2023}.
NGC~1784 has a systemic recessional velocity of
$2291 \pm 11$~\kms \citep{Koribalski2004}, when corrected for the influences
of the Virgo Cluster, the Great Attractor, and the Shapley Supercluster 
\citep{Mould2000}. Detailed \ion{H}{1} mapping of NGC 1784 reveals
an implied rotational velocity of $41 \pm 20$~kms at the site of \xkq
\citep{Ratay2004}. Throughout this work, spectra have been corrected for 
the combined line-of-sight velocity (recessional plus rotational) of
$2332 \pm 23$~\kms ($z = 0.007773 \pm 0.000077$). The \ion{H}{1} maps
also reveal a warped disk

\subsection{JWST Observations}

\xkq was observed on 2023 February 19.0 by JWST with the Mid-Infrared 
Instrument (\citealp[MIRI;][]{Rieke2015}) and Medium Resolution Spectrograph (MRS)
as part of program JWST-GO-2114 (PI: C. Ashall). Full details of our observational
set-up can be found in \autoref{tab:obs}. Observations began at 2023 February 19 00:30:30 UT
(MJD$=$59994.02) and ended at 2023 February 19 04:06:28 UT (MJD$=$59994.17). Throughout 
this work, we adopt the midpoint of the observations (MJD$=$59994.1), as the 
epoch of our observations, equivalent to 114.2 rest-frame days after $B$-band maximum 
(MJD$=$59879.03) and 128.1 rest-frame days after explosion (MJD$=$59865.0) \citep{Pearson_etal_2023}.

Based on the in-flight performance report from \citet{Argyriou2023}, MIRI/MRS
is accurate to 2-27 \kms depending on wavelength, and has a spectrophotometric
precision of ~5.6\%. The corresponding values for MIRI/LRS observation are
0.05 - 0.02~\mic; corresponding to errors of $\sim$1400-500~\kms, again dependent 
on wavelength \citep{DerKacy2023}, and a spectrophotometric precision of 
$\sim$2-5\% \citep{Kwok_etal_2023_21aefx}.

\begin{deluxetable}{lccc}
  \tablecaption{JWST MIRI/MRS Observation Details. \label{tab:obs}} 
  \tablehead{\colhead{Parameter} & \colhead{Value}& \colhead{Value}& \colhead{Value}}
  \startdata
    \multicolumn{4}{c}{MIRI Acquisition Image}  \\
    \hline
    Filter & F1000W & \nodata & \nodata \\
    Acq. Groups/Exp & 10 & \nodata & \nodata \\
    Exp Time (s) & 28 & \nodata & \nodata \\
    Readout Pattern & FAST & \nodata & \nodata \\
    \hline
    \multicolumn{4}{c}{MIRI/MRS Spectra}  \\
    \hline
    Wavelength range & Short & Medium & Long\\
    Groups per Integration & 36 & 36 & 36 \\
    Integrations per Exp. & 1 & 1 & 1 \\
    Exposures per Dither & 1 & 1 & 1 \\
    Total Dithers & 4 & 4 & 4 \\
    Exp Time (s) & 3440.148 & 3440.148 & 3440.148 \\
    Readout Pattern & SLOWR1 & SLOWR1 & SLOWR1 \\
    \hline
    $T_\text{obs}$ [MJD] & \multicolumn{3}{c}{59994.1} \\
    Epoch\tablenotemark{a} [days] & \multicolumn{3}{c}{114.2} \\   
  \enddata
  \tablecomments{~\tablenotemark{a}~Rest frame days relative to 
  time of $B$-band maximum \citep[MJD$=$59879.03;][]{Pearson_etal_2023}.}
\end{deluxetable}

\subsection{JWST Data Reduction}

\begin{figure*}
    \centering
    \includegraphics[width=0.99\textwidth]{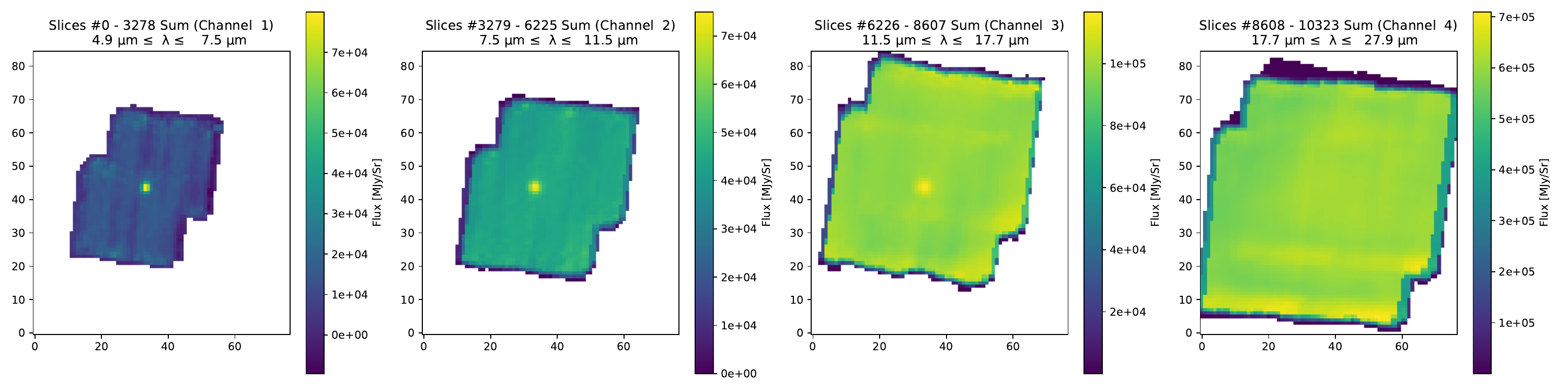}
     \includegraphics[width=0.99\textwidth]{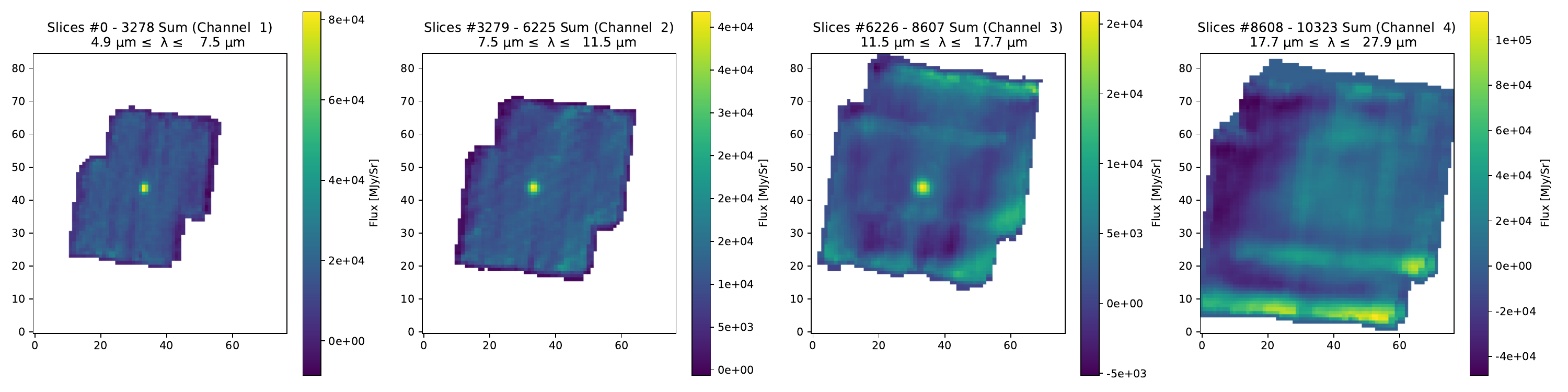}
    \caption{\textit{Top row:} MIRI/MRS cube of \xkq before background 
    subtraction, split into the four MIRI/MRS channels. The image shown
    for each channel is the collapsed sum of all its slices.
    \textit{Bottom row:} MIRI/MRS cube of \xkq after background 
    subtraction. We note that although the SN cannot be seen in the stacked 
    Channel 4 cube, it is visible in specific wavelength slices, as can be 
    seen from the spectra in \autoref{fig:22xkq_MRS}.}
    \label{fig:cubes}
\end{figure*}

\begin{figure*}
    \centering
    \includegraphics[width=\textwidth]{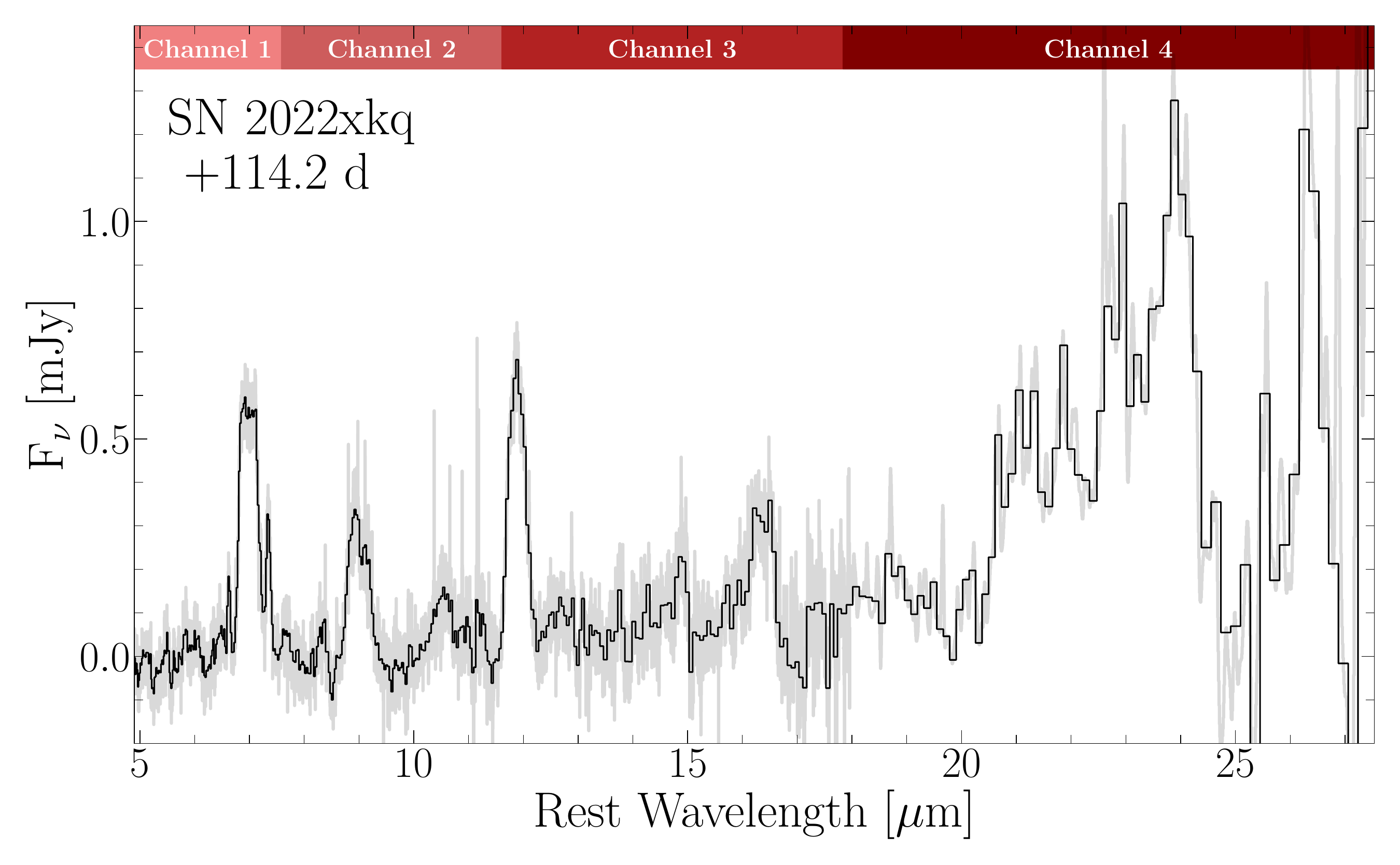}
    \caption{JWST MIRI/MRS observations of \xkq at $+114.2$~days relative to 
    $B$-band maximum, corrected for Milky Way extinction. Raw data points are
    shown in grey, with the \texttt{Spextractor} \citep{Burrow2020} smoothed 
    spectrum in black. The channel labels at along the top axis are also 
    corrected to the rest-frame of the SN.} 
    \label{fig:22xkq_MRS}
\end{figure*}

The data were reduced using a custom-built pipeline\footnote{\url{https://github.com/shahbandeh/MIRI_MRS}} 
designed to extract point source observations with a highly varying and/or 
complex backgrounds from MRS data cubes. This pipeline is discussed in detail 
in Appendix A of Shahbandeh et al. 2023 (in prep). In short, the pipeline 
creates a master background based upon 20 different positions in the data cube. 
This master background is then subtracted from the whole data cube before the 
aperture photometry is performed along the data cube, using the \textsc{Extract1dStep} 
in stage 3 of the \textit{JWST} reduction pipeline. The reduction shown
here utilized version 1.12.1 of the JWST Calibration pipeline \citep{Bushouse2023}
and Calibration Reference Data System files version 11.17.1. The resulting MIRI/MRS cube 
is shown in \autoref{fig:cubes} and the spectrum is shown in full in \autoref{fig:22xkq_MRS}. 
The raw data associated with this reduction can be found at
\dataset[DOI:10.17909/kfvh-wb96]{\doi{10.17909/kfvh-wb96}}.
The extracted spectrum has been smoothed with \texttt{spextractor} \citep{Burrow2020} 
channel-by-channel to properly account for the differences in resolution across 
the full MIRI/MRS wavelength coverage. 

\section{Line Identifications} \label{sec:line_ids}

\begin{figure*}
    \centering
    \includegraphics[width=0.85\textwidth]{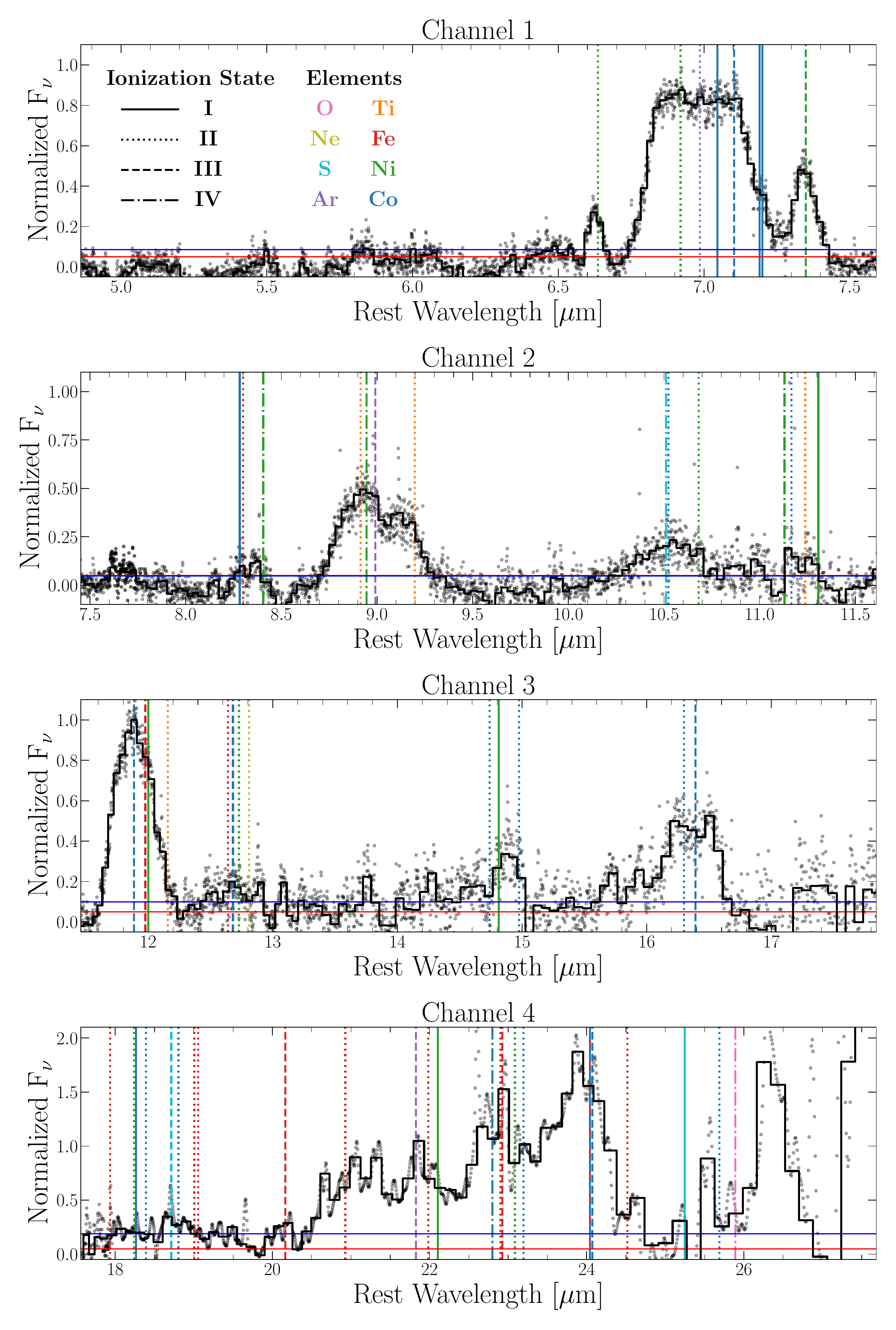}
    \caption{Spectrum of \xkq with line identifications based on known lines
    from previous MIR \sneia observations and models of under-luminous \snia models.
    The horizontal lines denote the two threshold values of 5\% of the overall flux 
    (red) and 10\% of maximum flux in each individual channel (blue); the larger of 
    which must be exceeded for a feature to have line identifications.}
    \label{fig:line_ids}
\end{figure*}

To identify the lines present in our $+114.2$~day spectrum of \xkq, we 
are guided by lines previously identified in other MIR spectra of 
normal-luminosity \sneia, as well as using the full non-LTE radiation-hydrodynamic 
models of the under-luminous \xkq presented in \autoref{sec:models}.
The line identifications are presented by MIRI/MRS channel 
in \autoref{fig:line_ids}, with individual lines specified in \autoref{tab:line_ids}. 
We only identify features strong enough to exceed a flux level of 5\% of the maximum 
flux of the smoothed spectrum at $\lambda < 20$~\mic, or 10\% of the maximum flux in
the channel the feature is found, whichever is greater. Weaker features at or slightly 
exceeding the threshold level are considered tentative identifications.

At this phase, the ejecta is not yet fully nebular; meaning there is
a combination of both permitted lines, forbidden emission, and an
underlying continuum to the spectrum, creating a pseudo-photosphere 
(see \autoref{sec:models}), seen previously in the analysis of both 
SNe~2005df and 2014J at similar epochs 
\citep{Gerardy2007,Diamond2015,Telesco2015}. This results in many of 
the features arising from a combination of blends as well as radiative 
transfer effects, making the identification of individual components 
difficult. Below we discuss clear feature detections, but note that 
many weak lines have no cross sections. Work to determine these 
missing cross-sections using the MRS spectra of \aefx in 
combination with detailed non-LTE simulations is in progress 
(Ashall et al. in prep).

\subsection{Channel 1 (4.9--7.65~\mic)} \label{sec:channel1}
Channel 1 is dominated by a large, complex, seemingly box-shaped profile near 
$\sim 7.0$~\mic, with more isolated peaks on either side. These isolated peaks 
are easily identified as [\ion{Ni}{2}]~6.636~\mic, and [\ion{Ni}{3}]~7.349~\mic
lines arising from stable \Nife present in the ejecta. Both peaks appear  
blue-shifted relative to their rest wavelengths and are further explored in 
\autoref{sec:characterization}.

As seen in the top panel of \autoref{fig:line_ids}, the dominant box-like
feature reveals several peaks. This feature is primarily due to the
[\ion{Ar}{2}]~6.985~\mic line seen in other \sneia spectra, with additional
weak peaks across the box-like profile matching  [\ion{Ni}{2}]~6.920~\mic, 
[\ion{Co}{1}]~7.045~\mic, and [\ion{Co}{3}]~7.103~\mic lines. A shoulder on 
the red side of this feature likely arises from a blend of 
[\ion{Co}{1}]~7.191~\mic and 7.200~\mic lines. The [\ion{Ni}{3}]~7.349~\mic 
line also shows a weak shoulder in its red wing, which is likely attributable 
to the quasi-continuum \citep{fesen2014,Hoeflich2023}.

No other strong features are present in the channel. Hints of weak, broad 
features appear in the $\sim$5--6.5~\mic region, but none are well-matched
by known spectral lines. The broad feature spanning $\sim$5.8--6.1~\mic may arise in part 
from a blend of [\ion{Ni}{1}]~5.893 and [\ion{Ni}{2}]~5.953~\mic, lines, but 
it is unclear how much of the flux is due to the emission from forbidden Ni 
lines. As such, we leave this feature unidentified in both 
\autoref{fig:line_ids} and \autoref{tab:line_ids}.

\subsection{Channel 2 (7.51--11.7~\mic)}

Similar to Channel 1, Channel 2 is also dominated by a single prominent 
feature, with additional weak features spread across the channel. The strongest 
feature is the blended feature near $9.0$~\mic. This blend shows a strong blue 
peak at $\sim$8.9~\mic, likely from [\ion{Ni}{4}]~8.945~\mic with some 
contribution of [\ion{Ti}{2}]~8.915~\mic. This peak is blended with the 
central line of the feature, [\ion{Ar}{3}]~8.991~\mic. A secondary peak in the 
red at roughly $\sim$9.1~\mic could be produced from [\ion{Ti}{2}]~9.197~\mic, 
however, due to the increased noise in this part of the feature, it may be more 
shoulder-like in appearance with stronger contributions from additional iron 
group elements and/or the quasi-continuum. Due to the combination of this 
blending and asymmetry in the feature the exact profile of the 
[\ion{Ar}{3}]~8.991~\mic is difficult to determine requiring detailed models 
(e.g. \citealp{Penney2014,Hoeflich2021}), like in \aefx
\citep{Kwok_etal_2023_21aefx,DerKacy2023}. 

We tentatively identify some weaker features in Channel 2. The double-peaked 
feature from  $\sim$8.2--8.4~\mic is due to [\ion{Co}{1}]~8.283~\mic,
[\ion{Fe}{2}]~8.299~\mic, and [\ion{Ni}{4}]~8.405~\mic. A broad but weak blend 
is seen extending from roughly  $\sim$10--11.4~\mic, with contributions 
likely originating from [\ion{S}{4}]~10.511~\mic, [\ion{Ti}{2}]~10.511~\mic, 
[\ion{Co}{2}]~10.523~\mic, and [\ion{N}{2}]~10.682~\mic. A narrower, but equally
weak feature exists between $\sim$11.1--11.4~\mic, potentially arising from
[\ion{Ni}{4}]~11.130~\mic, [\ion{Co}{2}]~11.167~\mic, [\ion{Ti}{2}]~11.238~\mic,
and [\ion{Ni}{1}]~11.307~\mic lines.

Channel 2 terminates at roughly 11.7~\mic, within the blue edge of the feature 
dominated by the [\ion{Co}{3}]~11.888~\mic resonance line. As the bulk
of the feature is observed within Channel 3, we discuss the feature in the 
following section.

\begin{deluxetable}{lc|lc|lc}
  \tablecaption{Line Identifications \label{tab:line_ids}}
  \tablehead{\colhead{Line} & \colhead{$\lambda$ (\mic)} & 
  \colhead{Line} & \colhead{$\lambda$ (\mic)} & 
  \colhead{Line} & \colhead{$\lambda$ (\mic)}}
  \startdata
    \hline
    \multicolumn{6}{c}{Lines in Strong Features} \\
    \hline
    {[\ion{Ni}{2}]} & 6.636 & {[\ion{Ni}{4}]}$^{\dagger}$ & 8.945 & {[\ion{Co}{2}]}$^{\dagger}$ & 16.155 \\
    {[\ion{Ni}{2}]}$^{\dagger}$ & 6.920 & {[\ion{Ar}{2}]}$^{\dagger}$ & 8.991 & {[\ion{Co}{2}]}$^{\dagger}$ & 16.299 \\
    {[\ion{Ar}{2}]}$^{\dagger}$ & 6.985 & {[\ion{Ti}{2}]}$^{\dagger}$ & 9.197 & {[\ion{Co}{3}]}$^{\dagger}$ & 16.391 \\
    {[\ion{Co}{1}]}$^{\dagger}$ & 7.045 & {[\ion{Co}{3}]}$^{\dagger}$ & 11.888 & {[\ion{Fe}{2}]}$^{\dagger}$ & 17.936 \\
    {[\ion{Co}{3}]}$^{\dagger}$ & 7.103 & {[\ion{Fe}{3}]}$^{\dagger}$ & 11.978 & {[\ion{Ni}{2}]}$^{\dagger}$ & 18.241 \\
    {[\ion{Co}{1}]}$^{\dagger}$ & 7.191 & {[\ion{Ni}{1}]}$^{\dagger}$ & 12.001 & {[\ion{Co}{1}]}$^{\dagger}$ & 18.265 \\
    {[\ion{Co}{1}]}$^{\dagger}$ & 7.200 & {[\ion{Ni}{1}]}$^{\dagger}$ & 14.814 & {[\ion{Co}{2}]}$^{\dagger}$ & 18.390 \\
    {[\ion{Ni}{3}]} & 7.349 & {[\ion{Co}{2}]}$^{\dagger}$ & 14.977 & {[\ion{S}{3}]}$^{\dagger}$ & 18.713 \\
    {[\ion{Ti}{2}]}$^{\dagger}$ & 8.915 & {[\ion{Co}{2}]}$^{\dagger}$ & 16.152 & {[\ion{Co}{2}]}$^{\dagger}$ & 18.804 \\
    \hline
    \multicolumn{6}{c}{Lines in Weak Features} \\
    \hline
    {[\ion{Co}{2}]}$^{\dagger}$ & 8.283 & {[\ion{Fe}{3}]}$^{\dagger}$ & 12.642 & {[\ion{Fe}{2}]}$^{\dagger}$ & 22.902 \\
    {[\ion{Fe}{2}]}$^{\dagger}$ & 8.299 & {[\ion{Co}{3}]}$^{\dagger}$ & 12.681 & {[\ion{Fe}{3}]}$^{\dagger}$ & 22.925 \\
    {[\ion{Ni}{2}]}$^{\dagger}$ & 8.405 & {[\ion{Ni}{2}]}$^{\dagger}$ & 12.729 & {[\ion{Ni}{2}]}$^{\dagger}$ & 23.086 \\
    {[\ion{S}{4}]}$^{\dagger}$ & 10.511 & {[\ion{Ne}{2}]}$^{\dagger}$ & 12.811 & {[\ion{Co}{2}]}$^{\dagger}$ & 23.196 \\
    {[\ion{Ti}{2}]}$^{\dagger}$ & 10.511 & {[\ion{Co}{2}]}$^{\dagger}$ & 14.739 & {[\ion{Co}{4}]}$^{\dagger}$ & 24.040 \\
    {[\ion{Co}{2}]}$^{\dagger}$ & 10.523 & {[\ion{Co}{2}]}$^{\dagger}$ & 15.936 & {[\ion{Fe}{1}]}$^{\dagger}$ & 24.042 \\
    {[\ion{Ni}{2}]}$^{\dagger}$ & 10.682 & {[\ion{Fe}{3}]} & 20.167 & {[\ion{Co}{3}]}$^{\dagger}$ & 24.070 \\
    {[\ion{Ni}{4}]}$^{\dagger}$ & 11.130 & {[\ion{Fe}{2}]}$^{\dagger}$ & 20.928 & {[\ion{S}{1}]} & 25.249 \\
    {[\ion{Co}{2}]}$^{\dagger}$ & 11.167 & {[\ion{Ar}{3}]}$^{\dagger}$ & 21.829 & {[\ion{Co}{2}]} & 25.689 \\
    {[\ion{Ti}{2}]}$^{\dagger}$ & 11.238 & {[\ion{Fe}{2}]}$^{\dagger}$ & 20.986 & {[\ion{O}{4}]} & 25.890 \\
    {[\ion{Ni}{1}]}$^{\dagger}$ & 11.307 & {[\ion{Ni}{1}]}$^{\dagger}$ & 22.106 & & \\
    {[\ion{Ti}{2}]}$^{\dagger}$ & 12.159 & {[\ion{Co}{4}]}$^{\dagger}$ & 22.800 & & \\
  \enddata
  \tablecomments{\xspace $^{\dagger}$denotes line is part of a blended feature.}
\end{deluxetable}

\subsection{Channel 3 (11.55--18~\mic)}

Channel 3 begins in the blue wing of the [\ion{Co}{3}]~11.888~\mic blended 
feature. When examined as a whole, this feature is clearly dominated by 
[\ion{Co}{3}]~11.888~\mic, with the  [\ion{Fe}{3}]~11.978~\mic and 
[\ion{Ni}{1}]~12.001~\mic lines making weaker contributions to the red 
shoulder, as previously seen in \aefx \citep{DerKacy2023}, with additional
weak contributions due to [\ion{Ti}{2}]~12.159~\mic. Also similar to
\aefx, a broad, multi-peaked blend of weak features spanning $\sim$12.4--13~\mic 
potentially results from [\ion{Fe}{2}]~12.642~\mic, [\ion{Co}{3}]~12.681~\mic, 
[\ion{Ni}{2}]~12.729~\mic, and [\ion{Ne}{2}]~12.811~\mic lines 
\citep{DerKacy2023,Blondin2023}. 

Beyond $\sim$14~\mic, hints of 
strong features are seen between 14.6--15~\mic and 15.6--16.8~\mic. 
The former is likely the result of a complex of [\ion{Co}{2}]~14.739~\mic, 
[\ion{Ni}{1}]~14.814~\mic, and [\ion{Co}{2}]~14.977~\mic. In the latter, we 
identify the primary lines contributing to the main peak as
[\ion{Co}{2}] 16.152~\mic, 16.155~\mic, and 16.299~\mic, and 
[\ion{Co}{3}]~16.391~\mic. These are the first line identifications 
in this wavelength range for a \snia.
 
In much of the remaining parts of the channel, small peaks are seen at low
significance and remain unidentified. Many of these peaks are expected to 
be contributions from weak iron-group lines with no measured cross-sections. 
This is important for the astro-atomic physics community to address, with 
cross-validation between astronomical and atomic physics methods. 

\subsection{Channel 4 (17.7--27.9~\mic)} \label{sec:channel_4}

We see a clear point source present in the SHORT sub-band of Channel 4 (17.70--20.95~\mic). 
Between $\sim$18--19~\mic, a broad, possibly multi-peaked feature is discernible, 
arising from [\ion{Fe}{2}]~17.936~\mic, [\ion{Ni}{2}]~18.241~\mic, 
[\ion{Co}{1}]~18.265~\mic, [\ion{Co}{2}]~18.390~\mic, 
[\ion{S}{3}]~18.713~\mic, [\ion{Co}{2}]~18.804~\mic,
[\ion{Fe}{2}]~19.007~\mic, and [\ion{Fe}{2}]~19.056~\mic lines. 
In the MEDIUM and LONG sub-bands, the background becomes much more variable 
possibly due to the decreased sensitivity in Channel 4. This results in a likely 
under-subtraction which appears as a growing continuum under the strong peaks.
This continuum should not be interpreted as due to any physical process.
Regardless of this we see a clear point source in specific wavelength slices.
Additionally, this
continuum pushes the entire flux above the 10\% strong line threshold, and 
to account for this uncertainty, we only tentatively identify lines associated
with clear peaks and group them with the weak detections from Channels 1--3.
Lines roughly corresponding to strong and/or broad peaks in the complex blends 
of the MEDIUM and LONG sub-bands include:
[\ion{Fe}{3}]~20.167~\mic, [\ion{Fe}{2}]~20.928~\mic,
[\ion{Ar}{3}]~21.829~\mic, [\ion{Fe}{2}]~20.986~\mic,
[\ion{Ni}{1}]~22.106~\mic, a possibly blue-shifted line of 
[\ion{Co}{4}]~22.800~\mic, [\ion{Fe}{2}]~22.902~\mic, 
[\ion{Fe}{3}]~22.925~\mic, [\ion{Ni}{2}]~23.086~\mic,
[\ion{Co}{2}]~23.196~\mic, [\ion{Co}{4}]~24.040~\mic,
[\ion{Fe}{1}]~24.042~\mic, and [\ion{Co}{3}]~24.070~\mic.
Relatively isolated peaks associated with [\ion{S}{1}]~25.249~\mic,
[\ion{Co}{2}]~25.689~\mic, and [\ion{O}{4}]~25.890~\mic are also tentatively 
identified. As in Channel 3, many of the unidentified peaks likely correspond
to weak iron-group lines lacking measured cross-sections. Therefore, in 
order to obtain more information about which lines may be expected at 
$\lambda \gtrsim 21$~\mic and at what fluxes in under-luminous \sneia, we 
turn to model synthetic spectra in \autoref{sec:models}.

\section{Characterization of under-luminous \snia} \label{sec:characterization}

Having identified the strong lines present in the MIR features of \xkq, 
we now attempt to measure the physical properties of these lines through 
spectral fitting. These fits allow us to estimate the central location and 
widths of the lines. However, due to line blending, the complex physics of 
line formation at these epochs (see again \autoref{sec:line_ids} and later,
\autoref{sec:models}), and the unknown strengths of many weak lines that
produce the observed features in the spectrum, the 
resulting fits to most of the strong features in the spectrum are arbitrary 
with little physical meaning. By this we mean that the peak of the fit,
or its individual sub-components, cannot be identified with any particular 
transition and thus it cannot directly probe the ejecta velocity.

Instead, we focus on fitting only those features that are relatively well 
isolated and result from lines with known cross-sections, such as the 
[\ion{Ni}{2}]~6.636~\mic, [\ion{Ni}{3}]~7.349~\mic, and [\ion{Co}{3}]~11.888~\mic
lines. For analysis of the more complex features, we take a data-only approach
to estimating parameters from the features as a whole. The results of these
measurements and their implications are discussed below.

\subsection{Velocities of Isolated Lines}

Among the strong features present in the observed spectrum of \xkq, only 
three are isolated enough with minimal blending to be fit with simple 
analytic functions. These features are those dominated by the 
[\ion{Ni}{2}]~6.636~\mic, [\ion{Ni}{3}]~7.349~\mic, and 
[\ion{Co}{3}]~11.888~\mic lines. In each case, the fits were performed 
using the modified trust-region Levenberg-Marquardt algorithm found in the 
\texttt{scipy.odr} package. Each spectral region was fit with a Gaussian 
function with amplitude, mean, and standard deviation taken as free parameters. 
In order to better estimate the fit uncertainties, a Monte Carlo (MC) method 
was used to re-sample the spectrum (i.e. bootstrapping) and repeat the fit 
500 times. The errors of the MC sample are then added to the known uncertainties 
of the data, such as spectral resolution, in quadrature to obtain the values 
presented below.

\subsubsection{Stable Ni Lines}

\begin{figure}
    \centering
    \includegraphics[width=\columnwidth]{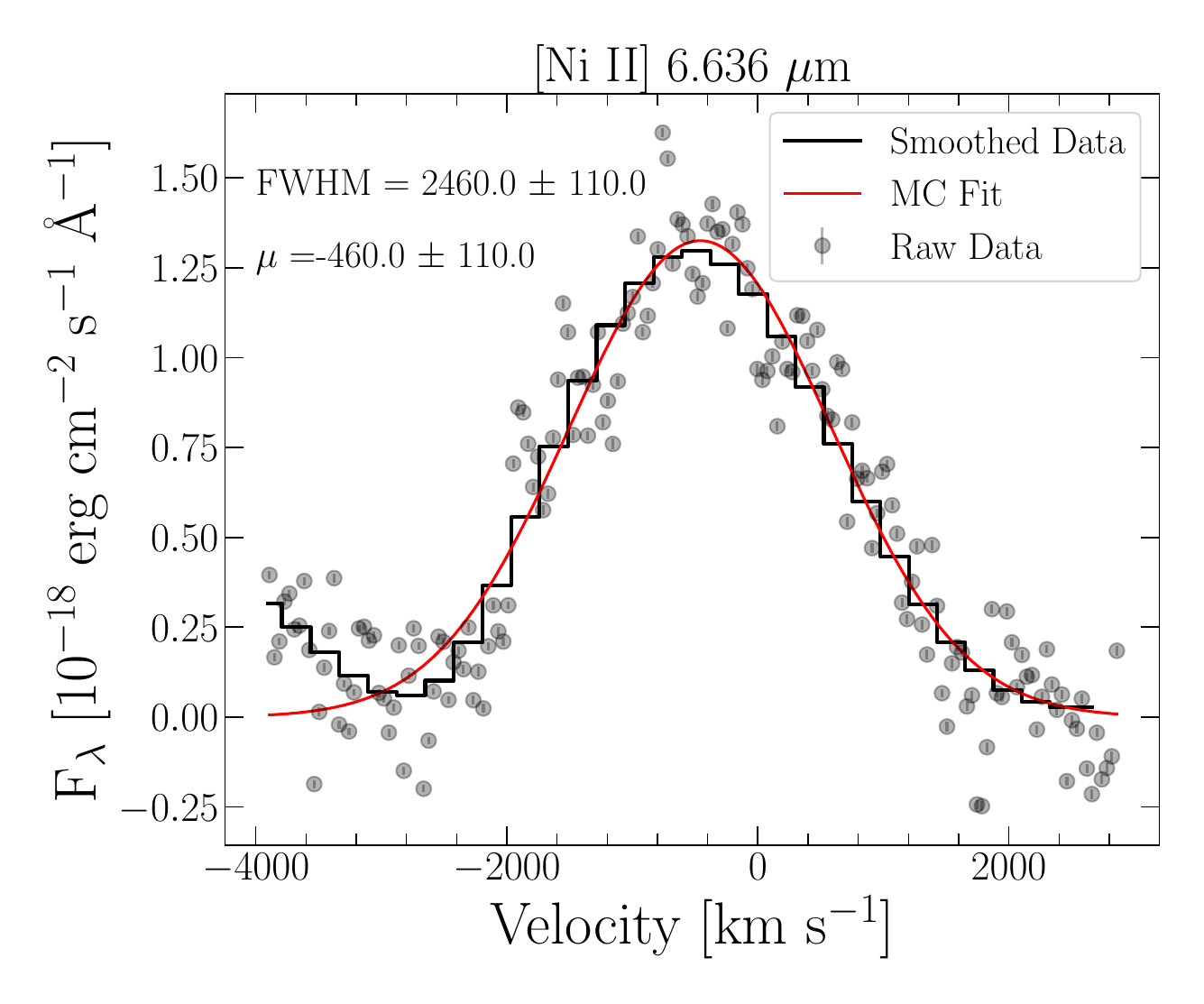}
    \caption{Single Gaussian MC fit to the [\ion{Ni}{2}]~6.636~\mic
    line. The best-fit parameters as determined from the MC results
    are: $\mu = -460 \pm 2$~\kms and FWHM~$= 2460 \pm 5$~\kms. The 
    resolution error of 110~\kms dominates over both the uncertainties 
    in the fit and in the wavelength calibration ($\sim$6~\kms; 
    \citealp{Argyriou2023}).}
    \label{fig:ni2_fit}
\end{figure}

In all previously published MIR spectra of \sneia with the corresponding 
wavelength coverage, the [\ion{Ni}{2}]~6.636~\mic line has either been
not detected \citep{Gerardy2007}, or identified as part of a blended feature 
\citep{Kwok_etal_2023_21aefx,DerKacy2023,Kwok_etal_2023_23pul}. In our 
MRS spectrum of \xkq, we see this feature as isolated and 
unblended --- suggesting it should be representative of the stable Ni distribution 
within the core of the ejecta. Given the shape of the line profile, it is 
assumed that the emission is coming from a region of the ejecta which has 
already reached the nebular state and can be modeled by a Gaussian function.
However, it should be noted that fitting emission features with Gaussian 
profiles makes implicit assumptions about the nature of the line formation 
and spectral features (see \autoref{sec:models}), the underlying chemical 
distribution of ions within the ejecta, and should therefore be interpreted 
with caution. The resulting fit is shown in \autoref{fig:ni2_fit}, revealing 
a shift in the line center of $-460 \pm 110$~\kms, and a full-width at 
half-maximum~(FWHM) of $2460 \pm 110$~\kms. 

Similarly, the [\ion{Ni}{3}]~7.349~\mic feature in \xkq is more isolated
and unblended than in previous MIR observations of \sneia. However, 
compared to the [\ion{Ni}{2}]~6.636~\mic feature, fitting the [\ion{Ni}{3}]
feature requires more careful treatment. There is a significant contribution 
to the blue wing of the profile either from the continuum or an unknown weak 
line, and a weak shoulder in the red wing also due to unidentified lines. Neither
wing component matches known nebular lines near these wavelengths. 

Initial attempts to fit the profile despite these complications yield a fit
which captures the width of the profile, but is unable to accurately measure
the peak of the flux. Attempts to simultaneously account for potential 
contributions from the continuum and weak blending similarly capture the 
width of the feature, but are also unable to reproduce the correct location
of the peak flux, and are strongly dependent on the choice of initial 
parameters defining the continuum. 

\citet{Penney2014} have noted that prior to the ejecta becoming optically 
thin through to the center, the locations of the peak fluxes will appear 
blue shifted relative to the true line center due to blocking of the red 
half of the line profile by the continuum, and that these effects are best 
measured after subtracting the continuum flux separately. After subtracting 
the continuum separately, the profile appears noticeably less Gaussian and 
requires at least a two Gaussian fit. The result of this fit captures the 
peak of the flux and width of the feature well, however the assumed [\ion{Ni}{3}] 
component is not the dominant line in the feature. Additionally, the strong
component is not near the correct location to fit the weak blending in the
wings from the possible [\ion{Co}{1}] lines. As there are no nearby lines 
resulting from ions commonly found in models of under-luminous \snia, we regard 
the presence of an unidentified, strong line as unlikely. 

\begin{figure*}
    \centering
    \subfigure[Single Gaussian MC fit]{\includegraphics[width=\columnwidth]{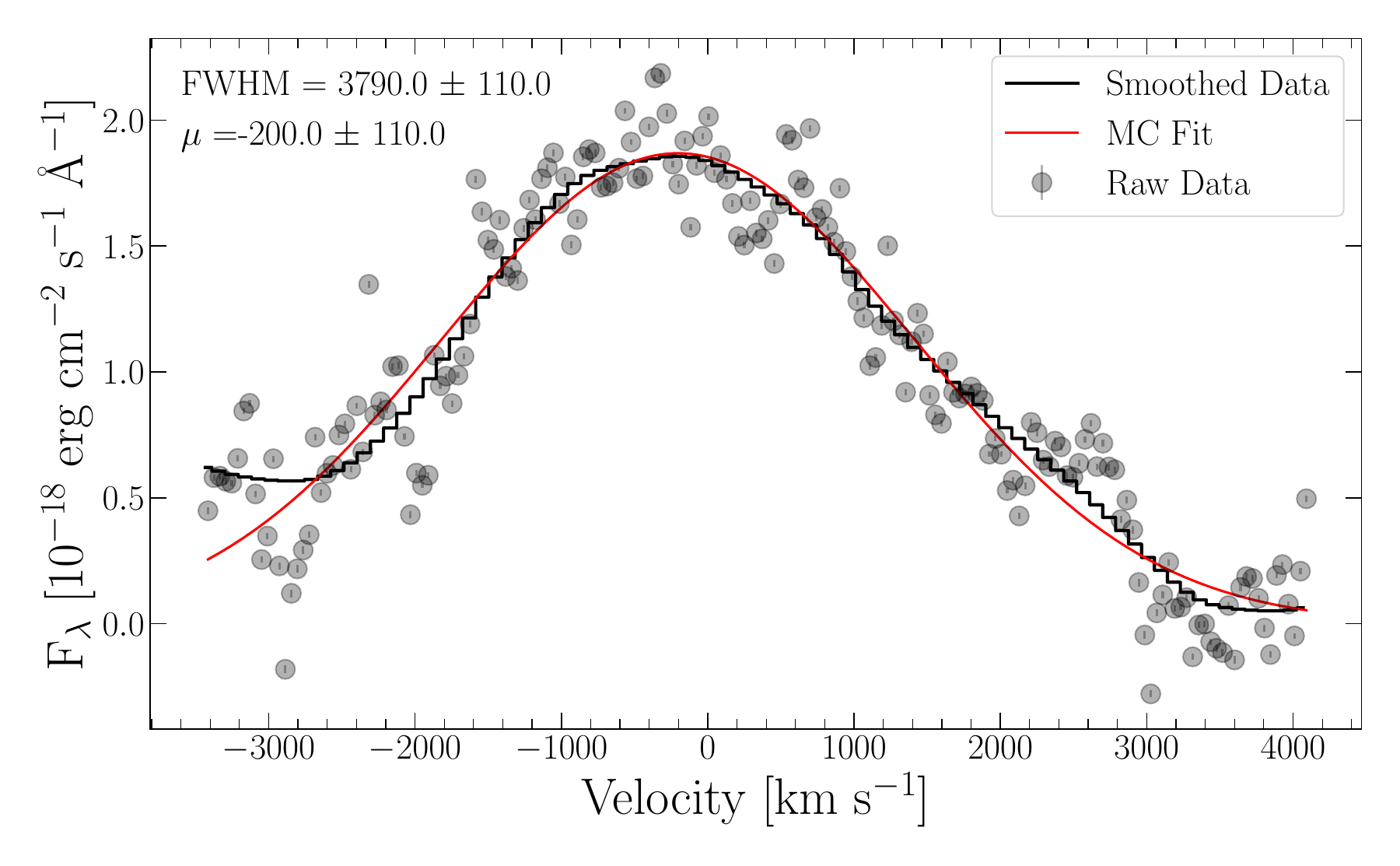}}
    \subfigure[Simultaneous multi-component fit]{\includegraphics[width=\columnwidth]{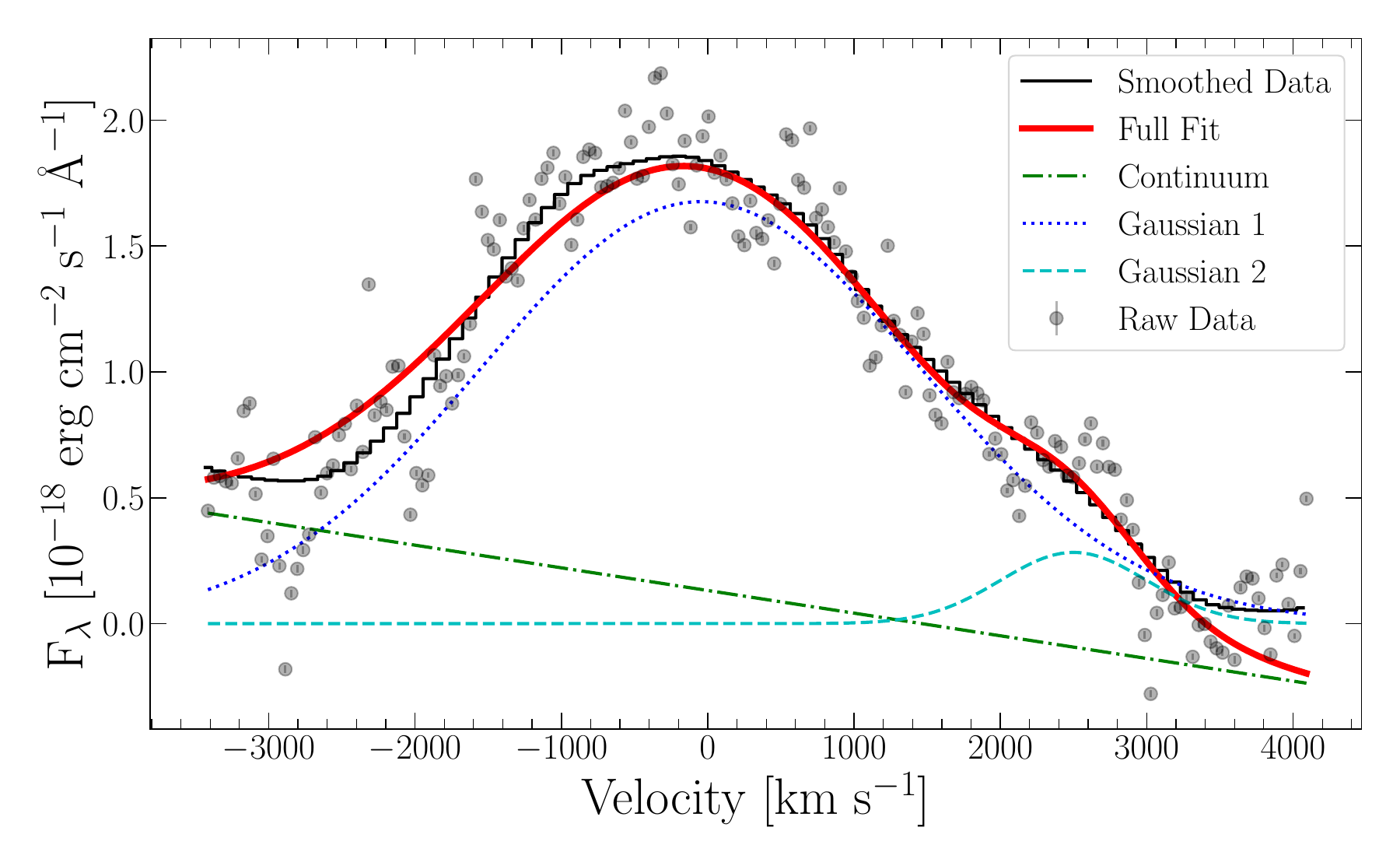}}
    \subfigure[Continuum-subtracted MC multi-Gaussian fit]{\includegraphics[width=\columnwidth]{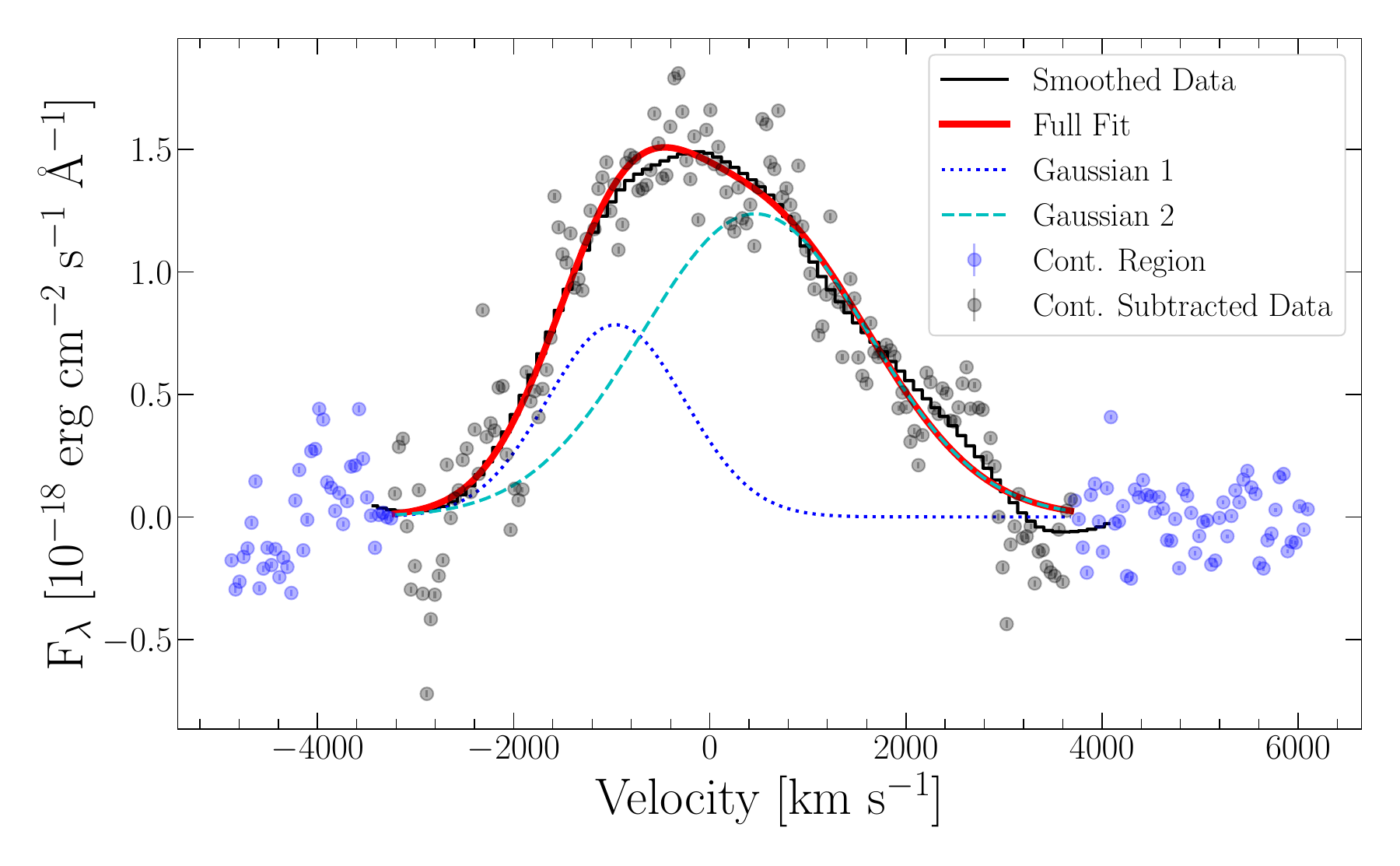}}
    \subfigure[Continuum-subtracted data derived values]{\includegraphics[width=\columnwidth]{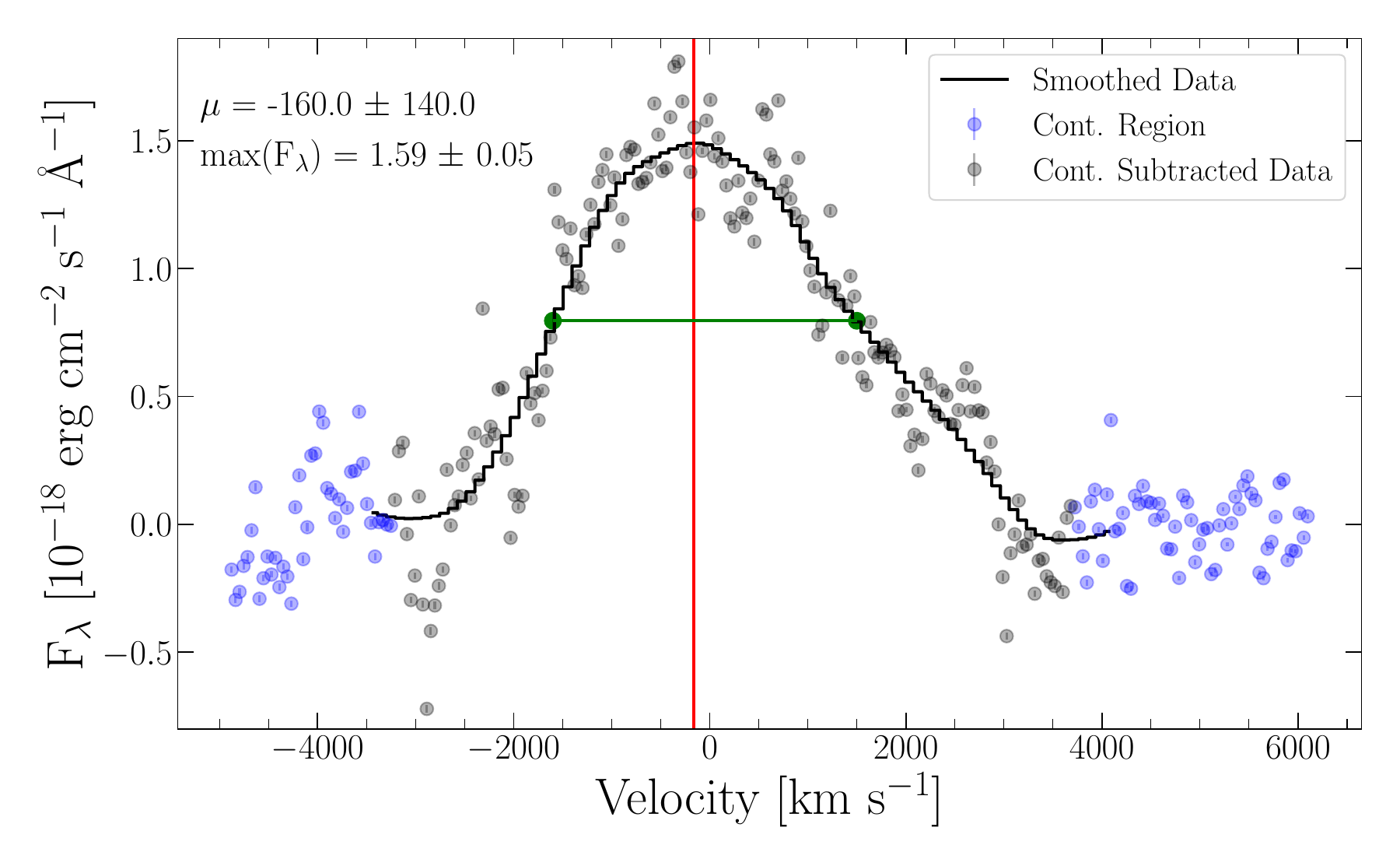}}
    \caption{Fits to the [\ion{Ni}{3}]~7.349~\mic feature. Panel (a) shows the 
    results of a single Gaussian fit similar to \autoref{fig:ni2_fit}. Panel (b)
    simultaneously fits the continuum component and multiple Gaussians, 
    illustrating the non-Gaussianity of the feature. Panel (c) fits multiple Gaussian 
    components after independently fitting and subtracting the continuum. Panel
    (d) shows the data derived values as measured from the continuum subtracted
    line profile. }
    \label{fig:ni3_fit}
\end{figure*}

Instead, we estimate the shift in the peak flux and the FWHM from a sample
of 1000 bootstrapped re-sampled MC realizations of the spectra, assuming the 
flux errors are normally distributed. We find that the peak flux is shifted 
by $-160 \pm 140$~\kms, with the FWHM of the feature measuring 3800~\kms. 
The results of this work are shown in \autoref{fig:ni3_fit}, and highlight the 
complexity of fitting even seemingly isolated and weakly blended features, and the 
effects of the photosphere present even during this late phase. Furthermore,
Gaussian fits to complex features need to be interpreted with caution,
as they fail to capture the physics of the line formation 
of these features, even at late times.

\subsubsection{[\ion{Co}{3}]~11.888~\mic Feature} 

Resonance lines such as [\ion{Co}{3}]~11.888~\mic are important tracers 
of the amount and distribution of ions in the SN ejecta, since most of 
the de-excitation and recombination of each species will pass through their
resonance transitions. Due to the decay chain of \Nifs $\rightarrow$ \Cofs 
$\rightarrow$ \Fefs which powers the emission in \sneia, the 
[\ion{Co}{3}]~11.888~\mic feature serves as a late-time tracer 
of the initial \Nifs distribution, which is located below the photosphere at early 
times. Previous low resolution observations of the features dominated by 
[\ion{Co}{3}]~11.888~\mic have found that the line profiles are well-captured by
a single Gaussian fit; however, spectral modeling reveals that up to 
$\sim$10 percent of the flux in the feature may result from weak blending with 
[\ion{Fe}{3}]~11.978~\mic and [\ion{Ni}{1}]~12.001~\mic lines in the red wing of 
the profile \citep{Telesco2015,Kwok_etal_2023_21aefx,DerKacy2023}.

\begin{figure}
    \centering
    \includegraphics[width=\columnwidth]{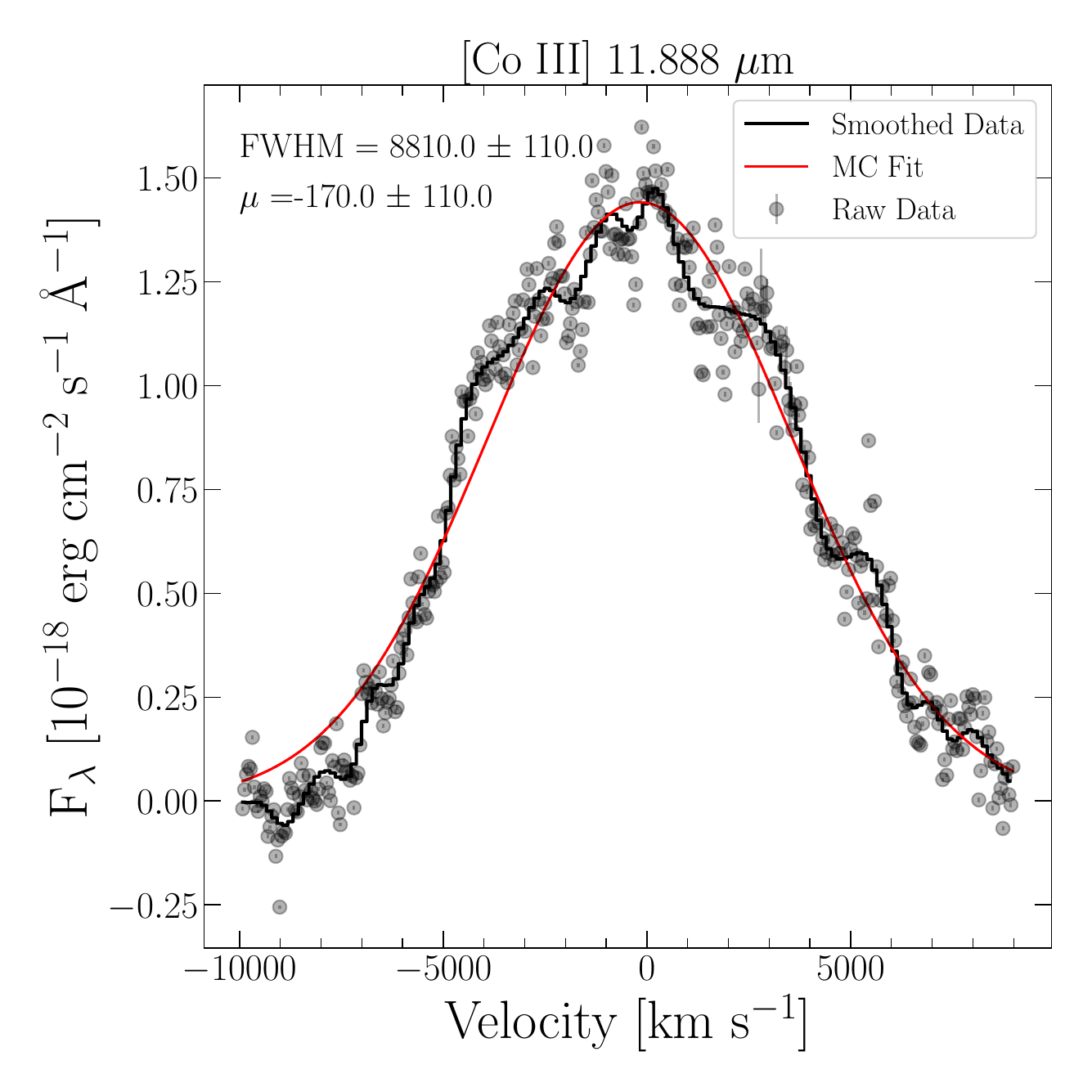}
    \caption{Same as \autoref{fig:ni2_fit} but for the [\ion{Co}{3}]~11.888~\mic-dominated 
    feature. The best fit parameters determined from the MC ensemble are: 
    $\mu = -170 \pm 5$~\kms and FWHM~$= 8810 \pm 5$~\kms. As before, the resolution 
    error of $\pm 110$~\kms dominates over the fit and wavelength calibration errors.}
    \label{fig:co3_fit}
\end{figure}

In \xkq, the effects of the blending are clearly seen as a series of shoulders 
in the red wing of the profile in \autoref{fig:co3_fit} that are not present 
in the blue wing of the feature. However, much like the previous lower-resolution 
observations, the velocity shift, peak flux, and width of the feature are 
well-represented by a single Gaussian fit, with the line center shifted by 
$-170 \pm 110$~\kms and a FWHM of $8810 \pm 110$~\kms. As in the fits to 
the [\ion{Ni}{3}]~6.636~\mic line, the resolution error dominates over the 
statistical error of the fit. Comparisons between these fits in \xkq and the sample
of other observed normal-luminosity \sneia are explored in further detail in 
\autoref{sec:compare}. This highlights the need to approach Gaussian fits
with caution, as the effects of blending are even stronger in the 
[\ion{Co}{3}]~11.888 feature than in the [\ion{Ni}{3}]~7.349 feature, despite 
the fact that a single Gaussian fit is sufficient to approximate the 
blended feature in this instance.

\begin{figure}
    \centering
    \includegraphics[width=\columnwidth]{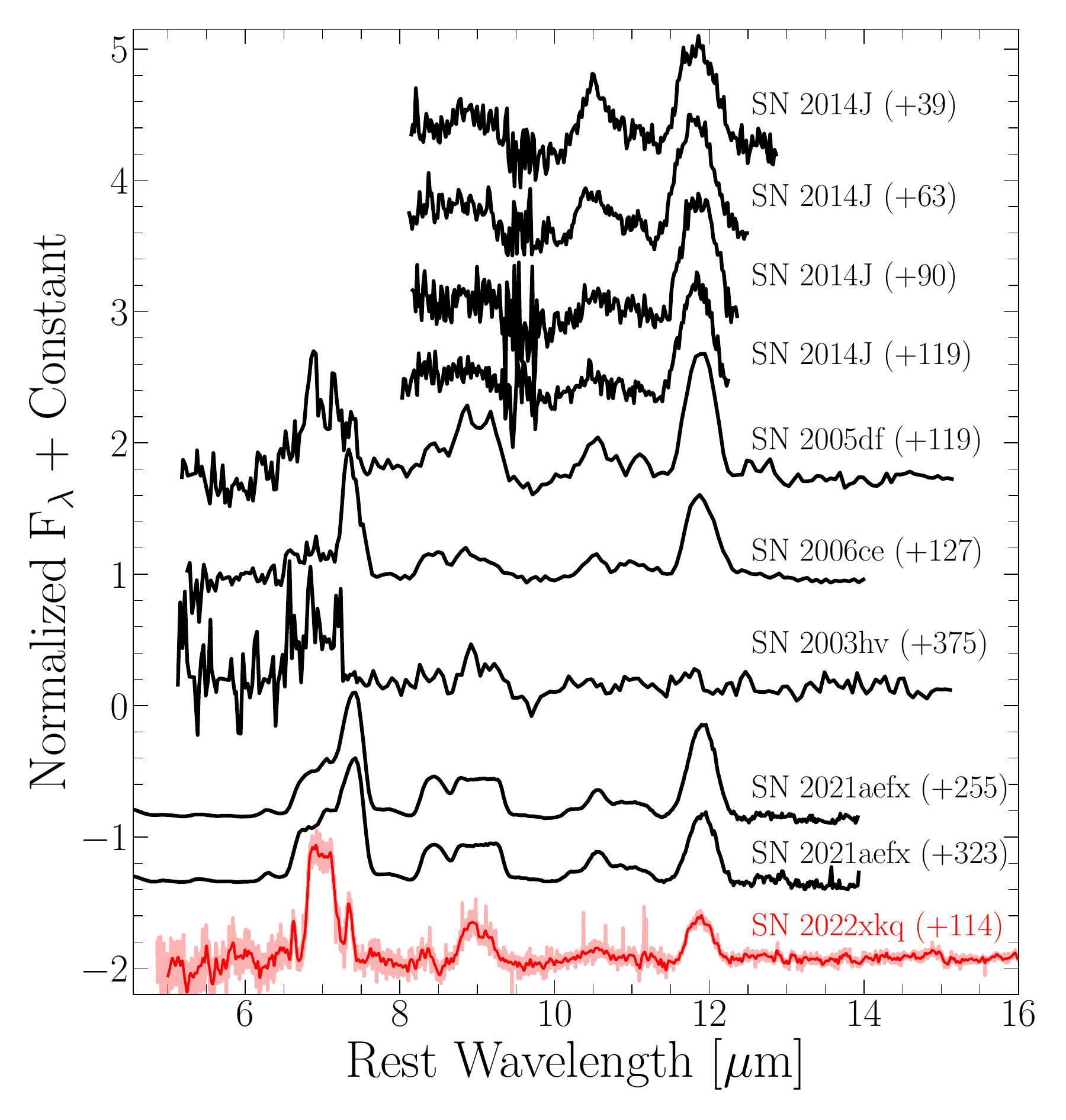}
    \caption{The published sample of normal-luminosity \sneia MIR spectral observations, including
    SNe 2003hv, 2005df \citep{Gerardy2007}, 2006ce \citep{Kwok_etal_2023_21aefx}, 2014J 
    \citep{Telesco2015}, and 2021aefx \citep{Kwok_etal_2023_21aefx,DerKacy2023}. Epochs 
    relative to $B$-band maximum are shown for each spectrum. For \xkq, a 
    smoothed spectrum is shown in the foreground while the raw data is shown
    in the background. All spectra have been corrected for MW extinction.}
    \label{fig:mir_sample}
\end{figure}

\begin{deluxetable*}{lcccc}
    \tablecaption{Properties of \sneia with MIR Spectra at $\sim$120 Days \label{tab:mir_sample}}
    \tablehead{\colhead{Parameter} & \colhead{SN~2005df} & \colhead{SN~2006ce} &
    \colhead{SN~2014J} & \colhead{\xkq} }
    \startdata
    Epoch (days) & 119 & 127 & 119 & 114.2 \\
    M$_{\rm B,peak}$ (mag) & $-19.27 \pm 0.31$ & \nodata & $-19.19 \pm 0.10$ & $-18.01 \pm 0.15$ \\
    \DmB (mag) & 1.12\tablenotemark{a} & \nodata & $1.12 \pm 0.02$ & $1.65 \pm 0.03$ \\
    \sBV & 0.95\tablenotemark{b} & \nodata & 0.95\tablenotemark{b} & 0.63 \\
    d (Mpc) & $19.8 \pm 2.8$ & $19.6 \pm 2.87$ & $3.4 \pm 0.1$ & $31.0 \pm 2.0$ \\
    \enddata
    \tablecomments{No optical photometry of SN~2006ce is publicly available. 
    \tablenotetext{a}{No error reported.} 
    \tablenotetext{b}{Value calculated from \DmB according to Eq. (4) of \citet{Burns2014}.}}
    \tablerefs{SN~2005df: \citet{Takats2015,Krisciunas2017}, SN~2006ce: \citet{Kwok_etal_2023_21aefx}; NED,
    SN~2014J: \citet{Marion2015}, \xkq: \citet{Pearson_etal_2023}, This work.}
\end{deluxetable*}

\begin{figure*}
    \centering
    \includegraphics[width=\textwidth]{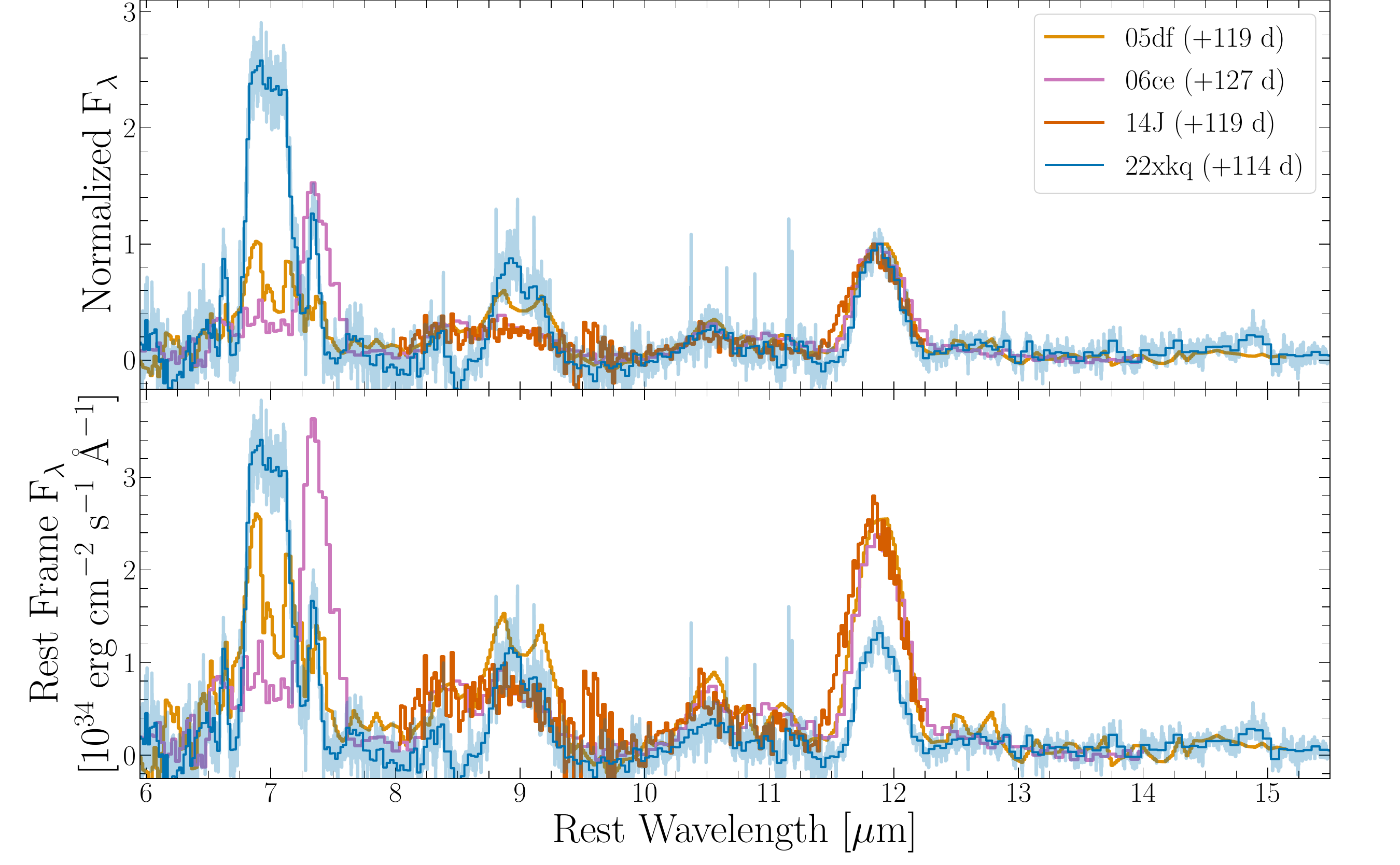}
    \caption{
    Comparison of \sneia spectra observed $\sim$120~days after maximum 
    light in the MIR. The top panel shows flux values normalized to the peak of 
    the [\ion{Co}{3}]~11.888~\mic features, while the bottom panel shows 
    the fluxes corrected to the SN rest frame based on the estimated 
    distances in \autoref{tab:mir_sample}. In \xkq, the singly ionized 
    features (e.g. [\ion{Ni}{2}]~6.636~\mic and [\ion{Ar}{2}]~6.985~\mic)
    are stronger relative to their doubly ionized counterparts 
    ([\ion{Ni}{3}]~7.349~\mic and [\ion{Ar}{3}]~8.991~\mic) as is
    expected in an under-luminous \snia.}
    \label{fig:mir_compare}
\end{figure*}

\subsection{Under-luminous vs. Normal-luminosity \sneia MIR Spectra} \label{sec:compare}

\subsubsection{Qualitative Comparisons}

The published MIR sample of normal-luminosity \sneia are compared to our spectrum 
of \xkq in \autoref{fig:mir_sample}. Four objects in the sample (SNe~2005df,
2006ce, 2014J, and 2022xkq) have MIR spectra taken roughly 120~days after 
$B$-band maximum light. These objects are directly compared 
 in \autoref{fig:mir_compare}.

The most prominent difference between the MIR spectrum of \xkq and 
the sample of normal-luminosity objects is the strength of the [\ion{Ar}{2}]
feature near 7~\mic relative to the [\ion{Ar}{3}] feature at $\sim$9~\mic. 
These features are known to vary based on the geometry of the Ar distribution 
and viewing angle of the SN \citep{Gerardy2007,DerKacy2023}, although 
distinguishing these effects from continuum effects in spectra taken 
before the ejecta become fully nebular is much more difficult. Yet, when 
comparing the line identifications in \xkq to those of the rest of 
the sample, we find the elements that dominate these features are similar,
and produce prominent features in the same wavelength regions. The 
differences in the [Ar] lines are explored further in both 
\autoref{sec:ar_lines} and \autoref{sec:models}. 

Both the under-luminous \xkq and the sample of normal-luminosity 
objects display weak, blended, broad features in the region 
between $\sim$10--11.5~\mic (\autoref{fig:mir_compare}). 
When compared to previous observations normalized to the peak 
of the [\ion{Co}{3}]~11.888~\mic resonance line, the relative flux in 
the features are similar. Spectra taken roughly 130 days after 
explosion have tentative identifications of [\ion{S}{4}] in combination 
with iron group elements (e.g. Fe, Co, and Ni) while the later spectra of
\aefx only show the iron-group lines \citep{Kwok_etal_2023_21aefx,DerKacy2023}. 
The low spectral resolutions of previous observations make further 
direct comparisons difficult.

As expected, due to the under-luminous nature of \xkq, the peak flux of
the [\ion{Co}{3}]~11.888~\mic resonance line is significantly weaker than
in the other objects in the sample at similar epochs, all of which are 
normal-luminosity \snia (see the bottom panel of \autoref{fig:mir_compare}). 
As a resonance line, the [\ion{Co}{3}]~11.888~\mic line directly traces 
the radioactive decay of the \Nifs produced in the explosion through the 
\Nifs$\rightarrow$\Cofs$\rightarrow$\Fefs decay chain. The [\ion{Co}{3}]
line is also noticeably narrower than the other normal-luminosity \sneia
in the sample, consistent with measurements of the [\ion{Co}{3}]~5890~\AA\
in the optical \citep{Graham2022}. 
This behavior is also expected for under-luminous SNe~Ia, as
nuclear burning occurs under lower density compared to normal SNe~Ia
resulting in lower \Nifs production and a shift of nuclear burning 
products towards the inner, slower expanding layers. Moreover, more 
quasi-statistical equilibrium (QSE) elements are formed at the expense 
of nuclear statistical equilibrium (NSE) elements leading to increasing 
[Ar II/III] emission and narrower lines of all QSE and NSE elements 
\citep{Hoeflich_etal_2002,Ashall18,Mazzali2020}. The differences in these 
[\ion{Co}{3}] lines are quantified in \autoref{sec:co3_lines}.

\subsubsection{Ar Lines} \label{sec:ar_lines}
\begin{figure}
    \centering
    \includegraphics[width=\columnwidth]{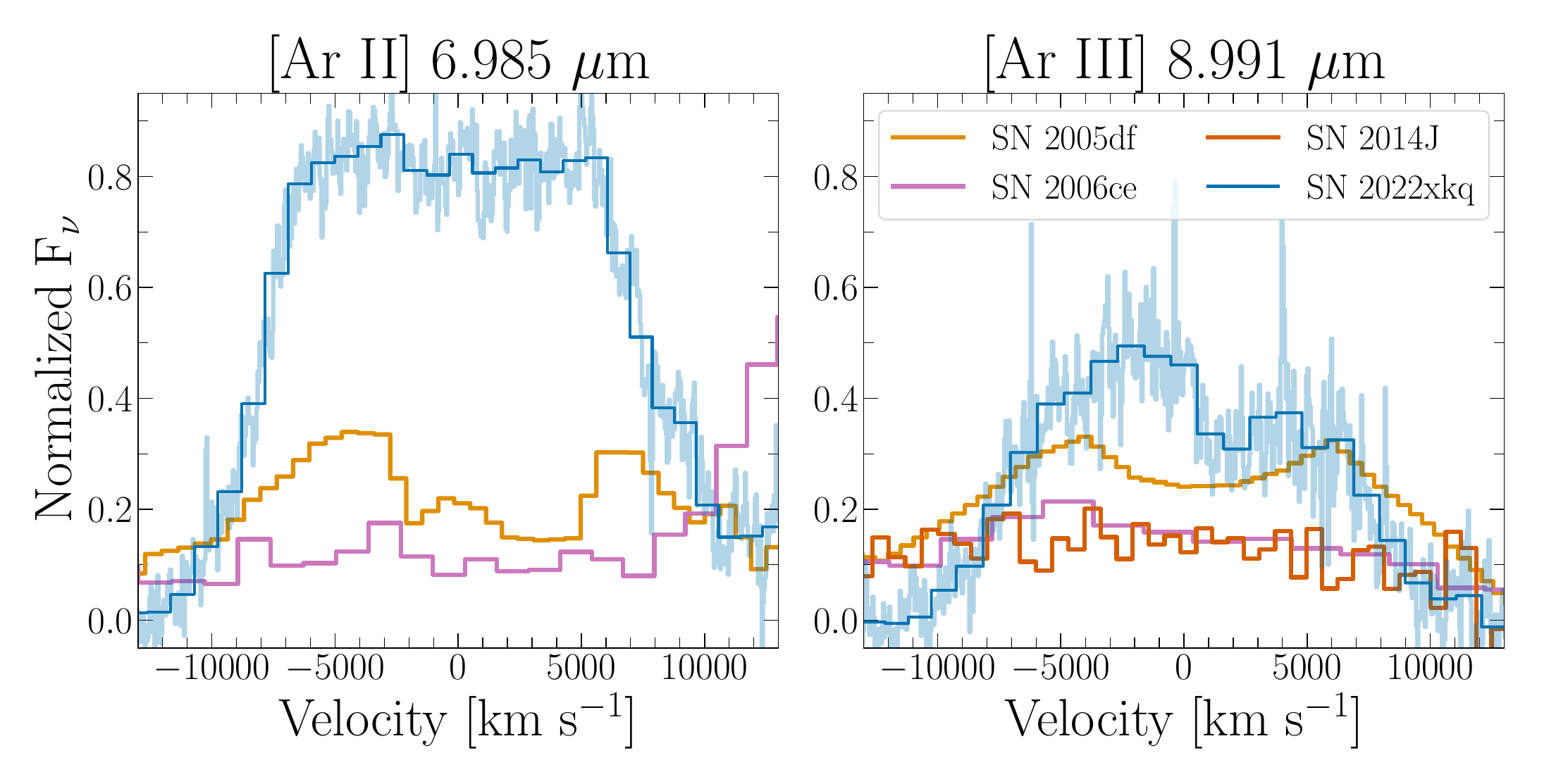}
    \caption{Comparison of Ar line profiles between \xkq and other 
    normal-luminosity \sneia roughly 120 days after $B$-band max.
    As expected, the overall fluxes in the [Ar]-dominated features
    are larger in \xkq than in the normal-luminosity objects, while
    the increased strength of the [\ion{Ar}{2}] feature relative to
    the [\ion{Ar}{3}] feature in \xkq indicates that the ionization 
    balance in the ejecta favors \ion{Ar}{2} over \ion{Ar}{3}
    \citep{Gerardy2007,Ashall18,Mazzali2020}.}
    \label{fig:d120_arlines}
\end{figure}

\autoref{fig:d120_arlines} shows the region of the two prominent argon
features, [\ion{Ar}{2}]~6.985~$\mic$ and [\ion{Ar}{3}]~8.991~$\mic$, 
in \xkq compared to the same features in the subset of normal-luminosity \sneia 
spectra taken near 120 days after maximum light. For the normal-luminosity 
\sneia, it is difficult to ascertain the exact extent of the emission in 
velocity space due to a combination of low resolution and low S/N 
in the individual observations. However, for \xkq it is clear that 
the wings of the emissions extend to $\sim$10,000~\kms in both argon
dominated features. The ionization balance of \xkq 
is noticeably different than that of the other objects. In \xkq, the bulk 
of the argon is singly ionized, resulting in an average ratio of 
the peak flux of $\sim$2 between the [\ion{Ar}{2}] and [\ion{Ar}{3}]
dominated features. In the normal-luminosity objects observed at similar
epochs, this ratio is $\sim$1. Furthermore, the total integrated flux 
in the Ar regions relative to the peak of the [\ion{Co}{3}] is higher in 
\xkq, demonstrating that there is a larger argon abundance in the ejecta 
relative to the \Nifs\ mass, as expected for an under-luminous SN~Ia 
\citep{Ashall18}. 

\subsubsection{[\ion{Co}{3}]~11.888~\mic Profiles} \label{sec:co3_lines}

\begin{figure}
    \centering
    \includegraphics[width=\columnwidth]{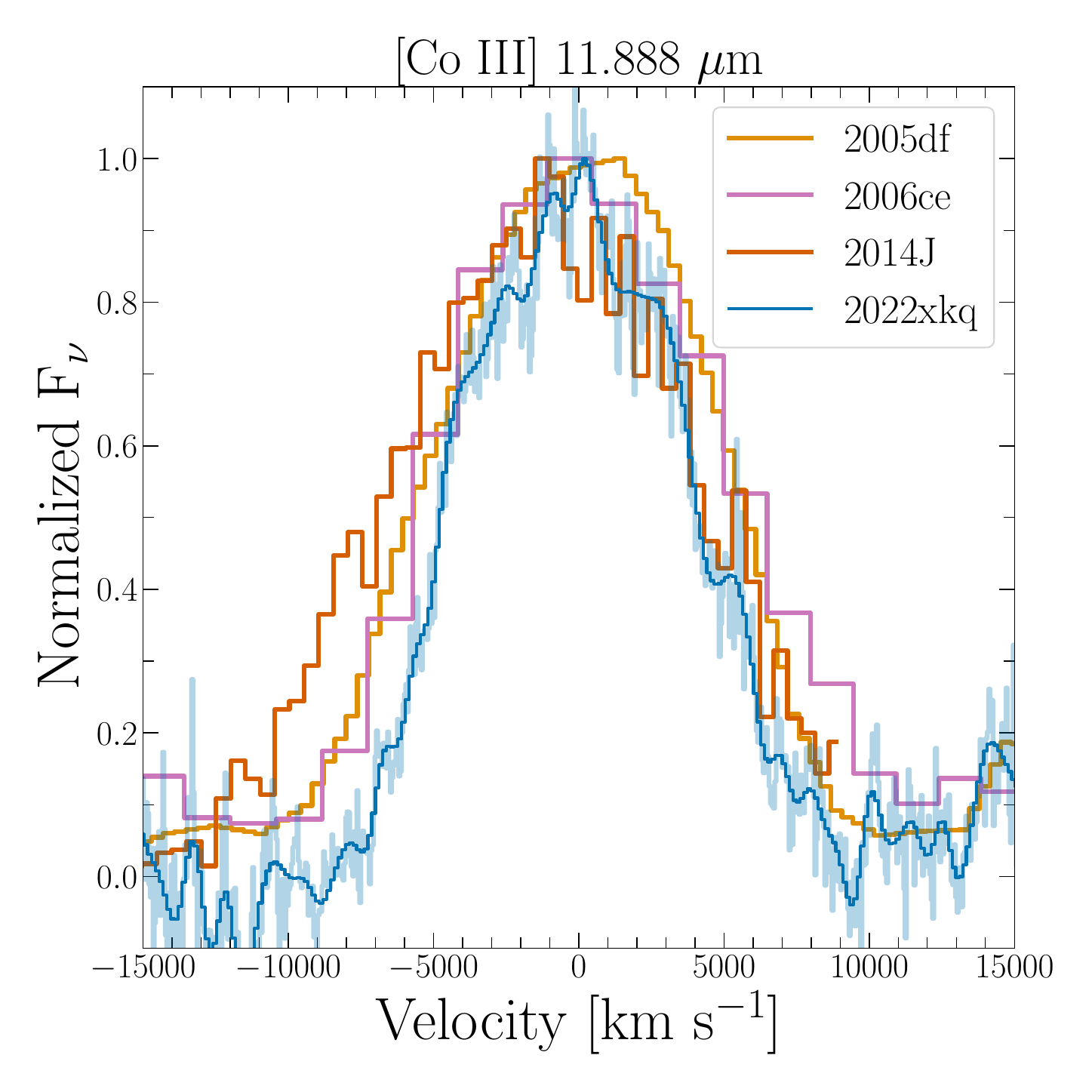}
    \caption{Same as \autoref{fig:d120_arlines}, but instead comparing 
    the [\ion{Co}{3}]~11.888~\mic dominated features. The unblended blue
    wings of the profile reveal \xkq has a much narrower distribution of
    \Cofs relative to the normal-luminosity objects.}
    \label{fig:d120_co3line}
\end{figure}

\autoref{fig:d120_co3line} shows the comparison of the [\ion{Co}{3}]~11.888~\mic
features across the sample of MIR spectra taken $\sim$120~days after $B$-max. 
Compared to the lower-resolution observations with {\it Spitzer} (SNe~2005df, 2006ce)
and CanariCam (SN~2014J) that smooth out much of the blending and continuum 
effects, the profile of \xkq is decidedly non-Gaussian. The blending effects 
primarily impact the red wings of the profile, with all four SNe showing similar 
full widths at zero-intensity (FWZI). These effects are not prominent in the blue
wing, where the FWZI of \xkq at $\sim$7,500~\kms is narrower than those of the 
normal-luminosity objects ($\gtrsim$10,000~\kms). 

As a consequence of the [\ion{Co}{3}]~11.888~\mic resonance line being 
formed by the radioactive \Cofs in the ejecta, it serves as a direct tracer of 
the original distribution of \Nifs in the explosion. As such, the width of the 
feature is characteristic of the burning conditions of the NSE region.
As previously discussed above for under-luminous \sneia, burning to \Nifs\ occurs 
more centrally in the ejecta, which has the effect of producing a narrower 
[\ion{Co}{3}]~11.888~\mic feature. 

\begin{figure}
    \centering
    \includegraphics[width=\columnwidth]{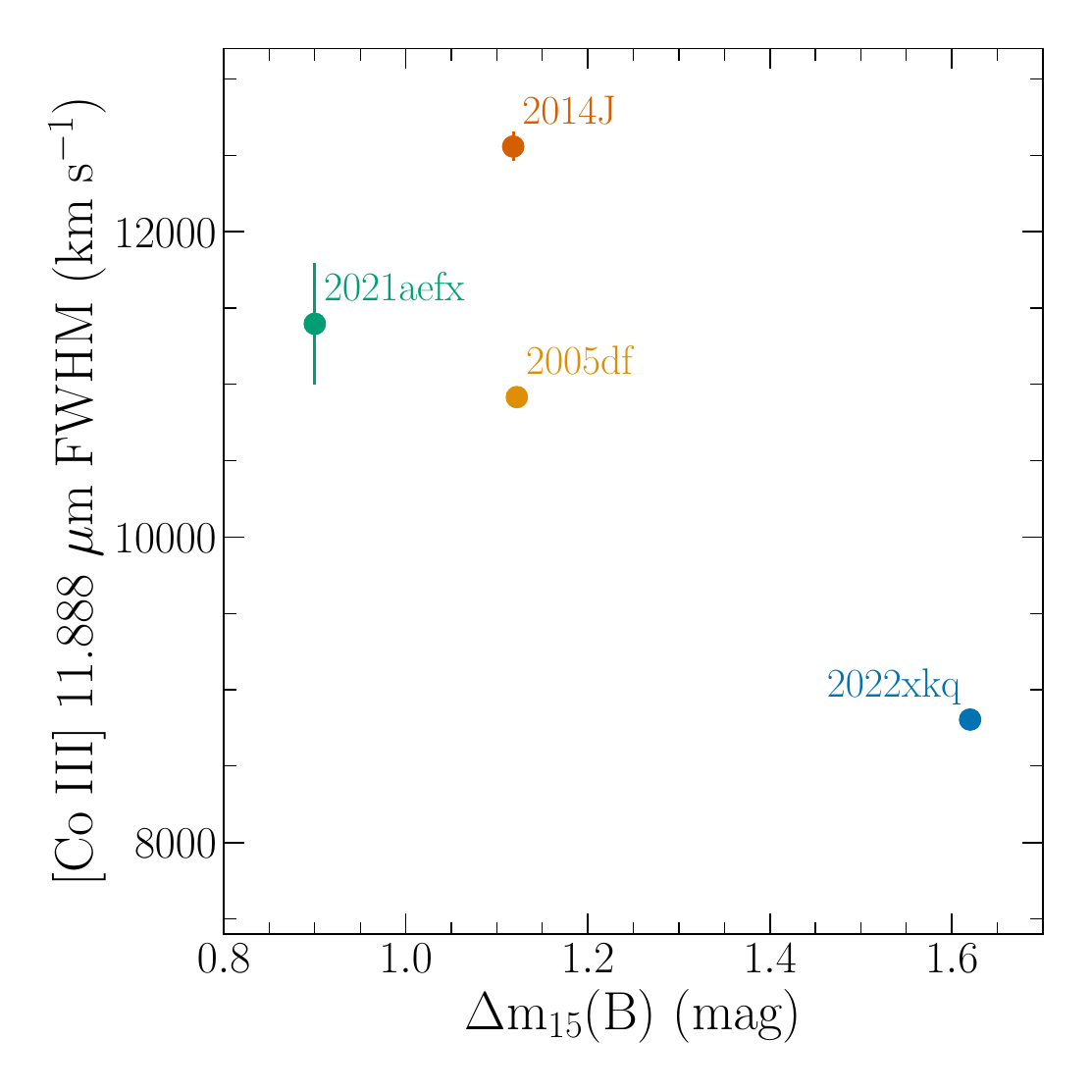}
    \caption{[\ion{Co}{3}]~11.888~\mic FWHM versus \DmB for \sneia with
    detected [\ion{Co}{3}] emission and well-characterized light curves.
    We include the value from \aefx obtained at $\sim+323$ days from 
    the explosion \citep{DerKacy2023}, but note that this value may be 
    uncertain due to the different physical regime in which the epoch is 
    measured and potential line blending. }
    \label{fig:mir_wlr}
\end{figure}

\autoref{fig:mir_wlr} shows the FWHM of this [\ion{Co}{3}]~11.888~\mic feature 
as a function of light curve shape, \DmB. We see a rough trend between light 
curve shape and width of the [\ion{Co}{3}]~11.888~\mic feature, where broader, 
more luminous \sneia have larger values of FWHM compared to the under-luminous 
\xkq. 
As the sample of MIR spectra of SNe~Ia increases, this figure can be 
populated with various types of SNe~Ia to understand how closely the 
[\ion{Co}{3}]~11.888~\mic feature traces the luminosity of the SNe, allowing us 
to see how closely this plot follows the luminosity-width relation \citep{Phillips1993}.
A similar trend has been found through spectral modeling of the optical data \citep{1998ApJ...499L..49M,2007Sci...315..825M}, as well as in the 
[\ion{Co}{3}]~5890~\AA\ line in optical nebular spectra \citep{Graham2022}. 
While both the optical and MIR [\ion{Co}{3}] lines are resonance lines, 
it is important to test this relation with the MIR line that is from a 
very low lying state from which the resonance transition is the only 
downward transition out of this low lying state. The more energetic upper 
state of the [\ion{Co}{3}]~5890~\AA\ transition in addition to decaying to 
the ground state, also decays 
to the upper state of the MIR line. The optical line from this
transition at 6190~\AA\ is not a prominent feature in the spectra
shown by \citet{Graham2022}, nor are other nearby [\ion{Co}{3}] resonance lines.
There is also disagreement in the literature about whether the feature
associated with [\ion{Co}{3}]~5980~\AA\ is due (in part) to Na I D 
\citep{Kuchner1994,Mazzali1997,Dessart2014}.

\section{Modeling} \label{sec:models}

We compare \xkq to new simulations of off-center \Mch
explosion models\footnote{These simulations use the updated atomic models 
and MIR cross-sections of C.~Ashall et al., 2023 (in preparation)},
with parameters based upon the spherical models of the Model~16-series 
of \citet{Hoeflich_2017}. The early light curve properties and
maximum light luminosity of the models are very similar to that 
of \xkq. No further fine-tuning of the basic parameters have been 
done, nor is it is necessary in light of uncertainties in distance 
and reddening. In order to move beyond profile fits, the synthetic spectra 
are modeled by detailed non-LTE radiation-hydrodynamic simulations, which
link the observations to the underlying physics 
identify and quantify the contribution of specific transitions to
observed spectral features. The observed features are dominated by a
combination of emission lines, a quasi-continuum of allowed and
forbidden lines in an envelope with varying element abundances, and
strong non-LTE effects. We choose our best-fit model by comparing
line profiles and line-ratios of noble gases like
Ar and of electron capture elements such as Ni.

\begin{deluxetable*}{ccccc}
  \caption{Model parameters and light curve parameters for various $\rho_c$. \label{tab:model_params}} 
  \tablehead{\colhead{Parameter} & \colhead{Value} &\colhead{Value}&\colhead{Value}&\colhead{Value}}
   \startdata
    $\rho_c [\times 10^9$ \gcm] & 0.5 & 1.0 & 2.0 & 4.0 \\
    \hline
    M$_{\rm ej}$ (\Msun) & $\sim$1.37 & $\sim$1.37 & $\sim$1.37 & $\sim$1.37 \\
    M$_{\rm tr}$ (\Msun) & 0.33 &0.34 &0.34 &0.36 \\
    B(WD, turbulent) (G) & $10^6$& $10^6$& $10^6$& $10^6$\\
    M(B/V)  (mag) &   -18.43/-18.57 &-18.29/-18.46 &-18.22/-18.37 &-17.98/-18.06 \\
    $\Delta$m$_{15}(B/V)$ (mag) & 1.61/1.15 &1.65/1.16 & 1.67/1.18 & 1.42/1.02\\
    B-V (V$_{max}$) (mag) & 0.150 &0.152 & 0.152& 0.158\\ 
    \enddata
    \tablecomments{The DDT is unlikely to occur between 0.3--0.5 $M_\odot$ because 
    overlapping of the Ca-rich layers with the bulge of the \Nifs would lead to a 
    narrow [\ion{Ca}{2}] doublet. The amount of mass burned during the deflagration 
    M$_{\rm tr}$ includes both \Nifs and EC elements.}
\end{deluxetable*}

\begin{figure*}[t]
    \includegraphics[scale=0.35]{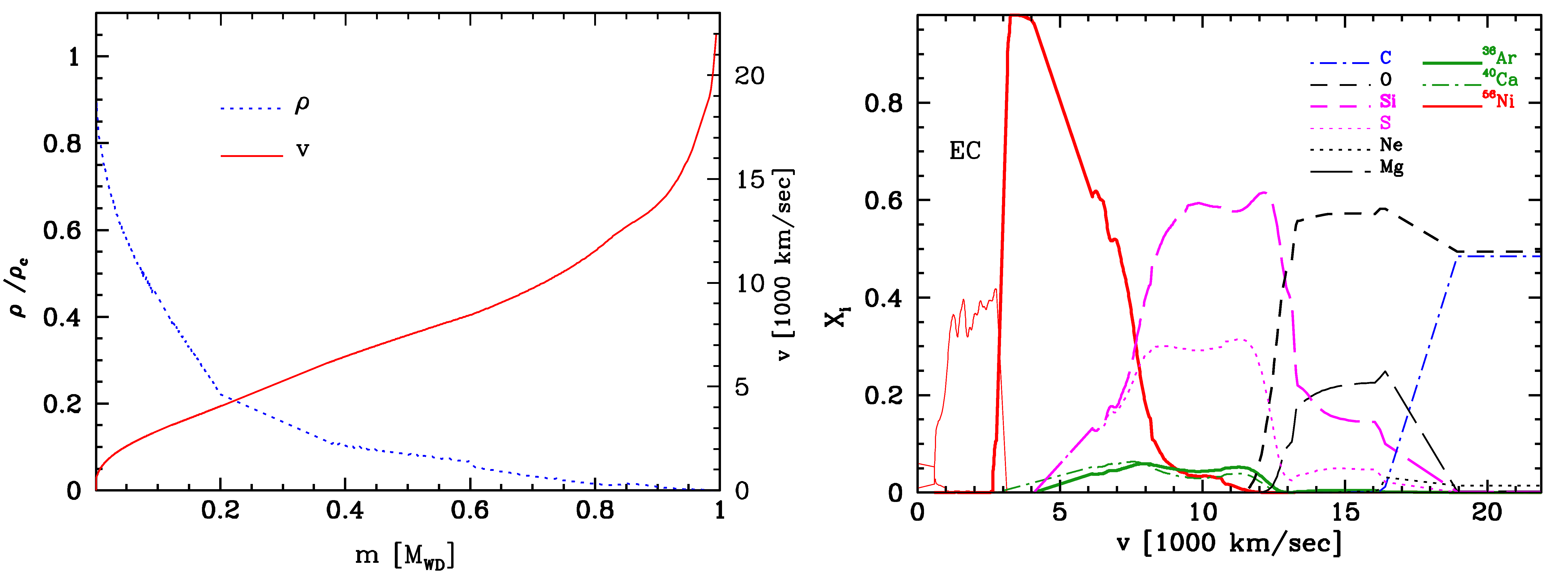}
    \caption{Density and velocity structure as a function of mass 
    coordinate (left) and distribution of elements for our reference 
    under-luminous models with $\rho_c=2 \times 10^{9}$~\gcm. Note 
    that M$_{\rm DDT}$ was chosen to be 0.2~\Msun. For the influence of 
    $\rho_c$ on the size electron capture core, see \cite{Diamond2015}.}
    \label{fig:models}
\end{figure*}

The simulations are parameterized explosion models, where we use the 
spherical delayed-detonation scenario to constrain the global parameters of 
the explosion. Fine-tuning these models is not necessary to achieve 
the goals of this study, as we focus on spectra rather than 
high-precision photometry. The reference model (\autoref{fig:models})
originates from a C/O~WD with a main-sequence progenitor
mass of $5$~\Msun, solar metallicity, and a central density 
$\rho_c=2.0 \times 10^9$~\gcm. The model produces $\sim$0.324~\Msun of \Nifs.
The resulting light curves have a peak brightness of $M(B) = -18.22$~mag and 
$M(V) = -18.37$~mag and light curve decline rates of \DmB$ = 1.67$~mag and 
\DmV$ = 1.18$~mag. These can be compared to $M(B) = -18.1 \pm 0.15$~mag, and 
\DmB$ = 1.65 \pm 0.03$~mag obtained from the early time light curve
\citep{Pearson_etal_2023}. The inferred absolute brightness 
at maximum light depends sensitively on the reddening, the distance
modulus, and their uncertainties. \citet{Pearson_etal_2023} estimated a 
\Nifs-mass of $0.22 \pm 0.03$~\Msun compared to the  $0.34$~\Msun we obtain
using nebular MIR spectra. Derived \Nifs masses are uncertain, in particular 
for under-luminous \sneia because the Q- or $\alpha$- value in Arnett's law 
varies from 0.8--1.6 over a small brightness range and similar differences 
occur among various methods \citep{Hoeflich_etal_2002,2006A&A...460..793S,Hoeflich_2017}. 
The value of the \Nifs mass obtained here is consistent with that of \citet{Pearson_etal_2023}
within the errors.

Our value of $\rho_c$ is chosen based upon the presence of stable Ni lines 
and to match the ionization balance in the observed MIR spectrum 
(see \autoref{fig:line_ids} and \autoref{sec:syntheticSpectra}). 
Burning starts as a deflagration front near the center and transitions 
to a detonation \citep{1991A&A...245..114K}. The DDT is triggered by
increasing the rate of burning by the Zeldovich-mechanism
\citep{Zeldovich_1940,von_Neumann_1942,Doering_1943}, that is, the mixing of
burned and unburned material. Various mechanisms have been suggested
to initiate the DDT transition, ranging from mixing by turbulence, shear 
flows at chemical boundaries, and differential rotation in the WD. Moreover, 
the mixing may depend on the conditions during the thermonuclear runaway 
\citep{1997ApJ...478..678K,1999ApJ...527L..97L,2004A&A...425..217Y,
2004ApJ...606.1029B,2006NuPhA.777..579H,2018ApJ...858...13H,
2013A&A...550A.105C,Poludnenko2019,2021MNRAS.501L..23B}.
Although the recently suggested mechanism of turbulent-driven DDT
in the distributed regime of burning \citep{2011AIPC.1376...38O,Poludnenko2019} 
is consistent with the amount of deflagration burning in our parameterized models
\citep{Hoeflich_etal_2002,Hoeflich2023}, other DDT mechanisms may be realized. 
We evaluate constraints from the observations of \xkq, treating the location
of the DDT as a free parameter.

In our simulations, the deflagration--detonation transition
is triggered `by hand' when $\sim$0.34~\Msun of the material has
been burned by the deflagration front and is induced by the mixing
of unburned fuel and hot ashes \citep{1991A&A...245..114K}. Our
simulations take into account magnetic fields for the
positrons. However, at 130 days, the mean free path of positrons is
small even for small initial magnetic fields
\citep{Penney2014}. We assume $B = 10^6$~G 
based upon our prior modeling of light curves and late-time nebular spectra
\citep{Diamond2015,2021ApJ...923..210H}. We have examined models
where $\rho_c$ was varied from $0.5-4.0 \times 10^{9}$~\gcm in order to
study the effect of the WD central density on the MIR spectra and
conclude that $\rho_c=2.0 \times 10^9$~\gcm best reproduces the
observations. The flux changes by $\sim 15$\% between models with our
fiducial value and those with $\rho_c=4\times 10^{9}$~\gcm,
because the photosphere masks the appearance of very neutron-rich isotopes.
However, lower central densities radically change the spectra.

Following the central carbon ignition and the propagation of the 
deflagration front, the deflagration transitions to a detonation. 
Rather than assuming that the DDT occurs in a spherical shell, it 
is assumed to begin as an off-center point as described by \citet{Livne1995}
and in our previous work
\citep{Penney2014,fesen2014,Hoeflich2021,DerKacy2023,Hoeflich2023}. 
The effective spectral resolution is $R \approx 300-600$ \citep{Hoeflich2021}.
We constructed models where the off-center location for the
delayed-detonation transition occurred at M$_{\rm DDT,off}$ of
0, 0.2, 0.5, and 0.9~\Msun. For computational expediency,only the 
model with 0.2~\Msun has been fully converged with complex model 
atoms. Note that we use an off-center DDT in a single spot because 
it does not produce a refraction wave.
Based on the discussion in \citet{Hoeflich2021}, the almost flat top
of the [\ion{Ar}{2}] 6.985~\mic feature suggests that \xkq is seen at
a low inclination angle ($\theta$) with $ |\theta| \la 20-30^{\circ}$.
The absence of a narrow ($< 1000$~\kms FWHM) [\ion{Ca}{2}] 
$\lambda\lambda 7293,7326$~\AA\ doublet in \xkq \citep{Pearson_etal_2023}
excludes that the DDT occurs close to the inner edge of the \Nifs region,
justifying a M$_{\rm DDT} \approx 0.2$~\Msun., or less. 
This is in contrast with our models for
SN~2020qxp where the large overlap between Ca-rich region and NSE region
results in a strong and narrow [\ion{Ca}{2}] feature appearing $\sim 100$ 
days after the explosion (see the inset of
Fig. 4 in \citealp{Hoeflich2021})\footnote{The simulations of SN~2022qxp
were developed from the same base-model as \xkq but with different central densities 
($4\times10^{9}$~\gcm) and DDTs (0.5~\Msun).}, regardless of the observation
angle. Theoretically, a central DDT 
cannot be excluded because unburned material could be dragged down during the 
deflagration phase. However, the observed narrow \Nife features exclude 
a central DDT. The DDT is triggered by increasing the rate of burning by 
the Zeldovich-mechanism \citep{Zeldovich_1940,von_Neumann_1942,Doering_1943}.
At the time of the DDT, the expansion of the inner electron 
capture region is subsonic 
\citep[see, for example, Figs.~9--10 in][]{2017hsn..book.1151H}. Thus, with 
an increased rate of burning and energy production the buoyancy will mix 
the electron capture elements out to higher velocity, inconsistent with 
the observed narrow \Nife lines.
For the best-fitting model (\autoref{fig:models}), we used an
off-center DDT with M$_{\text{DDT,off}}=0.2$~\Msun because we have no
strong evidence for asymmetry from the line profiles, either because
the observing angle is close to the equator or because the DDT is well
inside of the Ca/Ar region. 
Stronger constraints would require later time observations in the
optical to MIR. Nonetheless, off-center DDTs have been used to avoid
the artifact of low density burning due to the strong refraction wave
evident in spherical simulations \citep{khokhlov93,Hoeflich1996}.

The non-LTE atomic models and the radiation transport used in this
work were discussed in \citet{Hoeflich2021}. Detailed atomic
models are used for the ionization stages I-IV for C, O, Ne, Mg, Si,
S, Cl, Ar, Ca, Sc, Ti, V, Cr, Mn, Fe, Co, Ni. As before, the atomic
models and line lists are based on the database for bound-bound
($\approx $ 40,000,000 allowed and forbidden) transitions of
\citet{vanHoof2018}\footnote{Version v3.00b3
\url{https://www.pa.uky.edu/~peter/newpage/}} supplemented by
additional forbidden lines \citep{Diamond2015}, and lifetimes based on
the analysis by our non-LTE models of SN~2021aefx (C.~Ashall et al., in prep).  

\subsection{Model structure}\label{sec:models_structure}
 
The basic model parameters and light curve observables are given in
\autoref{tab:model_params}. The angle averaged structure of the best
fitting model is shown in \autoref{fig:models}. As expected for
`classical' delayed detonation models, the density and velocity distributions 
are smooth (e.g without a shell) because the WD becomes unbound during the
deflagration phase, in contrast to pulsating DD or pure deflagration
models \citep{Nomoto1984,Hoeflich_Khokhlov_Wheeler_1995,
Niemeyer_Woosley_1997,Bravo_etal_2009,Hoeflich_2017}. Within the \Mch 
scenario a low luminosity/low \Nifs mass is produced by an increased mass of
deflagration burning, which leads to a larger pre-expansion of the WD
and lower density burning. Our model for \xkq falls into the
$\Delta$m$_{15}(B/V)$ regime of rapidly dropping opacities soon after
maximum light, a regime of fast declining peak luminosity within a
relatively narrow range of $\Delta$m$_{15}(B/V)$ \citep{Hoeflich_etal_2002}. 
This is similar to SN~1991bg-like objects (as indicated by the [\ion{Ti}{2}] 
lines) and slightly more luminous than SN~2005ke or SN~2016hnk
\citep{Patat_etal_2012,Galbany2019}. As a result, the models show
unburned C/O, and products of explosive O-burning, and incomplete Si-burning layers
down to $\approx $ 16,000, 12,000 and 5000~\kms, respectively. For
\Mch models of under-luminous SNe~Ia, the element production is shifted
toward partial burned fuel corresponding to: $\sim$0.27--0.35~\Msun of
M(\Nifs), $\sim$0.65~\Msun of intermediate mass elements (Si and S),
and $\sim$0.1~\Msun of unburned carbon \citep{Hoeflich_etal_2002}. In
models of normal-luminosity \sneia, these values are 0.55--0.65~\Msun,
0.2~\Msun, and $1-2 \times 10^{-2}$~\Msun, respectively
\citep{Hoeflich_etal_2002}. In our reference model, the
layers expanding interior to $\approx 3000$~\kms are dominated by electron
capture elements, mostly $^{54}$Fe, $^{57}$Co and \Nife.

\begin{figure*}
    \centering
    \includegraphics[width=.99\textwidth]{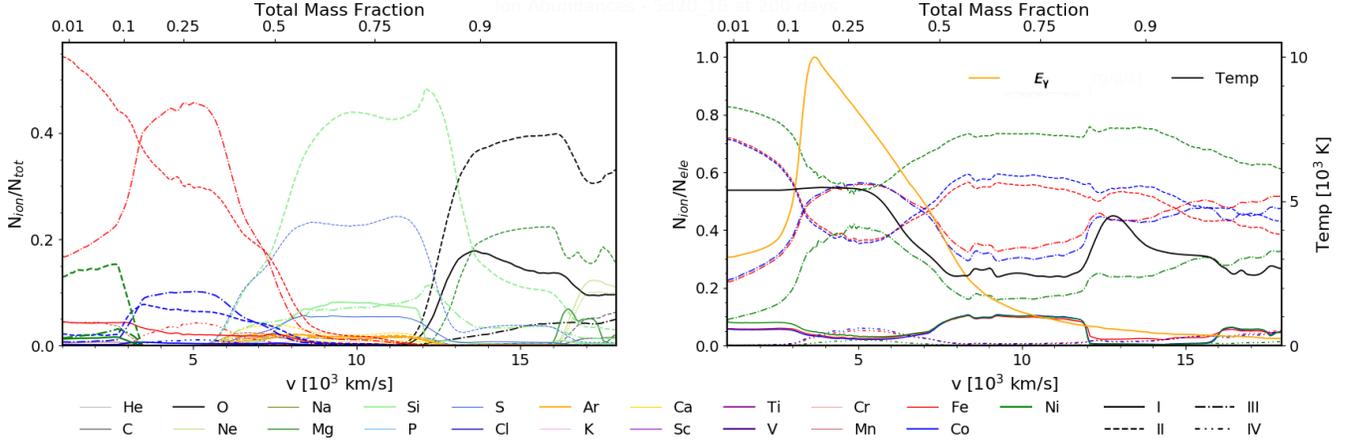} 
    \caption{Same as the right panel from \autoref{fig:models} but 
      angle-averaged ionization levels I-IV of Fe and Co as representative 
      of iron-group elements, specific energy input by $\gamma$-rays and
      non-thermal leptons normalized to its maximum, and temperature T
      as a function of expansion velocity. Note that the overall ionization
      structure is rather insensitive to $\rho_c$ because the shift in the
      balance is dominated by the recombination rate and, thus, the density. 
      $E_\gamma$ is dominated by $\gamma$-rays rather than positrons which 
      only is loosely correlated with the \Nifs distribution. Namely, 
      $E_\gamma$ escape the outer layers and we have significant heating in 
      the core resulting in overall peaked line profiles of the iron group 
      elements (see below and \autoref{fig:synthetic}).}
    \label{fig:ion}
\end{figure*}   

The ionization structures at day 130 are shown in \autoref{fig:ion}.
Overall, when compared to normal-luminosity SNe~Ia
\citep[e.g.,][]{Wilk20}, the global ionization balances 
are shifted towards lower ionization states, as has also been shown
for other under-luminous \sneia, \citep{Hoeflich2021}.
All the non-iron-group elements (e.g. C, O, Mg, Si, S, Ca) are dominated
by singly ionized features, with significant contribution from neutral
ions. For iron group elements, doubly ionized species dominate in the 
\Nifs region, but the equilibrium is shifted towards single ionized 
species further inwards due to the high densities. The region
dominated by double ionized iron group elements in \xkq, is
absent in SN~2020qxp. A further difference between \xkq and
SN~2020qxp is an extended region of singly ionized species in \xkq
rather than singly ionized species concentrated in the center.
The difference between the transitional spectrum of \xkq and the
nebular spectrum of SN~2020qxp obtained 190 days after explosion
can be understood as due to the higher densities (in the \Nifs region) 
and the energy input in \xkq that are dominated by non-local 
$\gamma$-ray heating. Positrons dominate at later times. The 
non-local $\gamma$-ray heating leads to significant energy 
deposition in the region of electron capture
elements \citep[see, \autoref{fig:ion}, and][]{Penney2014}.

\subsection{Spectral Analysis of the JWST Observation} \label{sec:model_spectra}

\begin{figure*}[t]
    \centering
    \includegraphics[width=0.99\textwidth]{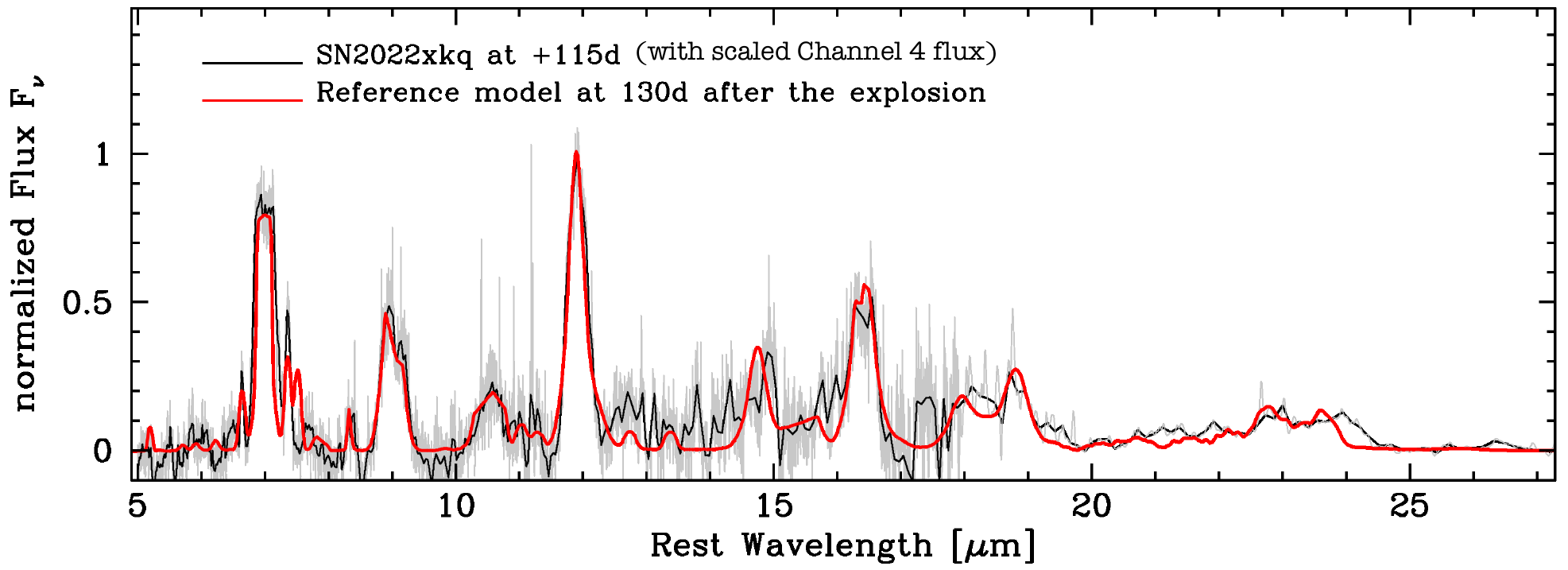}
    \caption{Comparison between the synthetic and observed MIR spectra of 
    \xkq (black) where the flux in Channel 4 ($\lambda \geq 18$~\mic ) has 
    been scaled to the model flux, with a wavelength-dependent scaling factor 
    (see text).}
\label{fig:synthetic_best}
\end{figure*}

The goal of this section is demonstrating constraints from the JWST
spectrum of \xkq, and to take full advantage of the reduced line
blending in the MIR compared to shorter wavelengths. Optical and NIR
spectral series of \xkq are presented in
\citet{Pearson_etal_2023}. Optical and NIR fits of our model to data
of other \sneia have been presented and discussed previously in detail
\citep{1995ApJ...443...89H,1998ApJ...496..908W,Hoeflich_etal_2002,Diamond2015,Telesco2015,Hoeflich2023}.
We focus on the MIR-spectra because interstellar reddening hardly
affect the MIR fluxes, the reddening corrections are below the
signal-to-noise ratio. Studies using observations from {\it JWST} 
have found that the effect of reddening on the
flux varies between $\sim 0.5-1$\% depending on wavelength
\citep{Gordon2023a}. However, the spectrophotometric accuracy of MIRI 
is 2-5\%. Thus, the detailed analysis of JWST MIR spectra allow the
uncertainties inherent to optical wavelength range
\citep{2003AJ....125..166K} to be avoided. For the theoretical
analysis, the focus is on line ratios between ions and the profile of
features that are mostly independent from uncertainties in the
distance and the reddening law. The absolute flux will be used as a
consistency check between observations and models. The overall 
spectrum is given in \autoref{fig:synthetic_best}. Our spectral 
analysis is mostly based on Channels 1--3 because, the Channel 4 
spectrum requires flux calibration based upon the model flux
and the highly variable, uncertain background at the longer
wavelengths make it unsuitable for using the ionization 
balance or line-profiles as diagnostics.
  
\subsubsection{Probing the Ionization and Abundance Structure}\label{sec:syntheticSpectra}

\begin{figure*}[t]
    \centering
    \includegraphics[width=0.99\textwidth]{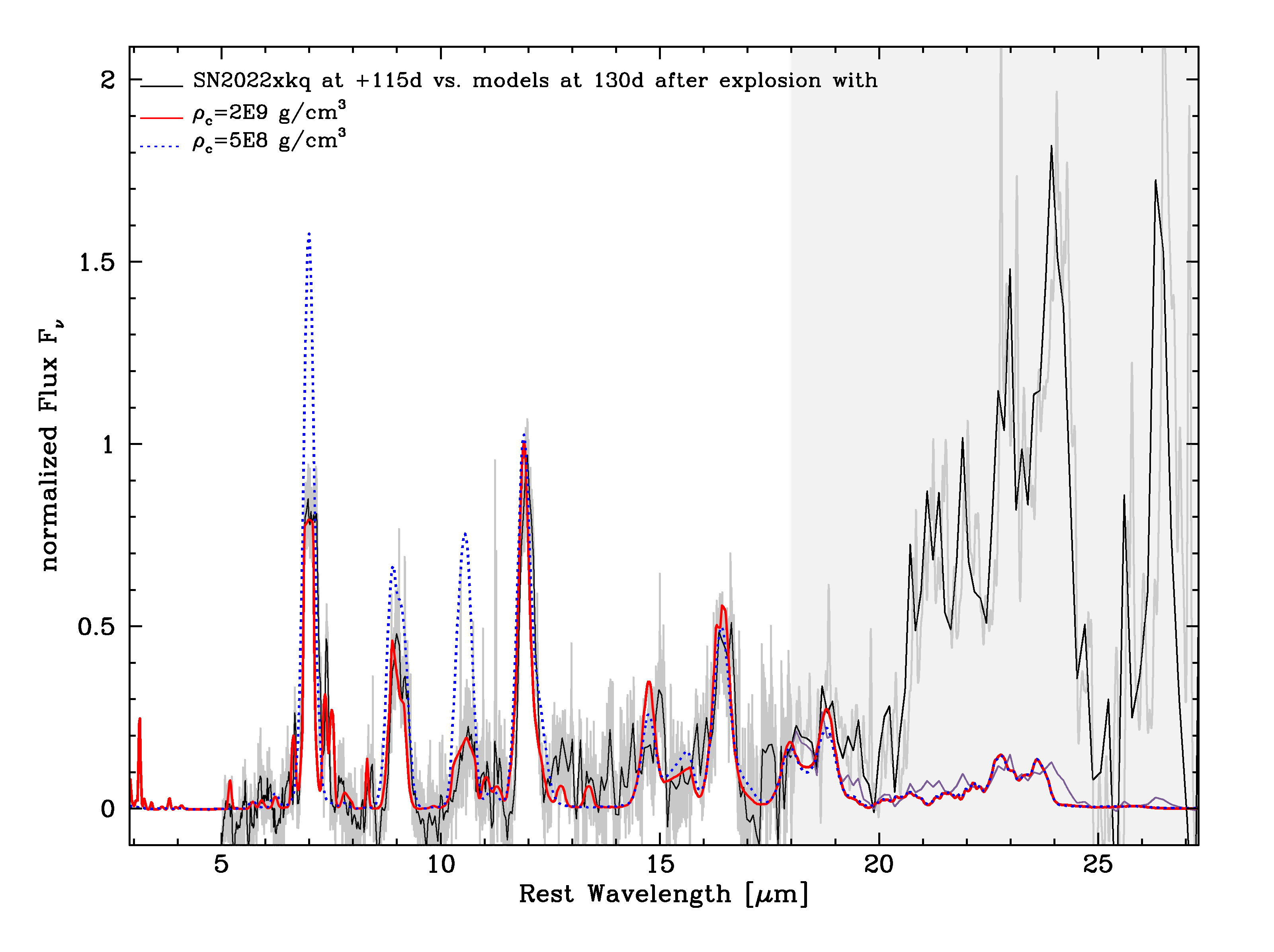}
    \caption{Comparison between the MIR spectra of \xkq in MIRI/MRS Channels 
    1--3 (solid black) and models with $\rho_c =~2\times 10^{9}$~\gcm (solid red) 
    and $5 \times 10^{8}$~\gcm (dotted blue) some 130 days after the explosion.
    The Channel 4 spectrum scaled to the flux level of the synthetic model 
    spectrum using 3~\mic-wide bins outside of the strong features is also
    shown (violet).
    The spectra in $F_\nu$ have been normalized to the peak of the
    [\ion{Co}{3}] at 11.888~\mic corresponding to a distance module of
    $32.52$~mag for our reference model. Note the strong [\ion{Ni}{1}] 
    feature at 3.12~\mic in the models (see text). The low central density 
    model strongly over-predicts the flux in the 7 and 9~\mic [Ar] 
    features as well as the 10.6~\mic [\ion{Ni}{2}]/[\ion{Co}{2}]
    blend because the energy input is shifted to higher density material.
    Higher density models show smaller effects on the spectrum than 
    low-density models (see text).}
\label{fig:synthetic}
\end{figure*}

\begin{figure*}[t]
    \centering
    \includegraphics[width=0.99\textwidth]{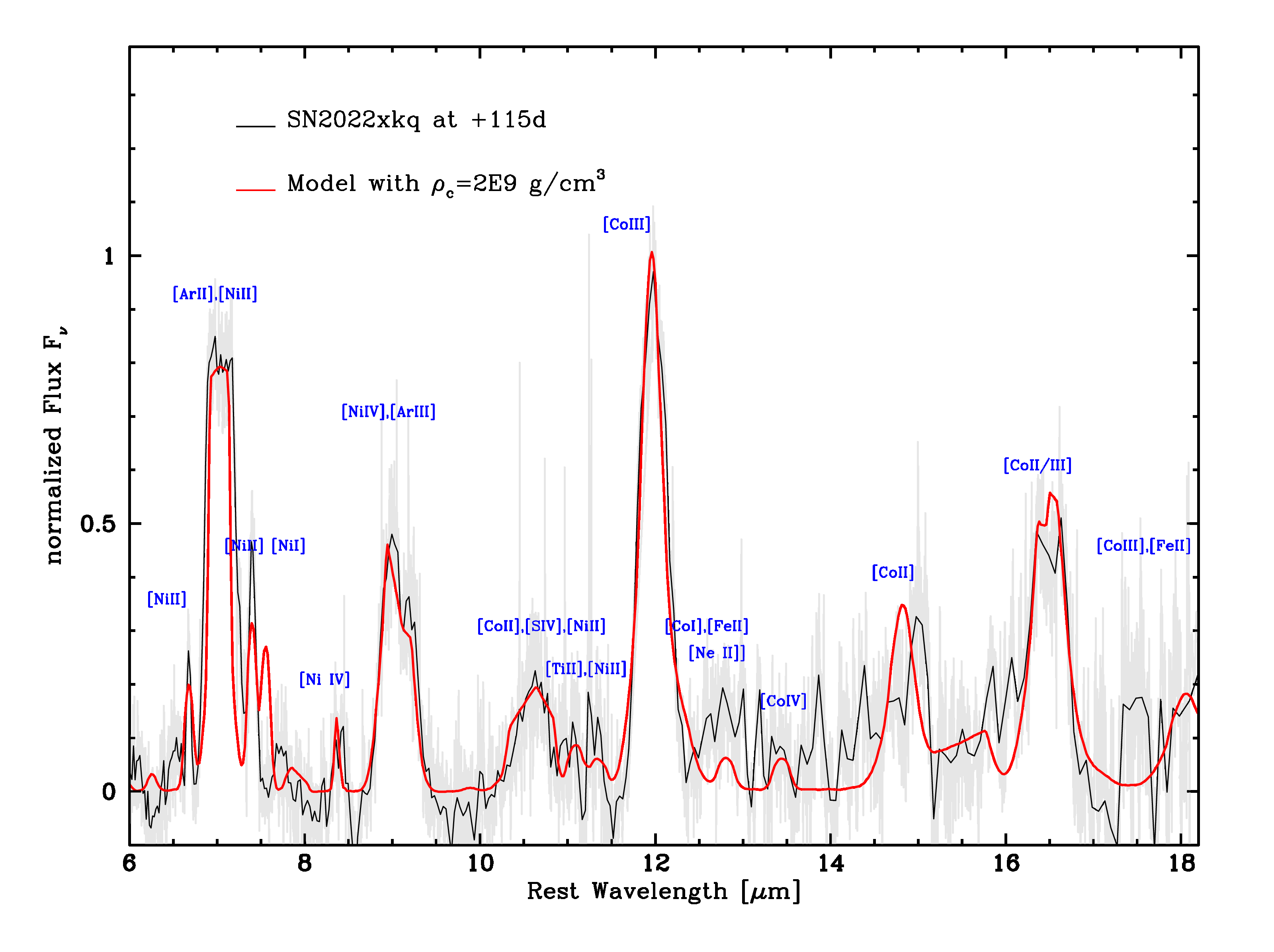}
    \caption{Same as \autoref{fig:synthetic} with the observed spectrum 
    for Channels 1-3 compared to the reference model, but with a tighter
    wavelength range. Strong features relevant for our interpretation
    are labeled. Note that many are blends (\autoref{tab:ir_lines}). Overall, 
    the profiles of the synthetic and observed features are in good agreement.  
    The peaks of many lines that form close to the photosphere are blue-shifted
    due to the blocking effect of the photosphere. However, the synthetic 
    [\ion{Ni}{1}] line at $\approx 7.5$~\mic and [\ion{Ni}{2}] at 
    $\approx 7.4~$\mic are too strong and slightly too weak, respectively 
    compared to the observations, suggesting a slightly higher non-thermal 
    excitation in the center of \xkq than predicted by the model, or 
    inhomogeneous mixing.}
    \label{fig:synthetic2}
\end{figure*}

The comparison between the observed MIR spectrum, the spectrum of our
reference model, and a low-density model is shown in \autoref{fig:synthetic}.
Only the reference model matches the observations.
As a reminder, the lines matching strong observed features are
already discussed in \autoref{sec:line_ids} and \autoref{tab:line_ids}.
The full list of significant optical and IR transitions in the
models are given in \Autoref{tab:ir_lines,tab:opt_lines}, and some are marked
in \autoref{fig:synthetic2}. We compare $F_\nu$ because the line features 
depend on the lifetimes thus, $F_\nu$ provides the proper scaling. The
spectra are normalized to the [\ion{Co}{3}]-dominated feature at
$11.888$~\mic. From the absolute flux in our models, the normalization
factor required to match the [\ion{Co}{3}]-dominated feature at
$11.888$~\mic corresponds to a distance modulus $\mu = 32.51$~mag. 
This is remarkably similar to the distance modulus derived
from SN independent methods, which find $\mu =  32.46 \pm 0.15$~mag
\citep{Pearson_etal_2023}. 

\begin{deluxetable*}{rcl|rcl|rcl|rcl|rcl}[ht]
  \tablecaption{Infrared Line Identifications at Day 130 in Reference Model \label{tab:ir_lines}}
  \tablehead{\colhead{\bf S} & \colhead{$\lambda$~[\mic]} & \colhead{Ion} 
    & \colhead{\bf S} & \colhead{$\lambda$~[\mic]} & \colhead{Ion}
    & \colhead{\bf S} & \colhead{$\lambda$~[\mic]} & \colhead{Ion}
    & \colhead{\bf S} & \colhead{$\lambda$~[\mic]} & \colhead{Ion}
    & \colhead{\bf S} & \colhead{$\lambda$~[\mic]} & \colhead{Ion}}
    \startdata
    {$\ast\ $} & 1.3209 & [\ion{Fe}{2}] & & 2.2187 & [\ion{Fe}{3}] & {$\ast\ $} & 5.6870$^{\dagger}$ & [\ion{V}{1}] & {$\ast\ \ast\ $} & 11.130 & [\ion{Ni}{4}] & {$\ast\ $} & 21.829 & [\ion{Ar}{3}] \\
    & 1.3210 & [\ion{Fe}{1}] & & 2.2425 & [\ion{Fe}{3}] & {$\ast\ $} & 5.7044 & [\ion{Co}{2}] & & 11.167 & [\ion{Co}{2}] & {$\ast\ $} & 22.297 & [\ion{Fe}{1}] \\
    {$\ast\ $} & 1.3281 & [\ion{Fe}{2}] & & 2.2443 & [\ion{Fe}{2}] & & 5.7391 & [\ion{Fe}{2}] & {$\ast\ $} & 11.238$^{\dagger}$ & [\ion{Ti}{2}] & & 22.80$^{\dagger}$ & [\ion{Co}{4}] \\
    & 1.3422 & [\ion{Fe}{1}] & & 2.3086 & [\ion{Ni}{2}] & & 5.8933 & [\ion{Ni}{1}] & & 11.307 & [\ion{Ni}{1}] & {$\ast\ $} & 22.902 & [\ion{Fe}{2}] \\
    & 1.3556 & [\ion{Fe}{1}] & & 2.3486 & [\ion{Fe}{3}] & & 5.9395 & [\ion{Co}{2}] & {$\ast\ \ast\ \ast\ $} & 11.888 & [\ion{Co}{3}] & {$\ast\ \ast\ \ast\ $} & 22.925 & [\ion{Fe}{3}] \\
    & 1.3676 & [\ion{Fe}{1}] & & 2.3695 & [\ion{Ni}{2}] & & 5.9527 & [\ion{Ni}{2}] & {$\ast\ \ast\ $} & 12.001 & [\ion{Ni}{1}] & & 23.086 & [\ion{Ni}{2}] \\
    {$\ast\ $} & 1.3722 & [\ion{Fe}{2}] & & 2.4139 & [\ion{Co}{1}] & {$\ast\ \ast\ $} & 6.2135 & [\ion{Co}{2}] & {$\ast\ $} & 12.1592$^{\dagger}$ & [\ion{Ti}{2}] & {$\ast\ $} & 23.196 & [\ion{Co}{2}] \\
    & 1.3733 & [\ion{Fe}{1}] & {$\ast\ $} & 2.5255 & [\ion{Co}{1}] & & 6.2730 & [ \ion{Co}{1}] & {$\ast\ \ast\ $} & 12.255 & [\ion{Co}{1}] & & 23.389 & [\ion{Fe}{3}] \\
    & 1.3762 & [\ion{Fe}{1}] & & 2.6521 & [\ion{Co}{1}] & & 6.2738 & [\ion{Co}{2}] & {$\ast\ $} & 12.2610 & [\ion{Mn}{2}] & & 24.04$^{\dagger}$ & [\ion{Co}{4}] \\ 
    & 1.4055 & [\ion{Co}{2}] & & 2.7256 & [\ion{Co}{1}] & & 6.3683 & [\ion{Ar}{3}] & {$\ast\ \ast\ $} & 12.286$^{\dagger}$ & [\ion{Fe}{2}] & {$\ast\ \ast\ $} & 24.042 & [\ion{Fe}{1}] \\
    {$\ast\ $} & 1.4434 & [\ion{Fe}{1}] & {$\ast\ $} & 2.8713 & [\ion{Co}{1}] & & 6.379 & [\ion{Fe}{2}] & {$\ast\ \ast\ $} & 12.642 & [\ion{Fe}{2}] & {$\ast\ \ast\ \ast\ $} & 24.070 & [\ion{Co}{3}] \\
    {$\ast\ $} & 1.4972 & [\ion{Co}{2}] & & 2.8742 & [\ion{Fe}{3}] & {$\ast\ \ast\ \ast\ $} & 6.6360 & [\ion{Ni}{2}] & {$\ast\ $} & 12.681 & [\ion{Co}{3}] & {$\ast\ \ast\ \ast\ $} & 24.519 & [\ion{Fe}{2}] \\
    {$\ast\ \ast\ $} & 1.5339 & [\ion{Fe}{2}] & & 2.9048 & [\ion{Fe}{3}] & & 6.7213 & [\ion{Fe}{2}] & {$\ast\ \ast\ $} & 12.729 & [\ion{Ni}{2}] & & 24.847 & [\ion{Co}{1}] \\
    {$\ast\ \ast\ \ast\ $} & 1.5474 & [\ion{Co}{2}] & {$\ast\ \ast\ $} & 2.9114 & [\ion{Ni}{2}] & {$\ast\ \ast\ $} & 6.9196 & [\ion{Ni}{2}] & {$\ast\ \ast\ $} & 12.729 & [\ion{Ni}{2}] & {$\ast\ \ast\ \ast\ $} & 25.249 & [\ion{S}{1}] \\
    & 1.5488 & [\ion{Co}{3}] & {$\ast\ $} & 2.9542 & [\ion{Co}{1}] & {$\ast\ \ast\ \ast\ $} & 6.9853 & [\ion{Ar}{2}] & {$\ast\ $} & 12.811 & [\ion{Ne}{2}] & {$\ast\ \ast\ $} & 25.689 & [\ion{Co}{2}] \\
    & 1.5694 & [\ion{Co}{2}] & & 3.0060 & [\ion{Co}{1}] & & 7.0454 & [\ion{Co}{1}] & & 13.058 & [\ion{Co}{1}] & {$\ast\ \ast\ $} & 25.890 & [\ion{O}{4}] \\
    {$\ast\ \ast\ $} & 1.5999 & [\ion{Fe}{2}] & & 3.0305 & [\ion{Co}{1}] & {$\ast\ $} & 7.103 & [\ion{Co}{3}] & & 13.820 & [\ion{Co}{3}] & & 25.986 & [\ion{Co}{2}] \\
    & 1.6073 & [\ion{Si}{1}] & & 3.0439 & [\ion{Fe}{3}] & {$\ast\ $} & 7.1473 & [\ion{Fe}{3}] & & 13.924$^{\dagger}$ & [\ion{Co}{4}] & {$\ast\ \ast\ \ast\ $} & 25.988 & [\ion{Fe}{2}] \\
    {$\ast\ $} & 1.6267 & [\ion{Co}{2}] & & 3.0457 & [\ion{Co}{1}] & & 7.2019 & [\ion{Co}{1}] & & 14.006 & [\ion{Co}{3}] & & 26.100 & [\ion{Co}{3}] \\
    {$\ast\ $} & 1.6347 & [\ion{Co}{2}] & {$\ast\ \ast\ \ast\ $} & 3.1200 & [\ion{Ni}{1}] & {$\ast\ $} & 7.3492 & [\ion{Ni}{3}] & & 14.356 & [\ion{Co}{1}] & & 26.130 & [\ion{Fe}{3}] \\
    {$\ast\ \ast\ \ast\ $} & 1.6440 & [\ion{Fe}{2}] & & 3.2294 & [\ion{Fe}{3}] & {$\ast\ \ast\ $} & 7.5066 & [\ion{Ni}{1}] & & 14.391 & [\ion{Co}{1}] & & 26.601 & [\ion{Fe}{2}] \\
    {$\ast\ $} & 1.6459 & [\ion{Si}{1}] & & 3.3942 & [\ion{Ni}{3}] & {$\ast\ $} & 7.7906 & [\ion{Fe}{3}] & {$\ast\ \ast\ $} & 14.739 & [\ion{Co}{2}] & & 27.530 & [\ion{Co}{2}] \\
    {$\ast\ \ast\ $} & 1.6642 & [\ion{Fe}{2}] & & 3.4917 & [\ion{Co}{3}] & {$\ast\ $} & 8.044 & [\ion{Co}{1}] & {$\ast\ $} & 14.8140 & [\ion{Ni}{1}] & & 27.550 & [\ion{Co}{1}] \\
    {$\ast\ $} & 1.6773 & [\ion{Fe}{2}] & {$\ast\ $} & 3.6334 & [\ion{Co}{1}] & {$\ast\ $} & 8.211 & [\ion{Fe}{3}] & {$\ast\ \ast\ $} & 14.977 & [\ion{Co}{2}] & & 28.466 & [\ion{Fe}{1}] \\
    {$\ast\ $} & 1.7116 & [\ion{Fe}{2}] & & 3.7498 & [\ion{Co}{1}] & & 8.2825 & [\ion{Co}{1}] & {$\ast\ \ast\ \ast\ $} & 15.459  & [\ion{Co}{2}] & & 29.675 & [\ion{Mn}{2}] \\ 
    {$\ast\ $} & 1.7289 & [\ion{Co}{2}] & & 3.8023 & [\ion{Ni}{3}] & {$\ast\ $} & 8.2993 & [\ion{Fe}{2}] & & 16.299 & [\ion{Co}{2}] & {$\ast\ \ast\ \ast\ $} & 33.038 & [\ion{Fe}{3}] \\ 
    {$\ast\ $} & 1.7366 & [\ion{Co}{2}] & {$\ast\ \ast\ $} & 3.9524 & [\ion{Ni}{1}] & {$\ast\ \ast\ \ast\ $} & 8.405 & [\ion{Ni}{4}] & {$\ast\ \ast\ \ast\ $} & 16.391 & [\ion{Co}{3}] & & 33.481 & [\ion{S}{3}] \\ 
    & 1.7413 & [\ion{Co}{3}] & {$\ast\ \ast\ \ast\ $} & 4.0763 & [\ion{Fe}{2}] & {$\ast\ $} & 8.6107 & [\ion{Fe}{3}] & {$\ast\ $} & 16.925  & [\ion{Co}{1}] & & 34.660 & [\ion{Fe}{2}] \\
    {$\ast\ $} & 1.7454 & [\ion{Fe}{2}] & & 4.0820 & [\ion{Fe}{2}] & {$\ast\ $} & 8.6438 & [\ion{Co}{2}] & {$\ast\ \ast\ \ast\ $} & 17.936 & [\ion{Fe}{2}] & {$\ast\ $} & 34.713 & [\ion{Fe}{1}] \\
    {$\ast\ $} & 1.7976 & [\ion{Fe}{2}] & {$\ast\ \ast\ $} & 4.1150 & [\ion{Fe}{2}] & {$\ast\ $} & 8.7325 & [\ion{Fe}{2}] & {$\ast\ $} & 18.2410 & [\ion{Ni}{2}] & {$\ast\ \ast\ \ast\ $} & 34.815 & [\ion{Si}{2}] \\
    {$\ast\ $} & 1.8005 & [\ion{Fe}{2}] & {$\ast\ $} & 4.3071 & [\ion{Co}{2}] & {$\ast\ $} & 8.9147$^{\dagger}$ & [\ion{Ti}{2}] & & 18.265 & [\ion{Co}{1}] & {$\ast\ \ast\ \ast\ $} & 35.349 & [\ion{Fe}{2}] \\
    {$\ast\ $} & 1.8099 & [\ion{Fe}{2}] & & 4.5196 & [\ion{Ni}{1}] & {$\ast\ \ast\ $} & 8.945 & [\ion{Ni}{4}] & {$\ast\ $} & 18.390 & [\ion{Co}{2}] & {$\ast\ $} & 35.777 & [\ion{Fe}{2}] \\ 
    & 1.8119 & [\ion{Fe}{2}] & {$\ast\ \ast\ $} & 4.6077 & [\ion{Fe}{2}] & {$\ast\ \ast\ \ast\ $} & 8.9914 & [\ion{Ar}{3}] & {$\ast\ \ast\ \ast\ $} & 18.713 & [\ion{S}{3}] & & 38.801 & [\ion{Fe}{1}] \\
    {$\ast\ $} & 1.9040 & [\ion{Co}{2}] & {$\ast\ \ast\ $} & 4.7881 & [\ion{Ni}{1}] & {$\ast\ $} & 9.1969$^{\dagger}$ & [\ion{Ti}{2}] & {$\ast\ \ast\ $} & 18.804 & [\ion{Co}{2}] & & 39.272 & [\ion{Co}{2}] \\
    {$\ast\ \ast\ $} & 1.9393 & [\ion{Ni}{2}] & & 4.8603 & [\ion{Fe}{3}] & & 9.279$^{\dagger}$ & [\ion{Co}{4}] & & 18.985 & [\ion{Co}{2}] & {$\ast\ \ast\ \ast\ $} & 51.301 & [\ion{Fe}{2}] \\
    & 1.9581 & [\ion{Co}{3}] & {$\ast\ \ast\ $} & 4.8891 & [\ion{Fe}{2}] & {$\ast\ $} & 9.618 & [\ion{Ni}{2}] & {$\ast\ $} & 19.0070 & [\ion{Fe}{2}] & {$\ast\ \ast\ \ast\ $} & 51.770 & [\ion{Fe}{3}] \\
    & 2.0028 & [\ion{Co}{3}] & & 5.0623 & [\ion{Co}{1}] & & 9.8195 & [\ion{Co}{1}] & {$\ast\ \ast\ $} & 19.056 & [\ion{Fe}{2}] & & 54.311 & [\ion{Fe}{1}] \\
    & 2.0073 & [\ion{Fe}{2}] & & 5.1635 & [\ion{Co}{1}] & {$\ast\ $} & 10.080 & [\ion{Ni}{2}] & & 19.138 & [\ion{Ni}{2}] & {$\ast\ $} & 56.311 & [\ion{S}{1}] \\
    & 2.0418 & [\ion{Ti}{2}] & {$\ast\ $} & 5.1796 & [\ion{Co}{2}] & {$\ast\ \ast\ $} & 10.1637$^{\dagger}$ & [\ion{Ti}{2}] & & 19.232 & [\ion{Fe}{3}] & & 60.128 & [\ion{Fe}{2}] \\
    {$\ast\ $} & 2.0466 & [\ion{Fe}{2}] & {$\ast\ \ast\ $} & 5.1865 & [\ion{Ni}{2}] & {$\ast\ \ast\ $} & 10.1890 & [\ion{Fe}{2}] & & 20.167 & [\ion{Fe}{3}] & & & \\
    & 2.0492 & [\ion{Ni}{2}] & & 5.2112 & [\ion{Co}{1}] & & 10.2030 & [\ion{Fe}{3}] & & 20.928$^{\dagger}$ & [\ion{Fe}{2}] & & & \\
    & 2.0979 & [\ion{Co}{3}] & & 5.3402 & [\ion{Fe}{2}] & {$\ast\ $} & 10.5105 & [\ion{Ti}{2}] &  & 21.17$^{\dagger}$ & [\ion{Fe}{1}] & & & \\
    & 2.1334 & [\ion{Fe}{2}] & & 5.4394 & [\ion{Co}{2}] & {$\ast\ $} & 10.5105 & [\ion{S}{4}] & & 21.986$^{\dagger}$ & [\ion{Fe}{2}] & & & \\
    & 2.1457 & [\ion{Fe}{3}] & & 5.4652$^{\dagger}$ & [\ion{V}{1}] & {$\ast\ \ast\ \ast\ $} & 10.523 & [\ion{Co}{2}] & & 22.106$^{\dagger}$ & [\ion{Ni}{1}] & & & \\
    & 2.1605 & [\ion{Ti}{2}] & {$\ast\ $} & 5.6739 & [\ion{Fe}{2}] & {$\ast\ \ast\ $} & 10.682 & [\ion{Ni}{2}] & {$\ast\ $} & 21.4810 & [\ion{Fe}{2}] & & & \\
    \enddata
    \tablecomments{The relative strengths are indicated by the number of $*$. For transitions without known 
    lifetimes (marked by $^{\dagger}$), $A_{i,j}$ are assumed from the equivalent iron levels.
    The transitions observed in \xkq above the threshold flux level are listed in \autoref{tab:line_ids}.}
\end{deluxetable*}

Note, that the underlying continuum flux is dominated by a
quasi-continuum 
\citep{1977ApJ...214..161K,1993A&A...268..570H,2012MNRAS.424..252H}
produced by allowed line transitions, electron scattering and
free-free radiation in a scattering-dominated MIR photosphere
expanding with a velocity of about 1,500 to 2,000~\kms, and particle
densities of $\approx 1-2\times 10^{7}$~cm$^{-3}$ which are close to the
critical density for collisional de-excitation. Note that, based on
non-LTE models, this effect was taken into account for the
interpretation of many nebular spectra
\citep[e.g.,][]{2004ApJ...617.1258H,Telesco2015,Diamond2015,Diamond_etal_2018,DerKacy2023}.
The high particle density and the presence of a photosphere leads to
weaker features in \Nife than expected otherwise and potentially,
blocks the emission of the most neutron-rich isotopes that form when
$\rho_c$ is high. These isotopes have been observed in SN~2016hnk
\citep{Galbany2019}, and may be observed in spectra of \xkq at later
epochs. Moreover, the presence of a photosphere blocks part of the
redshifted emission of forbidden lines formed in the outer optically
thin layers. This leads both to a blue-shift vs. the rest-wavelengths of
transitions formed close to the photosphere (compare
\autoref{fig:synthetic2} and \autoref{tab:ir_lines}) for \xkq. Due to the
photosphere receding with time and the decreasing blocking of the 
red-shifted emission components, the width of features by singly and 
doubly ionized transitions increases with time \citep{Penney2014} as 
been observed in many SNe, for example, [\ion{Fe}{2}]~1.644~\mic in 
SN~2014J and [\ion{Co}{3}]~11.888~\mic \citep{Diamond_etal_2018,Telesco2015}. 
This effect vanishes by day $\approx 200$. Note that, in general, the shift
of both the blended and unblended features of \xkq and the models are 
consistent within the spectral resolution (\autoref{fig:synthetic}). Here, 
the blue-shifts in the almost unblended [\ion{Ni}{2}] at 6.636~\mic are
$-460 \pm 110$~\kms in the observation (\autoref{fig:ni2_fit}) compared to 
$\sim -400$~\kms in the model (\autoref{fig:synthetic2}). \citet{Penney2014} 
found a shift of $\approx 2,000 $~\kms at day 100 for normal-luminosity models 
with a photosphere at $\approx 5,500$~\kms. The velocity shift is larger 
because the photosphere forms further out.

Another effect of the presence of the photosphere is a slight
asymmetry in the line profile, not to be mistaken with overall
asymmetries in the ejecta \citep{Penney2014}. Overall, the synthetic 
and observed spectra agree well (\autoref{fig:synthetic2}). Most
features are blended with weak transitions (\autoref{tab:ir_lines}).
The main features in the model are at 7~\mic ([\ion{Ar}{2}],
[Ni {\sc i}--~{\sc iii}]), 9~\mic ([\ion{Ar}{3}], [\ion{Ni}{4}]), a
group at 10.6~\mic ([\ion{S}{4}], [\ion{Co}{2}], [\ion{Ni}{4}]),
11.8~\mic ([\ion{Co}{3}], with some [\ion{Co}{1}] and [\ion{Fe}{2}]),
14.8--16.0~\mic ([\ion{Co}{2}], [\ion{Co}{3}]],
[\ion{Fe}{2}]), $16.3$ [\ion{Co}{3}], [\ion{Fe}{2}],
and 17.5--19~\mic ([\ion{Fe}{2}], [\ion{S}{3}]. Additionally, some weaker 
features of [\ion{Ti}{2}] can be seen at 9.19 \mic (\autoref{tab:ir_lines}),
expected for lower luminosity \sneia, and placing \xkq at the lower
end of transitional \sneia.

\begin{figure}[t]
    \centering
    \includegraphics[width=0.47\textwidth]{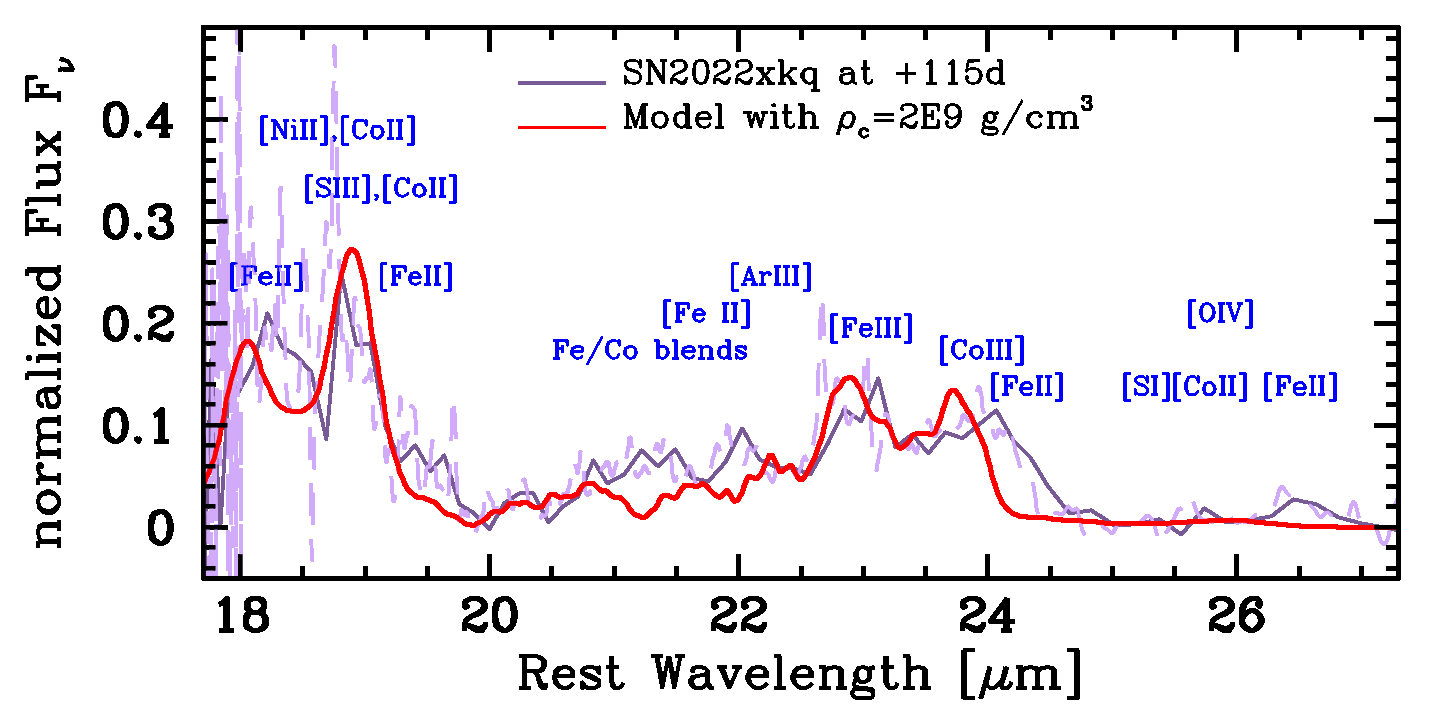}
    \caption{Same as \autoref{fig:synthetic} with the observed, calibrated 
    spectrum in Channel 4 with (violet) and without smoothing (light violet) 
    compared to the reference model (red). Strong features relevant for our 
    interpretation are labeled. Note that many are blends (\autoref{tab:ir_lines}). 
    Overall, the synthetic and observed strong features are in reasonable 
    agreement. Note the strong [\ion{S}{3}] is indicative for a high mass 
    WD (see text).}
    \label{fig:synthetic4}
\end{figure}

In Channel 4, the strong features only become apparent in the spectrum
of \xkq after scaling the flux by the synthetic spectrum
(\autoref{fig:synthetic4}). The scaling factor is wavelength dependent
and varies from a value of about 1 at 18~\mic to about 50 at 
26~\mic. Overall, the spectra in Channel 4 are dominated by blends of 
singly and doubly ionized Fe and Co with peaks at $\sim$18, 23 and 
24~\mic. In our models, the [\ion{S}{3}] at 18.713~\mic dominates 
the feature at 19~\mic. Mixing on scales of the positron mean free path, 
would further enhance this feature. The effect is similar to results on 
the photospheric \ion{S}{1} at 1.0821~\mic \citep{Diamond2015}. Even 
without mixing, the [\ion{S}{3}] feature is an indication of a massive 
progenitor. \Mch models produce a large amount of intermediate mass 
elements (IMEs) at the expense ejecta undergoing burning to NSE.
In the models, we see a hint of the [\ion{S}{1}] feature at 25~\mic. It 
is not prominent because the direct and significant heating by $\gamma$-rays 
leads to mostly ionized \ion{S}{2} (\autoref{fig:ion}). This feature is 
expected to grow with time as heating will transition from $\gamma$-rays 
to positrons by $\approx 200-300$ days. At wavelengths longwards of 18~\mic, 
the appearance of [\ion{S}{3}] is a direct consequence of the large
amount of Si and S characteristic of \Mch scenarios. Note the possible 
importance of high S/N observations in the MIRI/MRS Channel 4 to probe  
for $\rho_c $ larger than $4\times 10^{9}$~\gcm, because lines of 
[\ion{Cr}{5}] and [\ion{Mn}{4}] would appear at 19.62 and 22.08~\mic, 
respectively. 

For many of the elements, multiple ionization stages are present, which
lends support to the ionization structure discussed above. Within
\Mch scenarios, lower densities significantly degrade the fit by
boosting the emission by features dominated by [\ion{Co}{2}] at $10.6$~\mic.
In our models, lower $\rho_c$ results in more heating by \Cofs
at lower velocity as the size of the electron capture region recedes. 
This shifts a significant amount of energy input into the high density
region with the lower ionization stage dominating (\autoref{fig:ion})
shifting the [\ion{Co}{3}]/[\ion{Co}{2}] ratio.  With the spectra
normalized to [\ion{Co}{3}] and little effect due to Ar, the corresponding
features (e.g. 7, 9~\mic) appear more prominent.  

Our best fit model produces $\sim 6 \times 10^{-2}$~\Msun of \Nife.
About 50\% of it is located within the photosphere.  Mixing of the
innermost regions may expose a larger fraction of the \Nife but, as
discussed below, the narrow Ni lines put strong limits on extended
mixing. Note that we cannot exclude higher central density models as
used in \citet{Galbany2019} because the highest-density burning
layers, with the lowest electron/neutron ratio, $Y_e$ would be hidden
below the photosphere  \citep[see Figs. 25-26 in][]{Galbany2019}.  
The [\ion{Ni}{1}] line at about 
7.5~\mic is predicted by the model, but is clearly absent in the
observation. To reproduce the model prediction of [\ion{Ni}{1}] it 
would have to be blue-shifted by $\approx 2,500$~\kms and blended
(\autoref{fig:synthetic2}). The shift of [\ion{Ni}{1}] in the inner,
high-density layers is due to recombination rates that depend on
$\rho ^2$. Higher $\rho_c $ does not necessarily produce significantly
more \Nife because the nuclear statistical equilibrium shifts to
more neutron-rich isotopes \citep[see Fig. 25 in][]{Galbany2019}.
If the \Nife forms a shell it would be unstable to Rayleigh-Taylor~(RT) 
mixing. Without detailed simulations, the impact on the line profile
is hard to predict. The difference between the models
and data, can be due to several effects: 
(1) a higher $\rho_c$ and a corresponding shift towards more 
neutron-rich isotopes reducing $Y_e$,
\citep{Brachwitz2000,1998ApJ...495..617H,2021ApJ...923..210H}; 
(2) some moderate, inhomogeneous mixing \citep{fesen2014}; 
and (3) some passive drag by a turbulent field produced during the 
smoldering phase when the flame speed is low (close to the laminar 
speed) for $\approx 1-2$~s prior to central ignition
\citep{1995ApJ...449..695K,1997ASIC..486..475K,2000ApJ...528..854D}. 
See the velocity field shown in Fig. 9 of \citet{Hoeflich_Stein_2002}.
These effects may be reproduced in a finely tuned model, which is 
beyond the scope of this work.
We abstain from further tuning $\rho_c$ due to the lack of high S/N
spectra beyond 18~\mic needed to verify the presence of neutron rich
isotopes (see C.~Ashall et al. 2023, in prep.). 

\subsubsection{Line Profiles}\label{sec:lineprofiles}

\autoref{fig:synthetic2} shows the model and observed spectra, with
some of the stronger features identified. Ni lines appear from neutral
through triply ionized ions. Peaked profiles for iron-group elements 
are the direct result of the energy input being dominated by 
$\gamma$-rays (\autoref{fig:synthetic2}). The model predicts the 
correct line widths of [\ion{Co}{2}] and [\ion{Co}{3}]. 

The 7 and 9~\mic features are both dominated by Ar and blended
by [\ion{Ni}{2}], but the 9~\mic feature is heavily blended with 
[\ion{Ni}{4}]. Compared to normal-luminosity \sneia, the Ar layers 
are expanding at lower velocities. Within \Mch explosions, this can 
be understood as a consequence of the dominant production of 
quasi statistical equilibrium in low luminosity supernovae. As a 
consequence, in sub-luminous SNe, Ar (and Ca) are elements formed 
during the breakout to NSE and appear at lower velocities, resulting 
in narrower profiles. The profile is mostly flat, but narrower 
compared to previously observed normal-luminosity \snia 
(\autoref{fig:mir_sample}). As a result, the Ni features in the 
blue end of the 7~\mic profile are well separated from the strongest 
[\ion{Ni}{2}] line at 6.636~\mic though somewhat contaminated by 
the weak {\ion{Ni}{2}} at 6.920~\mic. Note that the 
[{\ion{Ar}{2}}]-dominated 7~\mic feature has a `dome-shaped' rounded 
top produced by a narrow line blending rather than asymmetry, with 
a velocity corresponding to $\approx 3,800 $~\kms, consistent with 
the central hole in the argon distribution (\autoref{fig:models}).
The 7~\mic feature is slightly too narrow compared to the
observation, which may indicate some RT driven mixing.
In the model, the dome shape is produced by [\ion{Ni}{2}].  
While our models cannot explicitly discriminate between whether the
dome shape is due to the [\ion{Ni}{2}] blend or if the model is 
slightly too dim, the presence of the small [\ion{Ni}{1}] feature 
suggests that it is slightly too dim.

In contrast, the 9~\mic [\ion{Ar}{2}] profile is heavily 
distorted/tilted by `low velocity' [\ion{Ni}{4}] (and some 
[\ion{Ti}{2}]). The synthetic and the observed profiles are 
due to blending. The result is an asymmetric, peaked profile, 
making it unusable as an indicator for chemical asymmetry in 
the explosion.

The narrower [\ion{Co}{3}] at 11.888~\mic found in \xkq as compared 
to normal-luminosity \sneia is consistent with the fact that the 
NSE region becomes more concentrated towards the inner region in 
under-luminous \sneia; or more precisely, the QSE region grows at 
the expense of NSE elements
\citep{Hoeflich1996,Mazzali2020,Hoeflich_etal_2002,Ashall18} (see
also \autoref{sec:co3_lines}). 
It is amplified by the blocking due to the photosphere that cuts 
off parts of the red-shifted contribution to the line emission
(see \autoref{sec:syntheticSpectra}).

The model successfully reproduces the narrow [\ion{Ni}{2}] and
[\ion{Ni}{4}] lines observed providing a possible path to probing the
physics of the thermonuclear runaway. Within the framework of \Mch
explosions, the ignition process and its location is highly debated,
ranging from central ignition \citep{khokhlov93,Gamezo2003} to multi-spot
ignition in $\approx 100$ spots distributed over more than a hundred km
\citep{2013MNRAS.429.1156S}. Detailed simulations of the simmering phase
suggest a single-spot ignition close to the center at $\approx 35-50$~\kms
\citep{Hoeflich_etal_2002,2011ApJ...740....8Z}, but possibly out
to 100~\kms \citep{2011ApJ...740....8Z}. For the normal bright SN~2021aefx,
\citet{Blondin2023} found that the delayed detonation model based on a
strong off-center multi-spot ignition \citep{2013MNRAS.429.1156S} fails
to reproduce the strength of Ni lines by factors of 2 to 3, whereas
delayed-detonation models starting from a single-spot ignition can
reproduce the Ni lines within the model uncertainties \citep{DerKacy2023}. 
For \xkq, we find good agreement in the Ni lines with single spot
ignition. Because the early deflagration phase may be similar, this may
indicate that single-spot ignition is common.

\begin{figure}
    \includegraphics[width=0.45\textwidth]{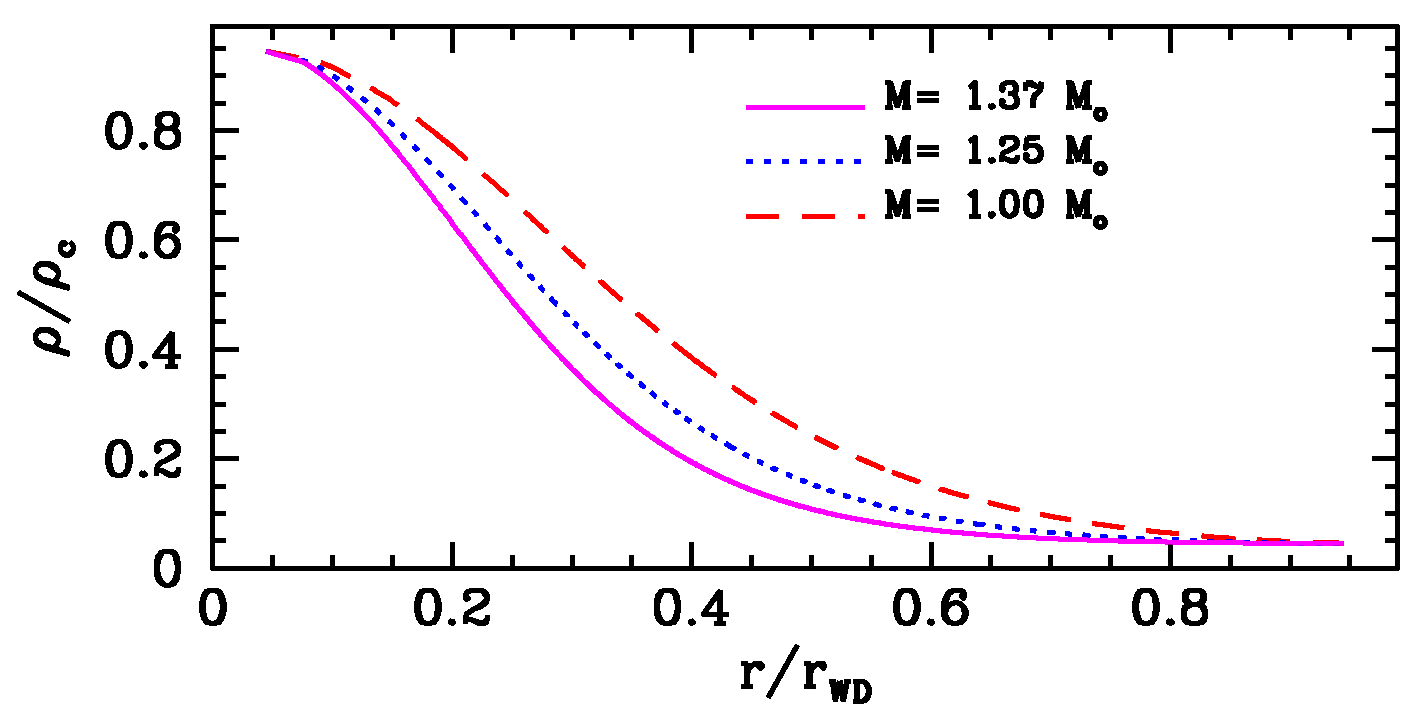}
    \caption{Normalized density as a function of radius. With increasing 
    mass, they become more centrally concentrated (from \citealp{2017hsn..book.1151H}).
    Note that M(WD) $\approx 1.25$~\Msun is the dividing 
    line between \Mch and sub-\Mch explosions.}
    \label{fig:wd}
\end{figure}

\subsubsection{Alternative Scenarios}

Detailed spectral fits of a wide variety of scenarios is beyond 
the scope of this paper, but constraints can be found by 
combining spectral indicators.

The observed [\ion{Co}{2}]/[\ion{Co}{3}] ratio requires high 
$\rho_c$, compared to normal-luminosity \sneia 
\citep{Diamond2015,Telesco2015,Diamond_etal_2018} in \Mch explosions 
because, more electron capture shifts the emission powered by 
radioactive decays towards lower densities. In contrast, for 
sub-\Mch mass explosions such as helium-triggered detonations 
\citep{Shen2018,Polin2019,Boos2021} or low density mergers
\citep{Garcia-Berro2017}, lowering M$_{\rm WD}$ and $\rho_c$ will 
decrease the average density of spectral formation, reducing the 
average [\ion{Co}{2}]/[\ion{Co}{3}] ratio. One of the effects that 
changes the ionization balance is the overall flatter density 
distribution in the initial sub-\Mch WD (see \autoref{fig:wd}), 
the other is the lowering of binding energy with decreasing M$_{\rm WD}$,
which changes the velocity distribution of the ejecta. 
Sub-\Mch explosions can produce an under-luminous \sneia like
\xkq for progenitors with $0.9 \le M_{WD} \le 1.0$~\Msun, but the
central densities are too low for the production of electron capture elements. 
However, stable Ni is seen in \xkq.

In the case of a sub-\Mch WD, a low $Y_e$ may be inherited from
the progenitor, due to an overabundance of $^{22}$Ne produced during 
the stellar burning in massive stars. As suggested by \citet{Blondin2022}, 
super-solar metallicity will shift $Y_e$ resulting in the production of 
\Nife in \Nifs in the QSE regions resulting in \Nife features as broad as 
the nebular \Cofs line. However, the Ni lines in \xkq are narrow 
(FWHM $\approx 2,500 $~\kms, \Autoref{fig:ni2_fit,fig:synthetic2}) rather 
having a FWHM of $\approx 10,000$~\kms \citep[see e.g.][]{Shen2018}, excluding 
this path. 

Alternatively, a low $Y_e$ and thus, some central Ni may be produced 
in sub-\Mch progenitor systems that contain very old WDs. Over time
scales of $\sim 5$ billion years, gravitationally driven diffusion may
allow $^{22}$Ne settling in the core of a crystallized WD \citep{Deloye_Bildsten_2002}. 
For an initial composition of solar metallicity, the amount may or
may not be sufficient to account for some [\ion{Ni}{2}] emission
observed at about 1.9~\mic in some other \sneia
\citep[e.g.,][]{Friesen2014,Diamond2015,Hoeflich2021,Blondin2023}.
However, in \xkq, we see Ni in all ionization stages. The amount of
stable Ni produced in old, low mass WDs is too small by a factor of
3--4 to account for the observed line strengths. The factor needs to be 
even larger because old stars in spiral galaxies have sub-solar 
metallicities. Moreover, the \Nife would settle at velocities below 
the photosphere at 130~days after the explosion, requiring mixing. 

Other alternative scenarios are dynamical mergers of two WDs where
the C/O is ignited on dynamical time-scales by interaction:
a) starting with grazing incidence (classical mergers),
b) violent mergers, or
c) direct collisions
\citep{Benz1990,Rosswog2009,Kushnir2013,Pakmor2013,Garcia-Berro2017}.
All three would lead to asymmetric envelopes in density or
abundance structures.

Unlike the 35 normal-bright \sneia \citep{2019MNRAS.490..578C} for
which polarization has been observed, the under-luminous SN~1999by 
and SN~2005ke show a qualitatively different polarization spectrum 
around peak brightness, indicating an overall rotationally symmetric 
photosphere with an axis ratio of $\approx 0.9$ based on detailed non-LTE
models \citep{Howell2001,Patat_etal_2012}.
Dynamical mergers have been suggested and discussed as
possible alternatives to \Mch explosions of rapidly rotating WDs
opening up the possibility that under-luminous \sneia are a population
distinct from normal-bright \sneia \citep{Patat_etal_2012,Hoeflich2023}.
So far, no late-time spectropolarimetry has been obtained
for under-luminous SNe~Ia that would produce the flip in the polarization
angle in configurations found in head-on collisions of two WDs
\citep{1995ApJ...443...89H,2016MNRAS.455.1060B}.

The intermediate state in classical mergers (for almost all mass 
ratios between the WDs) produces a puffed up, low density 
quasi-hydrostatic phase without the production of a significant amount 
of stable Ni \citep[for a review, see][]{Garcia-Berro2017}. However, 
violent mergers can produce a lot of \Nife and show strong MIR Ni 
features. In fact, the strength of the Ni features far exceed the 
observations of the over-luminous SN~2023pul \citep{Kwok_etal_2023_23pul} 
and the under-luminous \xkq. For both \sneia in the MIR, the observed MIR
[\ion{Ar}{2}] lines are too strong and the MIR Ni features are too weak
compared to the model by an order of magnitude. However, it is unclear
whether this is a generic property of this class of explosions.

A direct collision with parameters closer to those suggested for the
under-luminous SN~2007on may be more likely \citep{2018MNRAS.476.2905M}.
This scenario, however, resulted in the ignition of both WDs, a `double line'
pattern in all elements and a significant shift in Ni
\citep[$\approx 1,500-3,000$~\kms,][]{Dong2015,2018MNRAS.476.2905M,Vallely2020}.  
Neither of these effects are compatible with \xkq unless seen `equator on'. 
In addition, we see only one component in the narrow Ni line severely
limiting even small deviations from the equatorial viewing direction,
unless only one of the WDs is close to \Mch. We note that polarization
spectra and their evolution would be very unlike the two under-luminous
\sneia mentioned above. We consider WD collisions as the explosion 
mechanism for \xkq to be unlikely, but without detailed modeling (beyond 
the scope of this work) cannot definitely rule them out.
 
\section{Results} \label{sec:summary}

The main findings of our analysis are as follows:

\begin{itemize}

\item The spectrum of the under-luminous \xkq is distinct compared to
spectra of normal-luminosity \sneia. The MRS/MIRI spectrum permits 
many detailed inferences including the first lines identified beyond 
14~\mic in \snia observations. These newly identified lines include
[\ion{Co}{2}]-dominated blends between 14--17~\mic, and complex 
blends of several iron-group elements with contributions from [\ion{S}{3}],
[\ion{Ar}{2}], [\ion{S}{1}], and [\ion{O}{4}] at $\lambda > 18$~\mic.

\item The identification of [\ion{Ti}{2}] lines in the
blended MIR features are consistent with \xkq being an under-luminous
\snia.

\item The stronger ratio of [\ion{Ar}{2}]/[\ion{Ar}{3}] in \xkq
relative to the normal-luminosity spectra at similar epochs suggests a
shift in the ionization balance toward singly ionized species over
doubly ionized ones. 

\item In observations at these phases, line formation is a
complex mixture of allowed lines, forbidden transitions, a
pseudo-continuum, and other radiative transfer effects. The
combination of fits and data driven measurements to the
[\ion{Ni}{2}]~6.636~\mic, [\ion{Ni}{3}]~7.349~\mic, and
[\ion{Co}{3}]~11.888~\mic lines reveal: a) that the peaks of 
the IGEs are blue-shifted by $\sim 200-400$~\kms relative to 
the rest wavelength of the dominant line; and b) that these 
features are narrow in width when compared to the same lines in 
normal-luminosity \sneia. Fitting of the line profiles with 
simple analytic functions (e.g. Gaussians) must be approached 
with caution as they do not capture the complex line formation 
physics, even in some isolated lines with little
blending.  

\item A tentative correlation is seen between the FWHM of the
[\ion{Co}{3}]~11.888~\mic resonance line and the light curve shape
parameter \DmB. This relationship is consistent with the \Nifs being
produced in NSE burning occurring closer to the center in
under-luminous objects. Future MIR observations should continue to add
to this plot to see if it follows the luminosity-width relation
observed in the optical.

\item The light curve properties (\autoref{tab:model_params}) of the
model are consistent with the early-time data of \xkq \citep{Pearson_etal_2023}, 
as well as other under-luminous \sneia such as SN~2007on and SN~2011iv \citep{gall2018}.
While \xkq is at the lower luminosity end of the transitional \sneia 
distribution, it is significantly (1.5~mag), brighter, than 
SN~1991bg-like objects (\autoref{sec:models}).

\item The spectral characteristics can be understood within the 
framework of high-density burning in a delayed-detonation model 
in which somewhat more mass is consumed in the deflagration prior 
to the onset of the DDT, seen equator on (\autoref{fig:synthetic}). 
In particular, the shift from a predominately doubly ionized 
spectrum to a mix of ionization I-III is consistent with the low 
luminosity. The shift in the ionization balance towards lower stages 
is obvious (\autoref{fig:ion} and \autoref{sec:syntheticSpectra}).
Moreover, the strongest features in the models are also the strongest
features observed, and all weak features in the observations have their
equivalent in the model (compare \autoref{tab:line_ids} with
\autoref{tab:ir_lines}). 

\item Electron capture in the central region has a significant 
impact on the spectra: With increasing $\rho_c$, the higher average 
density shifts the ionization balance from [\ion{Co}{3}] to 
[\ion{Co}{2}]. As a result, the [\ion{Co}{2}] lines are stronger 
(\autoref{fig:synthetic}). Furthermore, the large number of Ni 
features indicates electron capture (\autoref{tab:ir_lines}).

\item The sensitivity of intermediate mass vs. iron-group elements 
and the important role of high resolution in the MIR accessible with 
{\it JWST} has been demonstrated as a key to constrain the underlying 
physics and to test models (\autoref{fig:synthetic}). 

\item At $\sim$130~days after explosion, the spectra form during the
onset of the nebular phase. In our models, the optical to MIR photosphere 
is formed by a quasi-continuum at $\approx 2,000~$\kms hiding the 
innermost layers. This photospheric blocking is consistent with the 
shift of the peak of the emission features relative to the rest wavelength 
by a few hundred \kms (\Autoref{fig:mir_sample,fig:synthetic2}).
As discussed in \autoref{sec:models_structure}, only some of the 
electron capture elements in the models are above the photosphere and 
potentially, electron capture elements such as Cr, and Mn are hidden. 
Within the framework of \Mch mass explosions, the central density of 
the WD $\rho_c = 2 \times 10^{9}$~\gcm and the \Nife mass may be 
regarded as a lower limit.

\item \xkq is likely to have been observed close to `equator-on' and/or
the point of the DDT is within the low-velocity distribution of \Nifs
(\Autoref{sec:syntheticSpectra,sec:lineprofiles}). Line polarization 
should be largest when seen from near to the equator, but vanishes when 
seen from other inclinations showing the importance of spectropolarimetry 
during the photospheric phase. 

\item The line profiles of [\ion{Ar}{2}] and [\ion{Co}{3}] show a
narrow flat top and are rounded, respectively. This is expected from 
under-luminous models for the explosion of a massive WD with large 
pre-expansion during the deflagration phase prior to the DDT, resulting 
in overall an increased production of IMEs (Si/S/Ca/Ar) compared to that
produced in normal-luminosity \sneia (\autoref{fig:mir_sample}).

\item The [\ion{Ar}{3}] profile at 9~\mic is strongly tilted because 
it is contaminated by Ti and Ni blends. In under-luminous \sneia, it 
cannot be used to infer asymmetry in the density distribution 
(\autoref{sec:lineprofiles}).

\item The line profile of [\ion{Ar}{2}] at 7~\mic is better suited 
to deciphering the location of the DDT and the viewing angle of 
under-luminous \snia. This is because line-blending is reduced due 
to the lower overall velocities. The size of the `Ar hole' is 
slightly smaller in our models than inferred from the observations, 
but suggests that \xkq is seen from the equator. The domed-shaped 
profile is a result of line-blending and not asymmetry  
(\autoref{sec:lineprofiles}).

\item The [\ion{Co}{3}] feature at 11.888~\mic shows a rounded, broad
profile characteristic for a central \Nifs-free region. This is also 
supported by and consistent with the numerous narrow \Nife lines 
(\autoref{sec:syntheticSpectra}).

\item The strong [\ion{S}{3}] at $\approx 19$~\mic is indicative of
a massive WD, as in under-luminous \sneia the IME production is 
increased compared to normal-luminosity \sneia (see 
\Autoref{fig:synthetic_best,fig:synthetic4}).

\item In under-luminous \sneia, the ionization balance is shifted 
towards lower states (\autoref{fig:ion}) and in the central region, 
neutral IGEs become significant. This is also seen as a the lower half
width of both electron capture lines and lines from neutral Fe, Co, Ni
when compared to normal-luminosity \sneia.

\item The lines of stable \Nife are clearly identified and are consistent
between models and observations. At $\sim 130$~d past the explosion, the
densities of the central region remained high, $> 10^7$~\gcm up to
$\approx 2,000$~\kms resulting in Ni emission less pronounced than
expected in the nebular phase (\autoref{sec:models}). The narrow
\Nife lines suggest a central ignition mixing of electron capture elements 
produced by pre-existing turbulent fields rather than by RT-instabilities
as would be expected from multi-spot, off-center ignition
(\autoref{sec:lineprofiles}).

\item A central DDT white C/O-rich material mixed-down can all but be
excluded because the \Nife lines are narrow. For the first time, this
may constrain the mechanisms possible for the DDT 
(\Autoref{sec:models,sec:lineprofiles}). 

\item The synthetic fluxes in MIR spectra (\autoref{sec:model_spectra}) 
are consistent with the observations of \xkq. The MIR has the advantage 
of insensitivity to reddening.

\end{itemize}

\section{Discussion} \label{sec:disc}

Compared to previous low-resolution observations, the additional
insight afforded by MIRI/MRS is readily apparent, especially as it
relates to the increased wavelength coverage of MRS relative to LRS
and the higher resolution. In \xkq, the additional wavelength
coverage reveals multiple [Co]-dominated features in the 
14--17~\mic range, as well as complex blends of select transitions 
from [O], [S] and [Ar] along with many blended features 
comprised of [Fe], [Ni], and [Co] lines beyond $\sim20$~\mic.

The decrease in velocity errors from $\gtrsim 2,000$~\kms to
$\sim 100$~\kms allows us to accurately measure small-scale velocity 
line shifts (see \Autoref{fig:ni2_fit,fig:ni3_fit,fig:co3_fit}), to
assess whether they originate from asymmetries (such as those originating
from off-center DDTs), line formation effects, or
temporal evolution from sources such as positron transport (at later
epochs). This increased resolution also affords us the ability to 
better determine the shape of line profiles throughout the spectrum. 
In particular, the width and shape of the [\ion{Ar}{2}] and
[\ion{Ar}{3}] lines from weak blends permits the estimation of the
viewing angle of the supernova if observed at the appropriate
epochs. Furthermore, these Ar lines probe the transition between the NSE and
QSE regions during burning and provide clear discriminators
between different explosion models. Finally, the increase in resolution
compared to LRS data allows for previously blended lines to be
resolved, such as the [\ion{Ni}{2}]~6.636~\mic and
[\ion{Ni}{3}]~7.349~\mic lines. This provides
tighter constraints on the physical conditions present in
the ejecta.

The apparent correlation between the FWHM of the
[\ion{Co}{3}]~11.888~\mic line and the light curve shape parameter
offer tantalizing hope for the existence of a MIR spectroscopic
luminosity-width relation that may be similar to  the
luminosity-width relation \citep{Phillips1993}.
As more data are added to this plot, the relationship may provide an
independent method to calculate the luminosity of a \snia with a single
nebular phase spectrum. Over the coming years, as the sample of MIR
spectra from {\it JWST} grows this relationship can be tested.

Unlike the traditional luminosity-width relation, determining 
whether a \snia meets the criteria for inclusion on the MIR 
luminosity-width plot (\autoref{fig:mir_wlr}) is less straightforward.
In order for such a measurement to accurately probe
this relation, the SN must be in the nebular phase with little to no
contamination from an underlying photosphere, \Cofs decay must be the
dominant source of energy deposition in the ejecta, and the strength of
the weak blends of [\ion{Fe}{3}] and [\ion{Ni}{1}] should be low. In
the context of \Mch explosions, this phase will occur earlier in less
luminous objects and later in more luminous ones. The specific timing 
of this window will vary from object to object due to differences in the
physical conditions of the ejecta, but spectroscopic
observations targeting the $\sim$100--300~day window should meet these
conditions in most \sneia.

\xkq is one of the most extensively observed under-luminous \snia ever, 
with an early time data set starting hours after explosion and spanning 
from the ultraviolet to the near-infrared \citep{Pearson_etal_2023}. Yet, this 
extensive data set does not conclusively point towards a specific origin 
for \xkq. With the addition of the MIR observations presented here, a 
clearer picture emerges. The early time presence of carbon points towards 
unburnt material in the outer layers, which naturally arise in the \Mch 
model presented here. The lack of bi-modality in the 
[\ion{Co}{3}]~11.888~\mic line and the MIR electron capture lines 
(e.g. \Nife) located interior to the \Nifs region disfavor a collision; 
The lack of late-time O lines seen in the MIR, disfavors a merger;
and the presence of strong electron capture lines favors high-density 
burning in the ejecta. This highlights the importance of multi-epoch, 
multi-wavelength observations of \sneia to determine the specific origins 
of individual objects.

Finally, we  mention some limitations that will be
overcome in the future. {\it JWST} is still in the early stages of
operation. The data reduction and calibration will improve, in
particular at longer wavelengths. Moreover, additional later-time
spectra will be needed to see the innermost region.
Due to both these effects, we could only obtain a lower limit
upon electron capture elements such as Cr, Mn that have strong transitions
beyond 18~\mic. Detecting these lines would further
refine the central density and the degree of mixing as these lines
will only be present in models with $\rho_c \gtrsim 4 \times
10^{9}$~\gcm. Estimates of the cross-sections for many transitions of iron
group elements will be obtained by calibration of bright
\sneia, as the cross-sections are not available for many transitions,
for example those of [\ion{Ti}{2}].

\section{Conclusion} \label{sec:conc}

We present a MIRI MIR spectrum of the under-luminous \snia, \xkq,
obtained at $+114.2$~d past maximum light. These are the first
\textit{JWST} observations of an under-luminous \snia and the 
first lines identified beyond 14~\mic in \sneia observations. The 
data show evidence of isolated narrow lines of stable Ni emission, and
strong emission from Ar and Co. In contrast to normal-luminosity \sneia,
\xkq shows lower ionization states; in particular, it contains more
[\ion{Ar}{2}] relative to [\ion{Ar}{3}]. Furthermore, the velocity
extent of the resonance [\ion{Co}{3}] feature is narrower compared to
more luminous \sneia observed at roughly the same phase. This indicates
the \Nifs\ is closer to the core of the explosion. We also find a
potential correlation between the FWHM of the [\ion{Co}{3}] resonance
feature and the light curve shape parameter \DmB. Observing more \sneia
at this phase will allow us to understand if this follows the 
luminosity-width relation for SNe Ia. 

By comparing the data to newly synthesized non-LTE models we determine
that the presence of narrow forbidden Ni lines is strong evidence for
the explosion of a massive WD. The presence of multiple [\ion{Ti}{2}]
lines is consistent with \xkq's location at the low luminosity end of 
this class. We show
that the MIR spectra and the absolute flux can be well understood
within an off-center delayed-detonation scenario seen equator on with
a peak luminosity $\sim 1.2$~mag below normal \sneia. The spectra
also hint that the SN had a central thermodynamic runaway and only
a moderate off-set of the DDT. The strong IME elements and low
luminosity suggest, at least in some \sneia, the lower luminosity
is not correlated with a lower total WD mass.  

This work demonstrates the power of using {\it JWST} data in 
combination with detailed simulations; especially, the wavelength
coverage beyond 14~\mic provided by MIRI/MRS. Looking ahead, 
harnessing this capability through forthcoming observations 
promises to usher us into a new era of \sneia physics.

\begin{acknowledgments}
J.D., C.A, P.H., E.B., and M.S. acknowledge support by NASA grants
JWST-GO-02114, JWST-GO-02122, JWST-GO-04436 and , JWST-GO-04522. 
Support for programs \#2114, \#2122, \#4436, and \#4522 were 
provided by NASA through a grant from the Space Telescope Science 
Institute, which is operated by the Association of Universities for 
Research in Astronomy, Inc., under NASA contract NAS 5-03127. 
P.H. and E.B. acknowledge support by NASA grant 80NSSC20K0538. 
P.H. acknowledges support by NSF grant AST-2306395.

I.D. is partially supported by the project PID2021-123110NB-I00 
financed by MCIN/AEI /10.13039/501100011033 / FEDER, UE.
L.G. acknowledges financial support from the Spanish Ministerio 
de Ciencia e Innovaci\'on (MCIN), the Agencia Estatal de 
Investigaci\'on (AEI) 10.13039/501100011033, and the European Social 
Fund (ESF) "Investing in your future" under the 2019 Ram\'on y Cajal 
program RYC2019-027683-I and the PID2020-115253GA-I00 HOSTFLOWS project, 
from Centro Superior de Investigaciones Cient\'ificas (CSIC) under 
the PIE project 20215AT016, and the program Unidad de Excelencia 
Mar\'ia de Maeztu CEX2020-001058-M.
K.M. is supported by STFC grant ST/T506278/1.
M.D.S. is funded by the Independent Research Fund Denmark 
(IRFD, grant number 10.46540/2032-00022B.

This work is based on observations made with the NASA/ESA/CSA 
James Webb Space Telescope. The data were obtained from the Mikulski 
Archive for Space Telescopes at the Space Telescope Science Institute, 
which is operated by the Association of Universities for Research 
in Astronomy, Inc., under NASA contract NAS 5-03127 for JWST. These 
observations are associated with program \#2114.
The specific observations analyzed in this work can be accessed via
\dataset[DOI:10.17909/kfvh-wb96]{\doi{10.17909/kfvh-wb96}}. 

\end{acknowledgments}

\vspace{5mm}
\facilities{JWST(MIRI/MRS). The simulations have been performed on the 
Beowulf system of the Astrophysics group at Florida State University.}

\software{HYDRA\citep{Hoeflich2003,Hoeflich2009,Penney2014,Hoeflich_2017,2021ApJ...923..210H},
          \texttt{SNooPy} \citep{Burns2011,Burns2014},
          \texttt{spextractor} \citep{Burrow2020},
          Astropy \citep{astropy:2013, astropy:2018, astropy:2022},
          NumPy \citep{numpy2020}, SciPy \citep{SciPy2020}, 
          Matplotlib \citep{matplotlib},
          \texttt{dust-extinction} \citep{Gordon2023b,Gordon2023a}}

\clearpage

\bibliographystyle{aasjournal}
\bibliography{ms}


\begin{appendix}

\section{Optical lines at day 130 after the explosion} 

\begin{deluxetable}{rcl|rcl|rcl|rcl|rcl}[h]
  \tablecaption{Optical Line Identifications at Day 130 in Reference Model \label{tab:opt_lines}}
  \tablehead{\colhead{\bf S} & \colhead{$\lambda$~[\mic]} & \colhead{Ion} 
    & \colhead{\bf S} & \colhead{$\lambda$~[\mic]} & \colhead{Ion}
    & \colhead{\bf S} & \colhead{$\lambda$~[\mic]} & \colhead{Ion}
    & \colhead{\bf S} & \colhead{$\lambda$~[\mic]} & \colhead{Ion}
    & \colhead{\bf S} & \colhead{$\lambda$~[\mic]} & \colhead{Ion}}
    \startdata
      & 0.3689 & {[\ion{Co}{2}]} & & 0.4641 & {[\ion{Fe}{2}]} & & 0.5562 & {[\ion{Co}{2}]} & & 0.8030 & {[\ion{Co}{2}]} & & 0.9697 & {[\ion{Co}{1}]} \\
      & 0.4104 & {[\ion{Co}{2}]} & {$\ \ \ast\ $} & 0.4659 & {[\ion{Fe}{3}]} & {$\ \ \ast\ $} & 0.5890 & {[\ion{Co}{3}]} & {$\ \ast\ \ast\ $} & 0.8123 & {[\ion{Co}{2}]} & & 0.9705 & {[\ion{Fe}{3}]} \\
      & 0.4154 & {[\ion{Co}{2}]} & & 0.4703 & {[\ion{Fe}{3}]} & & 0.5908 & {[\ion{Co}{3}]} & & 0.8123 & {[\ion{Co}{2}]} & & 0.9946 & {[\ion{Co}{2}]} \\
      & 0.4178 & {[\ion{Fe}{2}]} & & 0.4729 & {[\ion{Fe}{2}]} & & 0.6072 & {[\ion{Co}{1}]} & & 0.8303 & {[\ion{Ni}{2}]} & {$\ast\ \ast\ \ast\ $} & 0.9946 & {[\ion{Co}{2}]} \\
      {$\ \ \ast\ $} & 0.4245 & {[\ion{Fe}{2}]} & & 0.4735 & {[\ion{Fe}{3}]} & & 0.6129 & {[\ion{Co}{3}]} & & 0.8336 & {[\ion{Co}{2}]} & {$\ast\ \ast\ \ast\ $} & 1.0191 & {[\ion{Co}{2}]} \\
      & 0.4246 & {[\ion{Fe}{2}]} & & 0.4749 & {[\ion{Co}{2}]} & & 0.6197 & {[\ion{Co}{3}]} & & 0.8466 & {[\ion{Co}{2}]} & {$\ \ast\ \ast\ $} & 1.0248 & {[\ion{Co}{2}]} \\ 
      & 0.4278 & {[\ion{Fe}{2}]} & & 0.4756 & {[\ion{Fe}{3}]} & & 0.6578 & {[\ion{Co}{3}]} & & 0.8469 & {[\ion{Co}{2}]} & & 1.0283 & {[\ion{Co}{2}]} \\
      {$\ \ \ast\ $} & 0.4289 & {[\ion{Fe}{2}]} & & 0.4771 & {[\ion{Fe}{3}]} & & 0.6586 & {[\ion{Co}{1}]} & & 0.8502 & {[\ion{Ni}{3}]} & {$\ \ \ast\ $} & 1.0283 & {[\ion{Co}{2}]} \\
      & 0.4307 & {[\ion{Fe}{2}]} & & 0.4804 & {[\ion{Co}{2}]} & & 0.6669 & {[\ion{Ni}{2}]} & & 0.8546 & {[\ion{Co}{1}]} & & 1.0611 & {[\ion{Fe}{3}]} \\
      & 0.4321 & {[\ion{Fe}{2}]} & & 0.4882 & {[\ion{Fe}{3}]} & & 0.6855 & {[\ion{Co}{3}]} & {$\ \ \ast\ $} & 0.8574 & {[\ion{Co}{2}]} & & 1.0718 & {[\ion{Ni}{2}]} \\
      & 0.4327 & {[\ion{Ni}{2}]} & & 0.4891 & {[\ion{Fe}{2}]} & & 0.6934 & {[\ion{Co}{2}]} & & 0.8583 & {[\ion{Co}{2}]} & & 1.0718 & {[\ion{Ni}{2}]} \\ 
      & 0.4348 & {[\ion{Fe}{2}]} & & 0.4932 & {[\ion{Fe}{3}]} & & 0.7138 & {[\ion{Ar}{3}]} & & 0.8597 & {[\ion{Co}{1}]} & {$\ \ast\ \ast\ $} & 1.0824 & {[\ion{S}{1}]} \\ 
      & 0.4354 & {[\ion{Fe}{2}]} & {$\ \ \ast\ $} & 0.5013 & {[\ion{Fe}{3}]} & & 0.7155 & {[\ion{Co}{3}]} & {$\ \ast\ \ast\ $} & 0.8619 & {[\ion{Fe}{2}]} & & 1.0885 & {[\ion{Fe}{3}]} \\ 
      & 0.4360 & {[\ion{Fe}{2}]} & & 0.5086 & {[\ion{Fe}{3}]} & {$\ \ast\ \ast\ $} & 0.7157 & {[\ion{Fe}{2}]} &  {$\ \ \ast\ $} & 0.8894 & {[\ion{Fe}{2}]} & {$\ \ \ast\ $} & 1.0976 & {[\ion{Co}{2}]} \\ 
      & 0.4361 & {[\ion{Fe}{2}]} & & 0.5113 & {[\ion{Fe}{2}]} & {$\ \ \ast\ $} & 0.7174 & {[\ion{Fe}{2}]} & & 0.9036 & {[\ion{Fe}{2}]} & & 1.0994 & {[\ion{Si}{1}]} \\ 
      & 0.4374 & {[\ion{Fe}{2}]} & {$\ \ \ast\ $} & 0.5160 & {[\ion{Fe}{2}]} & & 0.7249 & {[\ion{Co}{1}]} & & 0.9054 & {[\ion{Fe}{2}]} & {$\ \ \ast\ $} & 1.1283 & {[\ion{Co}{2}]} \\ 
      & 0.4415 & {[\ion{Fe}{2}]} & & 0.5222 & {[\ion{Fe}{2}]} & {$\ \ast\ \ast\ $} & 0.7293 & {[\ion{Ca}{2}]} & & 0.9071 & {[\ion{S}{3}]} & {$\ \ \ast\ $} & 1.1309 & {[\ion{S}{1}]} \\ 
      & 0.4418 & {[\ion{Fe}{2}]} & {$\ \ \ast\ $} & 0.5263 & {[\ion{Fe}{2}]} & {$\ \ast\ \ast\ $} & 0.7326 & {[\ion{Ca}{2}]} & & 0.9229 & {[\ion{Fe}{2}]} & & 1.1616 & {[\ion{Ni}{2}]} \\ 
      & 0.4453 & {[\ion{Fe}{2}]} & & 0.5270 & {[\ion{Co}{2}]} & {$\ \ast\ \ast\ $} & 0.7380 & {[\ion{Ni}{2}]} & & 0.9270 & {[\ion{Fe}{2}]} & & 1.2489 & {[\ion{Fe}{2}]} \\ 
      & 0.4459 & {[\ion{Fe}{2}]} & {$\ \ \ast\ $} & 0.5272 & {[\ion{Fe}{3}]} & & 0.7390 & {[\ion{Fe}{2}]} & {$\ \ast\ \ast\ $} & 0.9339 & {[\ion{Co}{2}]} & & 1.2525 & {[\ion{Fe}{2}]} \\ 
      & 0.4476 & {[\ion{Fe}{2}]} & & 0.5298 & {[\ion{Fe}{2}]} & {$\ \ \ast\ $} & 0.7414 & {[\ion{Ni}{2}]} & {$\ast\ \ast\ \ast\ $} & 0.9345 & {[\ion{Co}{2}]} & {$\ast\ \ast\ \ast\ $} & 1.2570 & {[\ion{Fe}{2}]} \\ 
      & 0.4490 & {[\ion{Fe}{2}]} & & 0.5335 & {[\ion{Fe}{2}]} & {$\ \ \ast\ $} & 0.7455 & {[\ion{Fe}{2}]} & & 0.9447 & {[\ion{Fe}{3}]} & {$\ \ \ast\ $} & 1.2707 & {[\ion{Fe}{2}]} \\ 
      & 0.4494 & {[\ion{Fe}{2}]} & & 0.5378 & {[\ion{Fe}{2}]} & & 0.7541 & {[\ion{Co}{2}]} & & 0.9474 & {[\ion{Fe}{2}]} & {$\ \ \ast\ $} & 1.2791 & {[\ion{Fe}{2}]} \\ 
      & 0.4501 & {[\ion{Co}{3}]} & & 0.5414 & {[\ion{Fe}{3}]} & & 0.7640 & {[\ion{Fe}{2}]} & & 0.9533 & {[\ion{S}{3}]} & {$\ \ \ast\ $} & 1.2946 & {[\ion{Fe}{2}]} \\ 
      & 0.4608 & {[\ion{Fe}{3}]} & & 0.5472 & {[\ion{Co}{2}]} & & 0.7689 & {[\ion{Fe}{2}]} & & 0.9642 & {[\ion{Co}{2}]} & & 1.2981 & {[\ion{Fe}{2}]} \\ 
      & 0.4624 & {[\ion{Co}{2}]} & & 0.5548 & {[\ion{Co}{2}]} & & 0.7892 & {[\ion{Ni}{3}]} & {$\ \ \ast\ $} & 0.9642 & {[\ion{Co}{2}]} & & & \\
    \enddata
\end{deluxetable}

\end{appendix}

\end{document}